\documentclass[acmsmall]{acmart}

\usepackage{nicefrac,mathtools}
\usepackage{algorithmic}
\usepackage[vlined]{algorithm2e}
\usepackage{textcomp}
\usepackage[tight,FIGBOTCAP]{subfigure}
\usepackage{listings}
\usepackage{hyperref}
\usepackage{multicol}
\usepackage{multirow}

\AtBeginDocument{%
  \providecommand\BibTeX{{%
    \normalfont B\kern-0.5em{\scshape i\kern-0.25em b}\kern-0.8em\TeX}}}

\setcopyright{acmcopyright}
\copyrightyear{2018}
\acmYear{2018}
\acmDOI{10.1145/1122445.1122456}

\acmJournal{JACM}
\acmVolume{37}
\acmNumber{4}
\acmArticle{111}
\acmMonth{8}

\newcommand{\modes}[0]{MoDeS3}
\newcommand{\mrt}[0]{\textsf{\textbf{models@run.time}}}

\newcommand{\infigure}[1]{{\upshape\textsf{#1}}}
\newcommand{\dslStyle}[1]{\texttt{\color{emphColor}{#1}}}

\newcommand{\dslObjectSet}[0]{\mathcal{O}}

\newcommand{\interpretation}[2]{{\llbracket #1 \rrbracket}^{#2}}

\newcommand{\dslObjectSetI}[1]{\dslObjectSet_{#1}}
\newcommand{\Obj}{\dslObjectSetI}
\newcommand*{\ScopeSymbol}{\mathcal{S}}
\newcommand*{\Scope}[1]{\ScopeSymbol_{#1}}
\newcommand*{\lVar}[1]{\mathfrak{#1}}
\newcommand*{\LVars}{\mathcal{X}}

\newcommand*{\abs}{\mathit{abs}}
\newcommand*{\refined}{\succcurlyeq}
\newcommand{\symbolClass}[2]{\dslStyle{#1}_{\color{emphColor}{#2}}}
\newcommand{\symbolRef}[2]{\dslStyle{#1}_{\color{emphColor}{#2}}}
\newcommand{\symbolExists}{\upvarepsilon}
\newcommand{\symbolEquals}{\sim}
\newcommand{\interpretationFun}[1]{\mathcal{I}_{#1}}
\newcommand{\Int}{\interpretationFun}
\newcommand{\true}[0]{1}
\newcommand{\false}[0]{0}
\newcommand{\unk}[0]{\nicefrac{1}{2}}
\newcommand{\interpretationVariable}[3]{\interpretation{#1}{#2}_{#3}}

\newcommand{\bb}[1]{\mathit{bb}_{#1}}
\newcommand*{\Path}{\pi}
\DeclareMathOperator{\Ran}{\operatorname{Ran}}

\definecolor{keywordcolor}{rgb}{0.5,0,0.1}
\definecolor{commentcolor}{rgb}{0,0.3,0.1}
\definecolor{stringcolor}{rgb}{0,0,1}
\definecolor{emphColor}{rgb}{0,0,0}
\definecolor{codecommentcolor}{rgb}{0.03,0.6,0}
\definecolor{codegreen}{rgb}{0,0.6,0}
\definecolor{codegray}{rgb}{0.5,0.5,0.5}
\definecolor{codepurple}{rgb}{0.58,0,0.82}

\lstdefinelanguage{vql}
{   
    morekeywords={@QueryBasedFeature,@Constraint,count,pattern,private, neg,find,import,true,false,or,check,job,action,state,severity,message,oclIsKindOf,self,exists,includes,invariant,class},
    sensitive=true, 
    morecomment=[l]{//}, 
    morecomment=[s]{/*}{*/},
	  morestring=[b]{"},
}
\lstdefinestyle{mystyle}{
    commentstyle=\color{codegreen},
    keywordstyle=\color{magenta},
    numberstyle=\tiny\color{codegray},
    stringstyle=\color{codepurple},
    basicstyle=\ttfamily\footnotesize,
    breakatwhitespace=false,
    breaklines=true,
    captionpos=b,
    keepspaces=true,
    numbers=left,
    numbersep=5pt,
    showspaces=false,
    showstringspaces=false,
    showtabs=false,
    tabsize=2
}
\theoremstyle{definition}
\newtheorem{proposition}{Proposition}[section]
\newtheorem{lemma}{Lemma}[section]

\begin{document}

\title[Worst-Case Execution Time Calculation for Query-Based Monitors by Witness Generation]{Worst-Case Execution Time Calculation for\\Query-Based Monitors by Witness Generation}

\author{M{\'a}rton B{\'u}r}
\email{marton.bur@mail.mcgill.ca}
\orcid{0000-0003-2702-6174}
\affiliation{%
  \institution{McGill University}
  \streetaddress{3480 Rue University}
  \city{Montreal}
  \state{Quebec}
  \postcode{H3A 2K6}
  \country{Canada}
}

\author{Krist{\'o}f Marussy}
\email{marussy@mit.bme.hu}
\orcid{0000-0002-9135-8256}
\affiliation{%
  \institution{Budapest University of Technology and Economics}
  \streetaddress{Magyar tud{\'o}sok k{\"o}r{\'u}tja 2}
  \city{Budapest}
  \postcode{1117}
  \country{Hungary}}

\author{Brett H. Meyer}
\email{brett.meyer@mcgill.ca}
\affiliation{%
  \institution{McGill University}
  \country{Canada}
}

\author{D{\'a}niel Varr{\'o}}
\email{daniel.varro@mcgill.ca}
\orcid{0000-0002-8790-252X}
\affiliation{%
  \institution{McGill University}
  \country{Canada}
}

\begin{abstract}
  Runtime monitoring plays a key role in the assurance of modern intelligent cyber-physical systems, which are frequently data-intensive and safety-critical. While graph queries can serve as an expressive yet formally precise specification language to capture the safety properties of interest, there are no timeliness guarantees for such auto-generated runtime monitoring programs, which prevents their use in a real-time setting. While worst-case execution time (WCET) bounds derived by existing static WCET estimation techniques are safe, they may not be tight as they are unable to exploit domain-specific (semantic) information about the input models. This~paper presents a semantic-aware WCET analysis method for data-driven monitoring programs derived from graph queries. The~method incorporates results obtained from low-level timing analysis into the objective function of a modern graph solver. This allows the systematic generation of input graph models up to a specified size (referred to as \textit{witness models}) for which the monitor is expected to take the most time to complete. Hence the estimated execution time of the monitors on these graphs can be considered as safe and tight WCET.\@ Additionally, we perform a set of experiments with query-based programs running on a real-time platform over a set of generated models to investigate the relationship between execution times and their estimates, and compare WCET estimates produced by our approach with results from two well-known timing analyzers, aiT and OTAWA.%

\end{abstract}


\maketitle

\section{Introduction}

Runtime monitoring has become a key technique in the assurance of safety-critical and intelligent cyber-physical systems (CPS) such as autonomous vehicles~\cite{Pek2020} (e.g., self-driving cars, drones) where traditional upfront design time verification is problematic due to the dynamically changing environment and the data-intensive nature of the system. 
However, it is an open challenge how to align real-time requirements with the ability to capture the highly dynamic operation context of the system. A promising line of research aims to leverage high-level, model-based techniques to manage system complexity and overcome current limitations~\cite{Jantsch2017,Tavcar2019}.

In traditional embedded systems, runtime monitoring programs are integral components of the system that analyze events and execution traces~\cite{Bartocci2018} in order to detect potentially critical situations that violate a requirement. Since this requires formal precision to capture safety requirements, logic-based formalisms (e.g., propositional logic, temporal logic) are frequently used to specify execution traces. 
Furthermore, monitoring programs can be automatically synthesized from such specifications that are ready to be used in traditional hard real-time systems without compromising task schedulability and real-time properties of the existing program~\cite{Pike2010,Havelund2002}.

However, existing runtime monitoring approaches used in safety-critical applications have certain limitations, which are increasingly problematic for the new generation of data-intensive, intelligent, and self-adaptive, yet safety-critical CPSs. First, the \emph{moderate expressiveness of the specification language}~\cite{Havelund2015} makes it difficult for engineers to capture and understand complex rules. Moreover, while \emph{safety-critical programs typically use statically allocated data with bounded input sizes} and they conservatively avoid many %
programming language constructs, dynamically evolving data and advanced language constructs are inherent parts of data-intensive programs.

Recent advances in runtime monitoring aim to overcome these limitations by (1) offering high-level and expressive query-based~\cite{BriandMonitors,FASE2018} or rule-based~\cite{Havelund2015} formalisms to capture the properties to be monitored, and (2) using runtime graph models as an in-memory knowledge base which capture dynamic changes in the system or its environment at a high-level of abstraction~\cite{FASE2018,Hartmann2019greycat}.
Such \emph{data-driven safety monitors} derived from high-level specifications can analyze aggregated changes triggered by complex sequences of atomic events by evaluating queries over a continuously evolving data model. 
For example, in the railway domain, queries can check if a path exists between two points along a railway track, or identify cargo waiting at stations for more than a specified duration. %
As such, query-based monitoring programs use a network of linked objects as data structures and \emph{exhibit heavily input-dependent and semantic-aware complex control and data flow}.

To enable the use of data-driven safety monitors in hard real-time systems, the computation of safe worst-case execution time (WCET) estimates is required. While recent research has investigated data-driven runtime monitors for intelligent and critical CPSs in a distributed environment~\cite{FASE2018,Hartmann2019greycat,BriandMonitors}, and various testing approaches have been proposed~\cite{BriandTest,Semerath2018}, the timeliness aspect of the problem has been neglected. In fact, only very few initial ideas are available~\cite{tichy2006story}, which suggest limiting the maximum graph size and  employing optimized query plans. 
A wide range of existing timing analysis techniques and tools (e.g., aiT~\cite{Ferdinand2004}, Chronos~\cite{Li2007}, OTAWA~\cite{Hugues2006}, and SWEET~\cite{Lisper2014}) can provide safe and tight WCET bounds for traditional critical embedded software. %
However, there is a high degree of inherent design-time uncertainty present in data-driven  monitoring programs. In particular, the unknown contents of the runtime knowledge graph capturing the system and its operating environment constitutes an enormously large input space which can compromise
the accuracy of existing techniques. 
Therefore, novel techniques are needed to complement existing WCET analysis techniques to efficiently incorporate domain-specific restrictions for program inputs on a high-level of abstraction and automatically incorporate this domain knowledge as flow facts.

In order to obtain safe and tight WCET bounds for data-driven runtime monitors, major challenges in timing analysis need to be tackled. 
(i) First, domain-specific flow constraints pose several complex restrictions on the program flow, but the respective flow facts need to be manually formulated and the program needs to be annotated by experts. Such additional program flow information largely helps to enhance the precision of safe WCET bounds in existing timing analyzers. However, there are no generally applicable methods to automatically obtain such flow facts by exploiting high-level, domain-specific information and constraints on program flow during timing analysis. Specifying the flow facts manually is highly error-prone and the resulting annotations need to be updated after subsequent modifications to the program~\cite{Abella2015}. %
(ii) Furthermore, runtime graph models have varying underlying structure and memory demands which makes WCET analysis problematic since the worst-case graph structure needs to be considered regardless of domain-specific constraints. While memory can be preallocated to allow the timing analyzer to produce WCET bounds~\cite{Herter2009}, but a WCET estimate from the size of the preallocated memory for graph data without appropriate flow facts would still be overly conservative.
Moreover, the contents of the runtime model are regularly updated at runtime. Therefore, value analysis has no upfront access (at design-time) to the data that the reserved memory space will store.
(iii) Finally, traditional WCET analysis challenges also need to be tackled: detailed information is required about the executable binary and execution platform, including precise memory, pipeline, and cache descriptions~\cite{Wilhelm2008}.%

\paragraph*{Contributions}
This paper aims to address WCET estimation in the challenging setting of query-based runtime monitoring programs. In particular, we present the following contributions.

\begin{enumerate}
    \item We adapt query-based runtime monitoring programs derived from high-level graph query specifications~\cite{FASE2018} to real-time platforms (\autoref{sec:query-program-structure}).
    
    \item We provide a novel high-level static analysis technique for query-based monitors to estimate execution time on a given runtime model. The approach provides precise flow facts by counting the number of basic block executions %
    during query evaluation w.r.t. the given model even if exact memory allocation information is unavailable. Moreover, it combines such flow facts with constraints obtained from existing low-level analysis tools (\autoref{sec:bb-predicates}).

    \item We estimate the WCET of query-based programs by estimating execution time over designated \emph{witness models}. Such witness models have the highest estimated execution times for graph query programs executing over any input models up to a predefined model size and domain-specific constraints, and they are derived by a state-of-the-art \emph{graph solver}~\cite{Semerath2018} (\autoref{sec:witness-generation}).

    \item We perform an extensive experimental assessment of query evaluation times over a variety of graph models executed on an industry-grade real-time platform, and we compare our WCET estimates with those provided by two popular timing analyzers (OTAWA and aiT)  (\autoref{sec:evaluation}). %
    
\end{enumerate}

\paragraph*{Novelty}

Our technique complements existing WCET estimation methods
by extracting program flow information from domain-specific constraints of the abstract input space on top of existing constraints derived by traditional timing analysis.
To our best knowledge, our approach is the first to provide safe and tight WCET bounds of real-time graph query programs
by abstraction refinement using state-of-the-art model generation techniques.
This enables the use of query-based runtime monitoring programs in a real-time context by providing safe WCET estimates for a desired set of input models satisfying domain-specific constraints. %

\section{Related Work}
\label{sec:relwork}

Numerous static and probabilistic WCET analysis methods have been discussed in surveys~\cite{Wilhelm2008, Kozyrev2016}. Abella et al.~\cite{Abella2015} compares the most common WCET estimation approaches for programs in real-time systems and highlights their strengths and limitations. 
Based on the categorization of approaches of this latter work, our approach is a \emph{high-level, static deterministic timing analysis (SDTA)} which provides \emph{safe} execution bounds for embedded programs executing complex graph queries. 
Furthermore, measurement-based WCET estimations~\cite{Wenzel2005,Law2016} and probabilistic methods~\cite{Hansen2009, Grosjean2012} are out of scope for our work. Nevertheless, we focus on \emph{semantic-aware WCET estimation}~\cite{Maiza2017}, which aims at providing safe and tight estimates for programs where there are some semantic limitations on the input data, which cannot be automatically explored and exploited by current analysis techniques, and often times manual annotations of the code are necessary. In our case, control flow is often constrained by complex rules which are based on various properties of graph models. %
We provide an overview of existing work related to \emph{graph-based programs}, \emph{runtime monitoring} and \emph{program flow analysis}.%

\paragraph*{\textbf{\upshape Graph query programs:} Existing platforms}
Graph models and queries have been often used in design models and tools of real-time systems~\cite{Jurjens2003, Giese2003, Burmester2004}. Furthermore, graph-based techniques are used in various IoT and edge computing applications~\cite{Cheng2021, Li2019}. However, due to the soft real-time requirements of such applications, the WCET analysis aspect is often neglected. The focus of our work is to provide safe and tight WCET of such programs, and thus extend their application area.  %

\paragraph*{\textbf{\upshape Graph query programs:} WCET estimation}
One of the few related works that investigates real-time properties of graph-based techniques is~\cite{Burmester2005worst}. Motivated by the expressiveness of \emph{story diagrams}~\cite{Fischer1998}, the authors evaluate the applicability of this high-level modeling formalism to recognize hazardous situations in real-time systems. Their work investigates worst-case execution times of imperative programs generated from such story diagrams by executing measurements of manually created worst-case inputs. In contrast, our work aims to automatically synthesize worst-case well-formed input models as part of static analysis. %

\paragraph*{\textbf{\upshape Runtime monitoring:} Hard real-time embedded systems} One of the earlier works in the field is The Temporal Rover~\cite{Drusinsky2000}. This framework can generate monitoring code from temporal logic formulae with low overhead, but the verification of properties is done in a large part on a powerful remote host, while our method does not rely on any external component.
The concept of \emph{predictable monitoring} was introduced in~\cite{Zhu2009} where static scheduling techniques were used to show that a monitor fits its allocated time frame, but the analysis of monitoring tasks is out of its scope which is the topic of this current paper. Finally, synchronous component execution and observable program states are the main assumptions made in~\cite{Pike2010} to support sampling-based monitoring of input streams in real-time systems, whereas our work targets monitors executing complex queries over a graph model capturing contextual information on a high-level of abstraction.

\paragraph*{\textbf{\upshape Runtime monitoring:} Real-time database queries} In real-time databases~\cite{Ozsoyoglu1995}, access to data has strict time constraints. The work in~\cite{Hou1989} presents a data sampling-based statistical method to evaluate aggregate queries in a database. There is a trade-off between time available for query execution and the precision of the estimate. Such estimations would not be acceptable in a monitoring setting where precise query results are expected. The real-time object-oriented database RODAIN~\cite{Taina1996}, which targets telecommunication applications, does not support \emph{hard real-time} transaction (i.e., query) types, because it is considered too costly for the target domain. However, our objective is exactly to provide such guarantees over graph models to support hard real-time applications.

\paragraph*{\textbf{\upshape Program flow analysis}} 
Timing analyzers for program flow analysis often employ some version of the implicit path enumeration technique (IPET)~\cite{Li1995}. The general idea behind this method is to use the  control flow graph (CFG) of the program to create an integer linear program (ILP) where each variable encodes the number of executions of a corresponding basic blocks, and the objective function is to maximize their total execution time. %
Besides the IPET method, several \emph{tree-based} methods exist which use a tree representation of the program (obtained from the source code or compiled binary) and apply some traversal to find the longest path in a program~\cite{Lim1995,Colin2002,Ballabriga2017}. %
In any case, the effectiveness of these methods rely on precise program \emph{flow facts} (e.g., loop bounds, infeasible paths) to be able to determine a safe and tight WCET estimate.
Although there are several advanced (semi-)automated techniques available today to derive additional constraints on the program flow and thus improve the precision of the WCET estimate~\cite{Gustafsson2006,Ermedahl2007loop,Chu2011,Knoop2013,Lisper2014}, there is still a significant manual effort needed to specify flow facts~\cite{Abella2015}. Most closely related to our current work is~\cite{Knoop2013}, which uses abstraction refinement to reduce WCET estimates by \emph{squeezing}. %
However, this approach cannot exclude longest execution paths from the program which are infeasible due to complex domain-specific constraints on the inputs.

\section{Query-Based Runtime Monitors}
\label{sec:query-based-monitoring}

In this section, we provide an informal overview of query-based runtime monitors, while their formal treatment is deferred to \autoref{sec:background}.

\subsection{Running Example: the \modes{} CPS Demonstrator}
\label{sec:example}
\begin{figure}[!b]
	\centering
	\includegraphics[width=.7\linewidth]{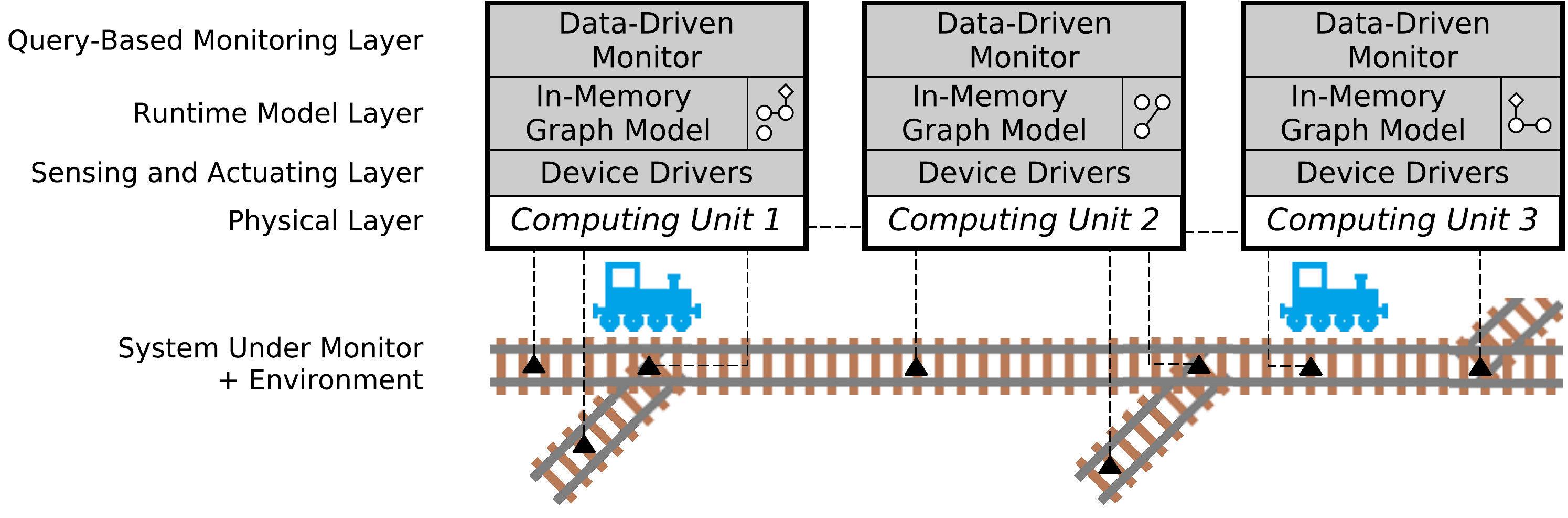}
	\caption{Runtime monitoring by graph queries}
	\label{fig:architecture-overview} 
\end{figure}

Our key concepts are illustrated %
in the context of the open source \emph{Model-Based De\-monstrator for Smart and Safe Cyber-Phys\-ical Systems} (\modes{})~\cite{NFM2018} platform, which showcases various  challenges of modern intelligent yet safety-critical CPS applications. The demonstrator (see \autoref{fig:architecture-overview}) is a model railway system with an added layer of safety to prevent train collision and derailment using runtime monitors. The railway track is equipped with several sensors and actuators, which are represented by black triangles in the lower part of \autoref{fig:architecture-overview}. Train shunt detectors can sense when trains pass by on a particular segment of the track, while direction of turnouts can be read and set.

The system is managed by a (distributed) monitoring service running on a network of heterogeneous computing units, such as Arduinos, Raspberry Pis, BeagleBone Blacks, etc. Relevant runtime information gained from sensor reads (e.g., the occupancy of a segment, or the status of a turnout) is uniformly captured in an in-memory \emph{runtime graph model}, which is also deployed on the platform.

Safety monitors are formally captured as \emph{graph queries}. %
Alerts from the monitoring services may trigger control commands of actuators (e.g., to change turnout direction) to guarantee safe operation.
The monitoring and control programs are running in a real-time setting on the computing units.

While the \modes{} platform can demonstrate various challenges of CPSs, this paper exclusively focuses on the real-time aspect of query-based runtime monitoring programs deployed to some embedded devices with limited resources (memory, CPU, etc).%

\subsection{Graph Models at Runtime}

\subsubsection{Graph Models}

The \mrt{} paradigm~\cite{BlairBF09} facilitates the capture of runtime knowledge about the system and its environment as a (typed and directed) \emph{graph model} continuously maintained at runtime for the system.
Such graphs are dynamically changing in-memory data structures which encode domain-specific instance models typed over a \emph{domain metamodel}, which captures core concepts (classes) in a domain and the relations (references) between those concepts.

\begin{figure}
	\begin{minipage}[b][][c]{.49\textwidth}
		\subfigure[Metamodel of the \modes{} domain%
		\label{fig:metamodel}]{\def\svgwidth{\linewidth}{
			\upshape
			\sffamily 
			\tiny
			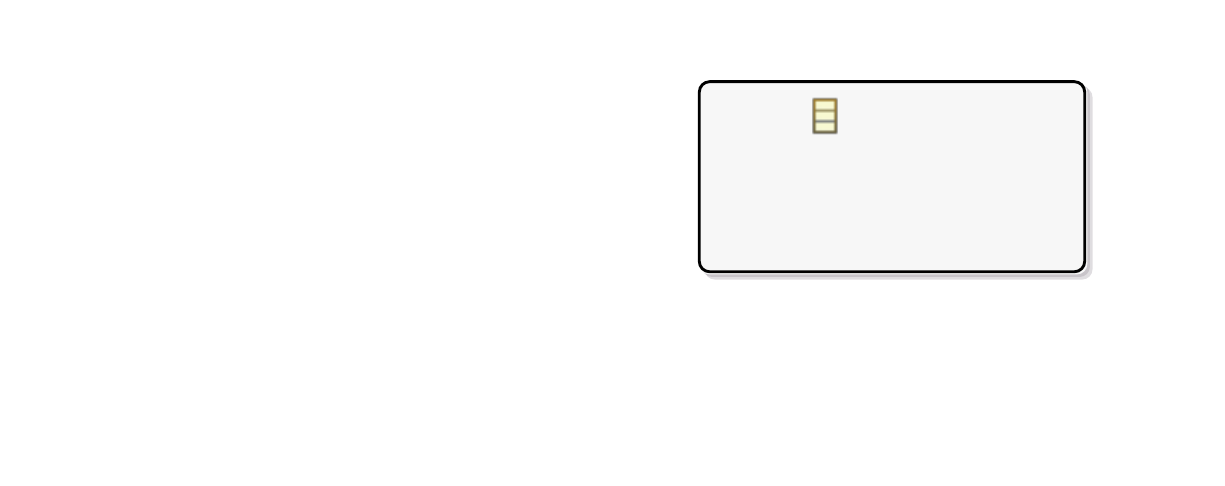}}
	\end{minipage}
	\hfill
	\begin{minipage}[b]{.49\textwidth}
		\subfigure[Concrete model representing a possible runtime snapshot%
		\label{fig:instance-model-modes3}]{\def\svgwidth{\linewidth}{
			\upshape
			\sffamily 
			\tiny
			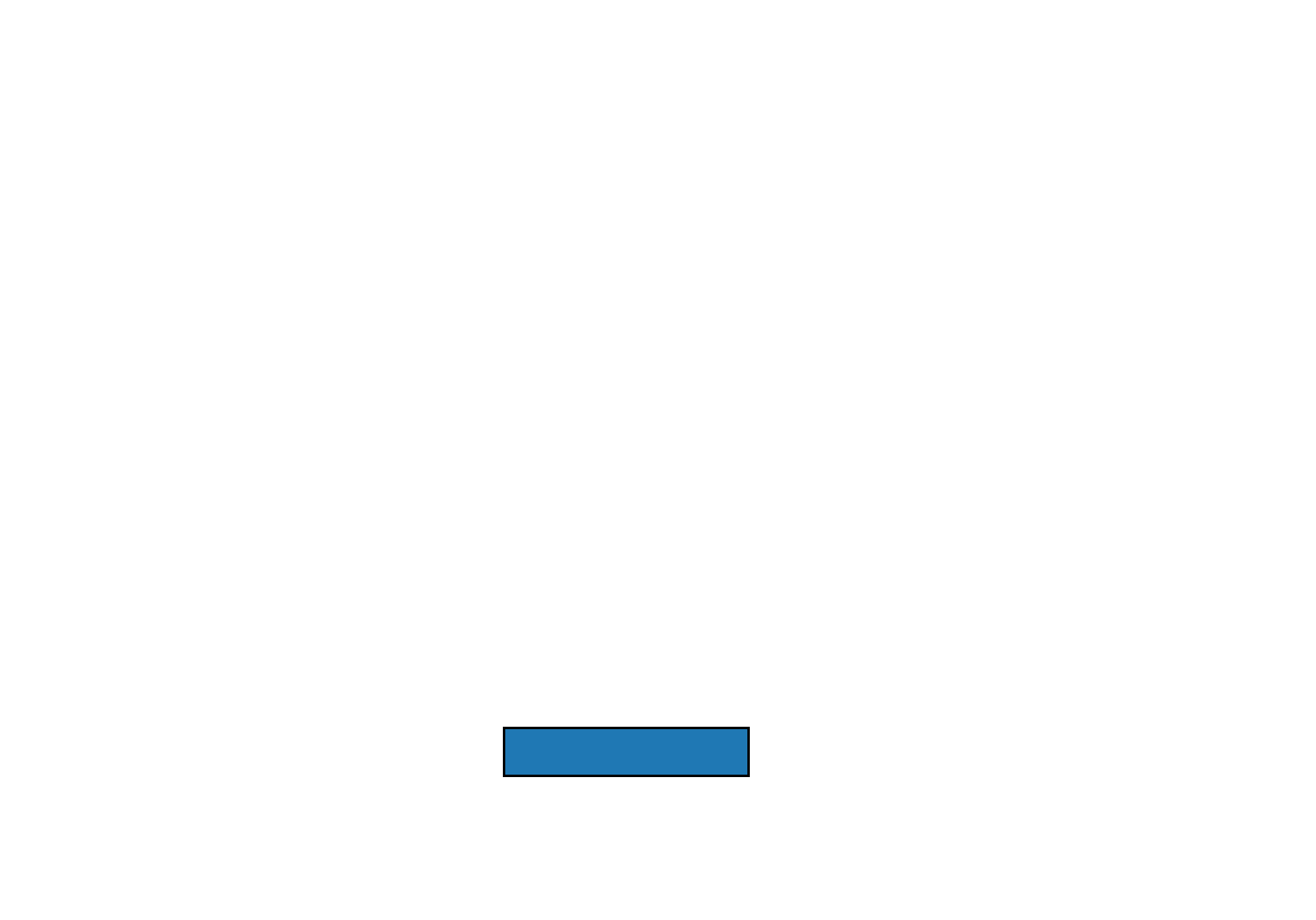}}
	\end{minipage}
	\hfill
	\caption{An instance (concrete) model and a partial model in the \modes{} domain}
	\label{fig:modes-models}
\end{figure}

\begin{example}
	The domain concepts of the \modes{} runtime model are captured in a metamodel excerpt shown in \autoref{fig:metamodel} using the Eclipse Modeling Framework (EMF) notation~\cite{EMF}. One domain concept is \infigure{Train}. Class \infigure{Segment} represents a section of the railway track with the \infigure{connectedTo} reference which describes what other segments it is linked to (up to two). 
	Moreover, each train maintains a \infigure{location} reference to a segment to describe its current position. Instances of class \infigure{Segment} record if they are occupied by a train with the \infigure{occupiedBy} reference. Moreover, \infigure{Turnout} is a special \infigure{Segment} that can change  its connections between \infigure{straight} and \infigure{divergent} segments. 
\end{example}

A runtime (instance) model captures a snapshot of the underlying system in operation~\cite{BlairBF09,Szvetits2013}. Relevant changes in the system are reflected in the runtime model and operations executed on the runtime model (e.g., setting values of controllable attributes of objects or updating links between objects) are reflected in the system itself (e.g., by executing scripts or calling services). In this work, we use \emph{concrete (graph) models} to formally capture runtime model snapshots.

\begin{example}
	\autoref{fig:instance-model-modes3} shows a concrete model in a graphical syntax. %
	The graph has six \infigure{Segment} objects (including two \infigure{Turnout}s) with their respective \infigure{connectedTo} links. %
	Turnouts  \(\mathsf{tu_0}\) and \(\mathsf{tu_1}\) can switch between segments \(\mathsf{s_1}\) and \(\mathsf{s_2}\) (see their \infigure{straight} and \infigure{divergent} edges). In the depicted state, both turnouts are switched to \(\mathsf{s_2}\) and the trains \(\mathsf{tr_0}\) and \(\mathsf{tr_1}\) are located on \(\mathsf{s_2}\) and \(\mathsf{s_2}\), respectively.
\end{example}

\subsubsection{Graph Data Structures in Embedded Systems}
\label{sec:graph-model-assumptions}

Runtime monitors captured by graph queries are continuously evaluated over the runtime models. %
This section informally summarizes our assumptions and requirements about such programs while the theoretical background is introduced in \autoref{sec:background}.

For data-driven monitors, the structure of the underlying graph model directly impacts the performance of query evaluation. Since an embedded device may have limited available CPU and memory resources, a lightweight data structure is needed to efficiently capture runtime graph models.
While the in-depth discussion of such a graph data structure is out of scope for this paper, we make the following assumptions about  the supported operations of the underlying graph:
\begin{itemize}
	\item \textbf{Dynamic element creation and deletion.} The runtime model serves as the knowledge base about the underlying system and its environment. For this reason, it needs to accommodate graph models without a theoretical a priori upper bound for model size. Based on~\cite{Herter2009}, one way to support this is to allocate the maximum amount of memory that is physically possible to be used for storing the graph. However, only the allocated memory is determined at compile time, the type (and distribution) of objects stored in the graph is runtime information. 
	\item \textbf{Indexing of objects by type using unique identifiers.} As query evaluation typically starts by iterating over all elements of a given type or accessing specific objects, it necessitates efficient object access, e.g. by maintaining a real-time index for memory resident data~\cite{Kong-RimChoi1996}.
	\item \textbf{Navigability along edges.} Query evaluation often navigates along the edges of selected objects to find further appropriate variable substitutions for unbound query variables. This feature can be supported by, e.g., maintaining direct pointers to reachable objects.
\end{itemize}

It is also important to note that the same graph model can be represented in memory in many ways, because different placements of the same data can cause different run times. For example, two memory images of the same graph may differ in the order the objects are stored in the array. For this reason, two different in-memory representations of the same graph may not necessarily yield identical run times, which must be considered when computing WCET of graph query programs. 

\begin{example}
    \autoref{lst:ds-segment-impl} and \autoref{lst:ds-train-impl} show a possible C implementation of data structures for \infigure{Segment} and \infigure{Train} classes in the metamodel of \autoref{fig:metamodel}. Line 2 in \autoref{lst:ds-segment-impl} and lines 2 and 3 in \autoref{lst:ds-train-impl} are fields created from respective attributes present in the metamodel, e.g., the \infigure{speed} attribute of class \infigure{Train} is represented by line 3. For each type, an \texttt{id} attribute encodes the type of the object for indexing and model manipulation purposes. Uniqueness of this attribute needs to be guaranteed at runtime to distinguish objects.
	Furthermore, in this example, we implement graph edges as pointers (line 4 in \autoref{lst:ds-train-impl}) or pointer arrays with sizes (lines 4 and 5 in \autoref{lst:ds-segment-impl}). Representing links between objects with pointers is highly efficient from a performance viewpoint.

	\begin{figure}
		\begin{minipage}[t]{.32\textwidth}
			\upshape %
			\small
\begin{lstlisting}[language=c,basicstyle=\ttfamily\scriptsize,caption={\infigure{Segment} class},label={lst:ds-segment-impl}]
typedef struct {
  uint16_t segment_id;
  Train *train;
  Segment *connected_to[2];
  uint8_t connected_to_count;
} Segment;
\end{lstlisting}
			\end{minipage}	
		\hfill
		\begin{minipage}[t]{.3\textwidth}
			\upshape %
			\small
\begin{lstlisting}[language=c,basicstyle=\ttfamily\scriptsize,caption={\infigure{Train} class},label={lst:ds-train-impl}]
typedef struct {
  uint16_t train_id;
  double speed;
  Segment *location;
} Train;
\end{lstlisting}
			\end{minipage}	
		\hfill
		\begin{minipage}[t]{.32\textwidth}
\begin{lstlisting}[language=c,basicstyle=\ttfamily\scriptsize,caption={Graph model root},label={lst:dg-graph}]
struct Modes3ModelRoot {
  Segment segments[SEGMENTS];
  uint16_t segment_count;
  Train trains[TRAINS];
  uint16_t train_count;
} runtime_model;
\end{lstlisting}
		\end{minipage} 
		\label{fig:graph-structure-impl}
	\end{figure}

	\autoref{lst:dg-graph} shows how a simple graph model container \texttt{Modes3ModelRoot}  can allocate static memory for graph objects in C. %
	The maximum memory used by the graph is preallocated in lines 2 and 4 by the \texttt{segments} and \texttt{trains} arrays which have a length of the  maximum expected number of trains (denoted by the constant \texttt{TRAINS}) and the maximum expected number of segments (\texttt{SEGMENTS}). 
    The \texttt{id} attribute of a given object used for indexing these arrays, i.e., encodes their positions in the arrays.

\end{example}

\subsection{Graph Query Programs in Real-Time Systems}
\label{sec:query-program-structure}

\emph{Data-driven runtime monitors} defined by graph queries can check structural properties of the runtime model representing a snapshot of the system. In other words, they focus on the most up-to-date data (maintained either by periodic updates with a certain frequency, or by certain event-driven triggers~\cite{BlairBF09,Szvetits2013}) available on the underlying system's state at a given point of time.

Classical \emph{event-based runtime monitors} rely on some temporal logic formalism to detect sequences of events occurring in the system at different points in time, while the underlying data model is restricted to atomic propositions. 
As such, data-driven and event-based monitors are complementary techniques.
While graph queries can be extended to express temporal behavior, our current work is restricted to (structural) safety properties where safety violations are expressible by graph queries.

\subsubsection{Graph Queries}

A \emph{graph query} is a declarative description of a model fragment to be identified by a set of variables and a set of constraints (type, reference, and equality assertions)~\cite{Varro2015}. A \emph{match} of the query is a binding of the query variables to objects in the model such that the constraints are satisfied.
In data-driven runtime monitors, such high-level descriptions allow us to automatically generate and optimize the monitor program by adaptive \emph{query planning}.

\begin{example}
	Trains are required to keep long headway distances to ensure that they can safely decelerate without collision~\cite{Emery2011}. The safety case captured by the \emph{closeTrains} (CT) graph query represents a situation when the headway distance between trains located on connecting segments is reduced below the safety limit. Any match of this query highlights segments where immediate action (e.g., stopping the trains) is required. The declarative query specification is presented in \autoref{fig:query-vql} in a textual syntax to identify the violating situation as a hazardous case. 
	The query returns pairs of segments \(s,e\) where there is a train located on segment \(s\) that is one segment away (i.e., there is a middle segment \(m\)) from a different segment \(e\), which is also occupied by a train. Any variables not appearing in the parameter list of the query are existentially quantified. 
	
	\autoref{fig:query-graphical} shows the same query in a graphical presentation often employed by modeling tools, and \autoref{fig:query-predicate} presents it as a first-order logic (FOL) formula \(\varphi_{\text{CT}}\) (discussed later in \autoref{sec:runtime-queries-background}).
	
	In the runtime snapshot \autoref{fig:instance-model-modes3}, the variable bindings \(\{s \mapsto \mathrm{s_2}, e \mapsto \mathrm{s_3}\}\) and \(\{s \mapsto \mathrm{s_3}, e \mapsto \mathrm{s_2}\}\) are matches of the \emph{closeTrains} query.

	\begin{figure}
	\begin{minipage}[b]{.40\textwidth}
		\upshape %
		\small
		\subfigure[Description of a query in VQL\label{fig:query-vql}]{
\begin{lstlisting}[language=vql]
pattern closeTrains(s, e) {
  Train.location(_t, s);
  Segment.connectedTo(s, m);
  Segment.connectedTo(m, e);
  Segment.occupiedBy(e, _ot);
  s != e
}
\end{lstlisting}}
	\end{minipage}		
	\hfill
	\begin{minipage}[c]{.54\textwidth}
		\subfigure[Graphical query presentation\label{fig:query-graphical}]{\def\svgwidth{\linewidth}{
		\upshape
		\footnotesize 
		\sffamily
		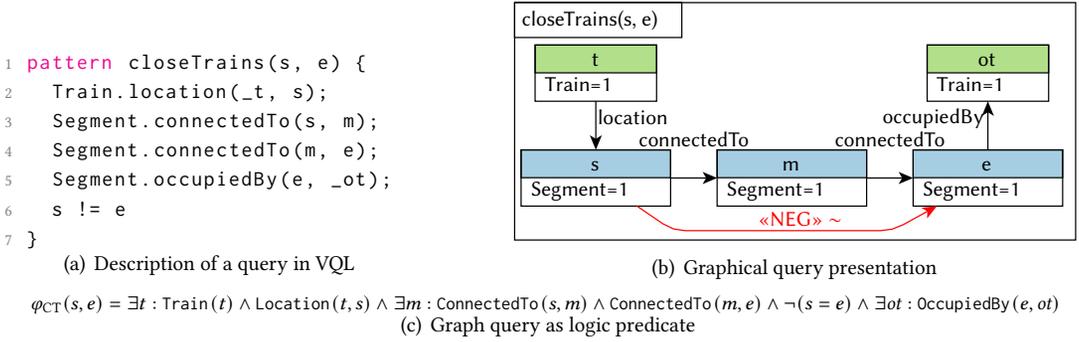
		}}
	\end{minipage}
		\begin{minipage}[c]{\textwidth}
			\centering
		{\scriptsize
		\subfigure[Graph query as logic predicate\label{fig:query-predicate}]{%
		\(\varphi_{\textnormal{CT}}(s, e) = \exists t : \symbolClass{Train}{}(t) \land \symbolRef{Location}{}(t, s)  \land \exists m : \symbolRef{ConnectedTo}{}(s, m) \land \symbolRef{ConnectedTo}{}(m,e) \land  \neg (s=e) \land \exists \mathit{ot} : \symbolRef{OccupiedBy}{}(e, \mathit{ot})\)
		}}
	\end{minipage} 
	\caption{Monitoring goal formulated as a graph query \(\varphi_{\textnormal{CT}}\) for \infigure{closeTrains}}
	\label{fig:query-defs}
	\end{figure}
\end{example}

\subsubsection{Local search-based graph query evaluation}

Among the many possible query evaluation strategies~\cite{Gallagher2006}, our runtime monitoring framework uses \emph{local search-based query evaluation}~\cite{Varro2015} to find matches of monitoring  over the entire runtime model. This strategy at its core uses a tailored depth-first search graph traversal. This keeps the memory footprint of the query evaluation algorithm constant. %
To obtain efficient performance at runtime, query evaluation is guided by a \emph{search plan}~\cite{Varro2015}, which maps each constraint in the query to a single pair of \(\langle \mathit{Step~index}, \mathit{Operation~type} \rangle\). In this tuple, \textit{Step index} specifies the order in which query evaluation should attempt to satisfy the respective constraint. \textit{Operation type} can be one of the followings: %
\begin{itemize}
	\item An \textbf{extend operation} evaluates a constraint with at least one free variable. Execution of such operations requires iterating over all potential variable substitutions and selecting the ones for which the constraint evaluates to $\true$.
	\item A \textbf{check operation} evaluates a constraint with only bound variables. Execution of such operations determines  if the constraint evaluates to $\true$ over the actual variable binding.
\end{itemize}

\begin{example}	
	\autoref{tab:close-trains-plan} shows a possible search plan for the \(\varphi_{\textnormal{CT}}\) query. Each row represents a search operation. The first column shows which constraint is enforced by the given step where free parameters at the start of the execution of the operation are underlined. The second column shows the ordering of steps, i.e., the step index, and the third column shows the search operation type (check or extend) which is based on the variable bindings prior to the execution of the search operation: if the constraint parameters are all bound, then it is a check, otherwise, it is an extend.

	\begin{table}[t]
		\begin{minipage}[t][][b]{.56\textwidth}
			\LinesNumbered
			\SetAlFnt{\scriptsize}
			\SetAlCapFnt{\small \normalfont \sffamily}
			\SetAlCapNameFnt{\small \normalfont \sffamily}
			\RestyleAlgo{ruled}
			\begin{algorithm}[H]
				\caption{Code generation from search plans}
				\label{alg:code-generation}
				\SetKw{TypeOf}{is}
				\SetArgSty{upshape}
				\SetKwProg{Fn}{Function}{ is}{end}
				\Fn{\upshape CompileSearchPlan(sp, idx) }{
					\lIf{idx $>$ sp.size()}{\Return{\emph{code for storing a match}}}
					step $=$ sp[idx]\\
					matcherCode $=$ ""\\
					\uIf{step \TypeOf extend}{
						\For{uv $\in$ step.getFreeVariables()}{
							matcherCode $+$$=$ \\\quad AddAssignmentFor(uv, step.getConstraintFor(uv))}
					} 
					\ElseIf{step \TypeOf check}{
						matcherCode $+$$=$ \\\quad AddIfFor(step.getAllVariables(), step.getConstraint())
					}
					\Return{ matcherCode $+$ CompileSearchPlan(sp, idx $+$ 1)}
				}
			\end{algorithm}
		\end{minipage}
		\hfill
		\begin{minipage}[t][][b]{.42\linewidth}
			\centering
			\caption{A possible search plan for query \infigure{close\-Trains} where free variables are underlined}
			\label{tab:close-trains-plan}
			{\footnotesize
			\begin{tabular}{@{}lcc@{}}
				\toprule%
				Constraint & Step\#& Op. type  \\ \midrule
				$\symbolClass{Train}{}(\mathit{\underline{t}})$ & 1 &extend\\ 
				$\symbolRef{Location}{}(t,\mathit{\underline{s}})$ & 2 & extend\\ 
				$ \symbolRef{ConnectedTo}{}(s,\mathit{\underline{m}})$ & 3 & extend\\ 
				$ \symbolRef{ConnectedTo}{}(m,\mathit{\underline{e}})$ & 4 & extend\\ 
				$\neg (s=e)$ & 5 & check\\ 
				$\symbolRef{OccupiedBy}{}(e,\mathit{\underline{ot}})$ & 6 &extend\\%
				\bottomrule
			\end{tabular}\par}
		\end{minipage}		
	\end{table}
\end{example}

\subsubsection{Implementations of Query Programs}\label{sec:query-based-monitoring:generated-code}

Although constructing effective search plans for graph queries is a complex challenge, it is outside of the scope of the current paper and has been formerly extensively studied (see, e.g.,~\cite{Varro2015} for a possible solution). However, we present pseudo-code that generates embedded query code from a search plan in \autoref{alg:code-generation}. The function \emph{CompileSearchPlan} takes a search plan and a search step index as parameters. Line 2 returns a code snippet to register a match if the provided index is beyond the index of the final search step. Otherwise, the search step is extracted (line 3) and the variable \emph{matcherCode} to hold the generated code is initialized to an empty string (line 4). 
Then, the different operation types of the query search plan are translated to structured imperative code:
\begin{itemize}
    \item Each \textbf{extend operation} binds all free variables of the respective constraint (lines 5--6). For each variable, this translates to either a single assignment or a \texttt{for} loop iterating over a set of candidate variable bindings, depending on the multiplicity of the respective navigation edge (reference constraint) (lines 7--8).
    \item Each \textbf{check operation} (line 9) is mapped to an \texttt{if} statement to check if the current variable binding satisfies a given condition created from the query constraint (lines 10--11).
\end{itemize}
Finally, in line 12, the generation continues recursively appending the code generated from the subsequent steps to the result.
The query code for the entire search plan \emph{sp} can be generated by calling  \emph{CompileSearchPlan$(\mathit{sp}, 1)$}. 
As a result, the source code contains a deep hierarchy of embedded for-loops and if-statements based on the ordering of constraints prescribed by the search plan. %

Besides obtaining a WCET, we also need to estimate the number of matches of a query to allocate appropriate space in memory in advance. In the case of runtime monitors of safety properties, we can assume that only a few violating matches will be detected~\cite{Varro2018}, thus the query result set is expected to be small and memory required for storing matches can be reserved at compile time.

\begin{example}
	\autoref{lst:ct-code} shows the C code generated from the query specification of  \infigure{closeTrains}. Assuming that a global variable \texttt{model} points to the root of the entire graph model including its up-to-date model statistics, calling the function \texttt{close\_trains\_matcher} with a 
	pointer to the result set structure \texttt{results} will compute and store all matches over the model in \texttt{results}. 

	In the example, initially all variables are assumed to be free, as indicated in line~2 with \texttt{NULL} values, because we aim to find all matches in the entire model. In line 3, the size of the result set is initialized to 0. The \texttt{for} loop in line 6 represents step 1 from the search plan (see \autoref{tab:close-trains-plan}) and iterates over all trains in the model, binding the variable \texttt{vars.t} to all possible objects in line 7. Lines 8--10 together represent search step 2. In line 9, \texttt{vars.s} is assigned a segment referred by \texttt{vars.t} via a single \texttt{location} link. If such a segment exists in line 10, execution continues with the third search operation that is mapped to lines 11--14, which iterates over segments connected to \texttt{vars.s} and assigns them to \texttt{vars.m}, one at a time. The next step in lines 15--18 does the same but with the connecting segments of \texttt{vars.m} and assigns them to \texttt{vars.e}. Search step 5 is a check, which is mapped to lines 19--20 to ensure that the segments referred by \texttt{vars.s} and \texttt{vars.e} are not the same. The final step of the search plan is mapped to lines 21--23. Here the train occupying the segment stored in \texttt{vars.e} is assigned to \texttt{vars.ot}. If such a train exists, a match is found and registered by assigning the corresponding variable values to parameter variables in a new match (lines 24--26) and incrementing the matches found counter \texttt{match\_cntr}. The execution concludes with saving the number of matches (line 28).
	\lstset{escapeinside={(*@}{@*)}}
	\begin{figure}[t]
		\begin{minipage}[t][][b]{.655\linewidth}
\begin{lstlisting}[language=c,basicstyle=\ttfamily\scriptsize,caption={Source code generated for query \infigure{closeTrains}},label={lst:ct-code}]
void close_trains_matcher(CloseTrainsMatchSet *results) {
  CTVars vars = {t=NULL, ot=NULL, s=NULL, m=NULL, e=NULL}
  int match_cntr = 0;
  // Constraint: (*@\color{codecommentcolor}\(\exists t: \mathtt{Train}(t)\)@*)
  int loop_bound0 = model->train_count;
  for (int i0 = 0; i0 < loop_bound0; i0++) {
    vars->t = model->trains[i0];
    // Constraint: (*@\color{codecommentcolor}\(\mathtt{Location}(t,s)\)@*)
    vars->s = vars->t->location;
    if (vars->s != NULL) {
      // Constraint: (*@\color{codecommentcolor}\(\exists m: \mathtt{ConnectedTo}(s,m)\)@*)
      int loop_bound1 = vars->s->connected_to_count;
      for (int i1 = 0; i1 < loop_bound1; i1++) {
        vars->m = vars->s->connected_to[i1];
        // Constraint: (*@\color{codecommentcolor}\(\mathtt{ConnectedTo}(m,e)\)@*)
        int loop_bound2 = vars->m->connected_to_count;
        for (int i2 = 0; i2 < loop_bound2; i2++) {
          vars->e = vars->m->connected_to[i2];
          // Constraint: (*@\color{codecommentcolor}\(\neg (s = t)\)@*)
          if (vars->s != vars->e) {
            // Constraint: (*@\color{codecommentcolor}\(\exists \mathit{ot} :\mathtt{OccupiedBy}(e,\mathit{ot})\)@*)
            vars->ot = vars->e->train;
            if (vars->ot != NULL) {
              // Register match
              results->matches[match_cntr].s = vars->s;
              results->matches[match_cntr++].e = vars->e;
  } } } } } }
  results->size = match_cntr; }
\end{lstlisting}
		\end{minipage}
		\hfill
		\begin{minipage}[t]{.27\textwidth}
				\vskip 6.25pt
				\centering 
            	\includegraphics[width=.99\linewidth]{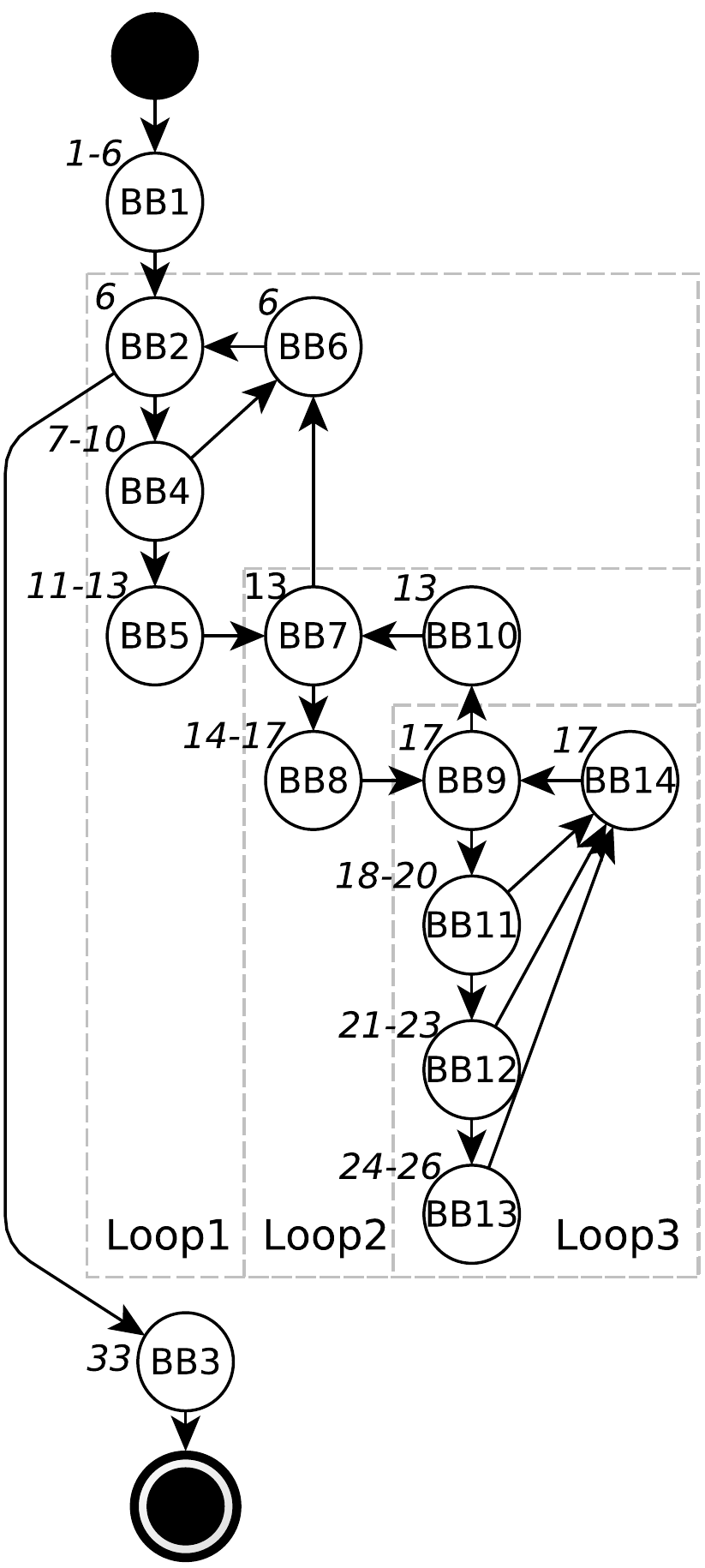}
				\caption{CFG of example query program}
				\label{fig:ct-cfg}
		\end{minipage}				
	\end{figure}
	
	Static analysis of the query code itself  in \autoref{lst:ct-code} would not impose any restrictions on line 20 despite the fact that the domain-specific constrains prescribe \infigure{connectedTo} links to be symmetrical. Therefore, at least every other execution of line 20 will jump back to line 17 instead of proceeding to line 22, yielding a flow constraint that is not discoverable by analysis of the code only.
\end{example} 

Cyclomatic complexity (CC) is frequently used as a metric in safety-critical software to estimate code complexity~\cite{Rierson2017}. As a general recommendation, code with high CC is traditionally avoided in a safety-critical system as it requires extra efforts to test and maintain. However, the derived imperative source code of data-driven monitoring programs is inherently complex even for small queries, which is largely attributed to the declarative nature of query specifications. For example, the CC of \autoref{lst:ct-code} is 7, which already indicates substantial complexity.

While modern WCET analyzers can  analyze complex code fragments, they heavily rely on manual annotation of the code (loop bounds, in particular) and design time information about variable values to be able to come up with an estimate that is both safe and tight. A key contribution of the current paper is to complement the existing WCET analysis by providing means to automatically exploit domain-specific restrictions of input data and tighten the resulting WCET estimate. For data-driven monitors, this is a key step in order to enable their use in a safety-critical context.

\section{Formal Background}\label{sec:background}

This section provides the formal background for the static analysis of data-driven runtime monitors, introduces definitions from traditional IPET-based approaches for WCET estimation, and revisits the state of the art of domain-specific graph modeling and graph model generation.

\subsection{Implicit Path Enumeration Technique for Estimating WCET}\label{sec:background:ipet}

Timing analysis frequently relies on the Implicit Path Enumeration Technique (IPET)~\cite{Li1995} that uses the control flow graph (CFG) of a program to estimate the WCET.\@ This section revisits definitions from~\cite{Puschner1997,Knoop2013} to introduce this classic WCET estimation approach.

\begin{definition}
    A \emph{program} is a pair \(\langle \mathit{BB}, \mathit{Loops} \rangle\), where \(\mathit{BB}\) is the set of \emph{basic blocks} and \(\mathit{Loops}\) is the set of (well-structured) loops. Each loop \(\ell \in \mathit{Loop}\) has a \emph{header} \(\bb{\ell,h} \in \ell\), while the rest of its blocks \(\bb{i} \in \ell\) constitute its \emph{body}.
\end{definition}

\begin{definition}
	 A \emph{weighted control flow graph} corresponding to a program \(\langle \mathit{BB}, \mathit{Loops} \rangle\) is a tuple \(\mathit{CFG} = \langle V, E, s, t, w, \mathit{tr} \rangle\), where
	 \begin{itemize}
	     \item \(V\) is a finite set of \emph{nodes};
	     \item \(E \subseteq V \times V\) is the set of \emph{edges};
	     \item \(s\) and \(t \in V\) are the \emph{program start} and \emph{end} nodes, respectively;
	     \item \(w\colon E \to \mathbb{N}\) is the \emph{weight function}, that assigns execution times (clock cycles) to the edges;
	     \item \(\mathit{tr}\colon V \to \mathit{BB}\) is the \emph{traceability} function that maps nodes to their originating program blocks.
	 \end{itemize}
\end{definition}

In the simplest case, \(V = \mathit{BB}\) and \(\mathit{tr}\) is the identity function. However, even on simple embedded processors, basic block execution times may vary due to microarchitectural effects (e.g., pipeline or cache state), which necessitates representing basic blocks with multiple nodes to encode context-sensitive execution times. The extended CFG encodes information from a \emph{low-level timing analysis}.
For example, the VIVU approach~\cite{Martin1998} addresses this concern by \emph{virtual loop unrolling} and creates additional nodes and edges in the CFG to explicitly model the first executions of loops.

\begin{definition}
    Every execution of the program (i.e., program trace) can be represented by a path \(\Path = \langle e_1,\ldots,e_m \rangle\) in the CFG from \(s\) to \(t\). %
    Let \(\Path\#e\) denote the frequency of the edge \(e\) appearing in \(\Path\).
    The program execution time \(\tau(\Path)\) can be estimated by the sum of the product of weights and frequencies of edges, i.e., \(\tau(\Path) = \sum_{e \in E} w(e) \cdot \Path\#e\). %
\end{definition}

The goal of timing analysis is to find the path with the longest possible  execution time of the program. Instead of explicitly enumerating all possible paths in the CFG, Puschner and Schedl~\cite{Li1995} create a system of linear inequalities whose solutions over-approximate the set of possible paths and define an integer linear programming (ILP) problem for estimating WCET.
First, we review some notations for systems of linear inequalities and ILP problems that will be used in this paper.

\begin{definition}
    Let \(\LVars = \{\lVar{x}_1, \ldots, \lVar{x}_{\lvert \LVars \rvert}\}\) be a large (but finite) \emph{reserve} of linear equation variable symbols.
    Then \(\ScopeSymbol = \bigl\{\sum_{\lVar{x}_{j} \in \LVars} a_{i,j} \cdot \lVar{x}_{j} \le y_{i} \bigr\}_{i=1}^{\lvert \ScopeSymbol \rvert}\) is a  \emph{system of linear equations}.
\end{definition}
    
\begin{definition}
	Let functions \(k\colon \LVars \to \mathbb{Z}\) be \emph{valuations}. A valuation \(k\) is a \emph{solution} of \(\ScopeSymbol\) (written as \(k \vDash \ScopeSymbol\)) if \(\sum_{\lVar{x}_j \in \LVars} a_{i, j} \cdot k(\lVar{x}_{j}) \le y_i\) for all inequalities \(\sum_{\lVar{x}_{j} \in \LVars} a_{i,j} \cdot \lVar{x}_{j} \le y_{i}\) in \(\ScopeSymbol\).
	The system of linear equations \(\Scope{1}\) \emph{entails} \(\Scope{2}\) (written as \(\Scope{1} \vDash \Scope{2}\)) if, for any solution \(k \vDash \Scope{1}\), we also have \(k \vDash \Scope{2}\).
\end{definition}

\begin{definition}
    An \emph{integer linear program} (ILP) is of the form \(\max \sum_{\lVar{x}_i \in \LVars} c_i \cdot \lVar{x}_i \textit{ subject to } \ScopeSymbol\), where \(\ScopeSymbol\) is a system of linear equations.
\end{definition}

While ILP with both maximization~(\(\max\)) and minimization~(\(\min\)) objectives can be considered, we restrict our attention to \(\max\) objectives w.l.o.g., since any \(\min\) objective can be transformed into a \(\max\) objective after multiplication by \(-1\).

\begin{definition}
    Let \(g(k) = \sum_{\lVar{x}_i \in \LVars} c_i \cdot k(\lVar{x}_i)\) denote the \emph{cost} of valuation \(k\). The valuation \(k^{*}\) is an \emph{optimal solution} of the ILP if \(k^{*} \vDash \ScopeSymbol\) and \(g(k^{*}) \ge g(k)\) for all \(k \vDash \ScopeSymbol\), i.e., its cost is maximal.
\end{definition}

An ILP can be derived from the CFG \(\langle V, E, s, t, \mathit{tr}\rangle\) as follows. Let \(\lVar{f}\colon E \to \LVars\) be a function that associates variable symbols to CFG edges. In the system of linear equations \(\ScopeSymbol\), each feasible execution path \(P\) is associated with a solution \(k_{\Path}\) such that \(k_{\Path}(\lVar{f}(e)) = \Path\#e\). Thus, linear constraints on variable obtained as \(\lVar{f}(e)\) restrict frequency of \(e\) in all feasible paths. Moreover, \(\sum_{e = \langle n_1, n_2 \rangle \in E, \mathit{tr}(n_1) = \bb{i}} k_{\Path}(\lVar{f}(e))\) is the number of times the basic block \(\bb{i} \in \mathit{BB}\) was executed in \(\Path\).
\begin{itemize}
    \item We add \(\sum_{e=\langle s,n\rangle \in E} \lVar{f}(e) = 1\) and \(\sum_{e=\langle n,t \rangle \in E} \lVar{f}(e) = 1\) to \(\ScopeSymbol\), because the program is entered and exited exactly once.
    \item Except for \(s\) and \(t\), each node is entered and exited the same number of times, so we add \(\sum_{e=\langle n_1,n_2 \rangle \in E} \lVar{f}(e) - \sum_{e=\langle n_2,n_3\rangle \in E} \lVar{f}(e) = 0\) for each \(n_2 \in V \setminus \{s, t\}\).
    \item Any additional \emph{flow facts} regarding the execution frequencies of program parts (e.g., loop execution counts) are added in the form \(\sum_{e_{i,j} \in E} a_{i,j} \cdot \lVar{f}(e_{i,j}) \leq y_j\).
 	\item For each edge \(e \in E\), we also have \(-\lVar{f}(e) \le 0\), since the execution frequency is non-negative.
\end{itemize}

We have an ILP \(\max \sum_{e \in E} w(e) \cdot \lVar{f}(e) \textit{ subject to } \ScopeSymbol\). 
The objective function \(g(k)\) overapproximates the execution time of \(P\). 
Therefore the value \(g(k^{*})\) of any solution \(k^{*}\) is a WCET bound.

\subsection{Metamodels and Partial Models}
\label{sec:runtime-queries-background}

In order to extend static analysis of data-driven graph query programs with domain-specific flow information, we will formally capture metamodels by a logic signature and their instance models as logic structures following~\cite{Semerath2018,Marussy2020}.

\begin{definition}
  A \emph{metamodel} is formally represented as a first-order logic signature \(\langle \Sigma, \alpha \rangle\), where
  \begin{itemize}
	\item \(\Sigma = \{ \symbolClass{C}{1}, \ldots, \symbolClass{C}{m_{\dslStyle{C}}}, \symbolRef{R}{1},\ldots, \symbolRef{R}{m_{\dslStyle{R}}}, \symbolExists, \symbolEquals\}\) is a finite set of \emph{symbols}, where \( \{\symbolClass{C}{i}\}_{i=1}^{m_{\dslStyle{C}}} \) are unary \emph{class symbols}, \(\{\symbolRef{R}{j}\}_{j=1}^{m_{\dslStyle{R}}} \) are binary \emph{relation symbols}, \(\symbolExists\) is the \emph{object existence} symbol, and \(\symbolEquals\) is the \emph{object equality}; 
	\item \(\alpha\colon \Sigma \to \mathbb{N}\) is the \emph{arity} function with \(\alpha(\symbolClass{C}{i}) = 1\) for all \(i = 1, \ldots, m_{\dslStyle{C}}\), \(\alpha(\symbolRef{R}{j}) = 2\) for all \(j = 1, \ldots, m_{\dslStyle{R}}\), \(\alpha(\symbolExists) = 1\), and \(\alpha(\symbolEquals) = 2\).
  \end{itemize}
\end{definition}

The definition of a metamodel may also include binary \emph{attribute symbols} such as in~\cite{FASE2018}. However, their handling is analogous to binary relation symbols, thus their discussion is excluded from here.

Partial models explicitly capture uncertainty in models as well as the design decisions yet to be made using 3-valued logic~\cite{Sagiv2002,Famelis2012}. In addition to the usual \(\true\) and \(\false\) truth values, the \(\unk\) truth value corresponds to uncertainties in the model. We also add systems of linear equations as \emph{scopes}~\cite{Marussy2020} to partial models to impose numerical constraints on the sizes of the models by \emph{polyhedron abstraction}. Later, variables \(\lVar{x}_j \in \LVars\) in the scopes will be connected to the number of model objects and graph pattern matches through \emph{theories} of 3-valued logic expressions.

\begin{definition}
  A \emph{scoped partial model} over a signature \(\langle \Sigma, \alpha \rangle\) is a triple \(P = \langle\Obj{P}, \Int{P}, \Scope{P}\rangle\), where
  \begin{itemize}
	\item \(\Obj{P}\) is a finite set of \emph{objects};
	\item \(\Int{P}\) is a 3-valued logical interpretation \(\Int{P}(\sigma)\colon \Obj{P}^{{\alpha(\sigma)}} \to \{\true, \false, \unk\}\) for all \(\sigma \in \Sigma\); and
	\item the \emph{scope} \(\Scope{P} = \bigl\{\sum_{\lVar{x}_{j} \in \LVars} a_{i,j} \cdot \lVar{x}_{j} \le y_{i} \bigr\}_{i=1}^{\lvert \Scope{P} \rvert}\) is a system of linear equations.
  \end{itemize}
\end{definition}

The existence symbol \(\symbolExists\) allows us to represent objects \(o\) that optionally appear in the model by setting \(\Int{P}(\symbolExists)(o) = \unk\). In contrast, objects with \(\Int{P}(\symbolExists)(o) = \true\) surely appear. Uncertain equality \(\Int{P}(\symbolEquals)(o, o) = \unk\) of an object \(o\) with itself denotes \emph{multi-objects} that can represent multiple concrete model objects. In contrast, objects with \(\Int{P}(\symbolEquals)(o, o)\) stand for single concrete model objects.

\begin{figure}[tb]
    \begin{minipage}{.48\linewidth}%
      {\def\svgwidth{\linewidth}{
          \upshape
          \sffamily
          \tiny
          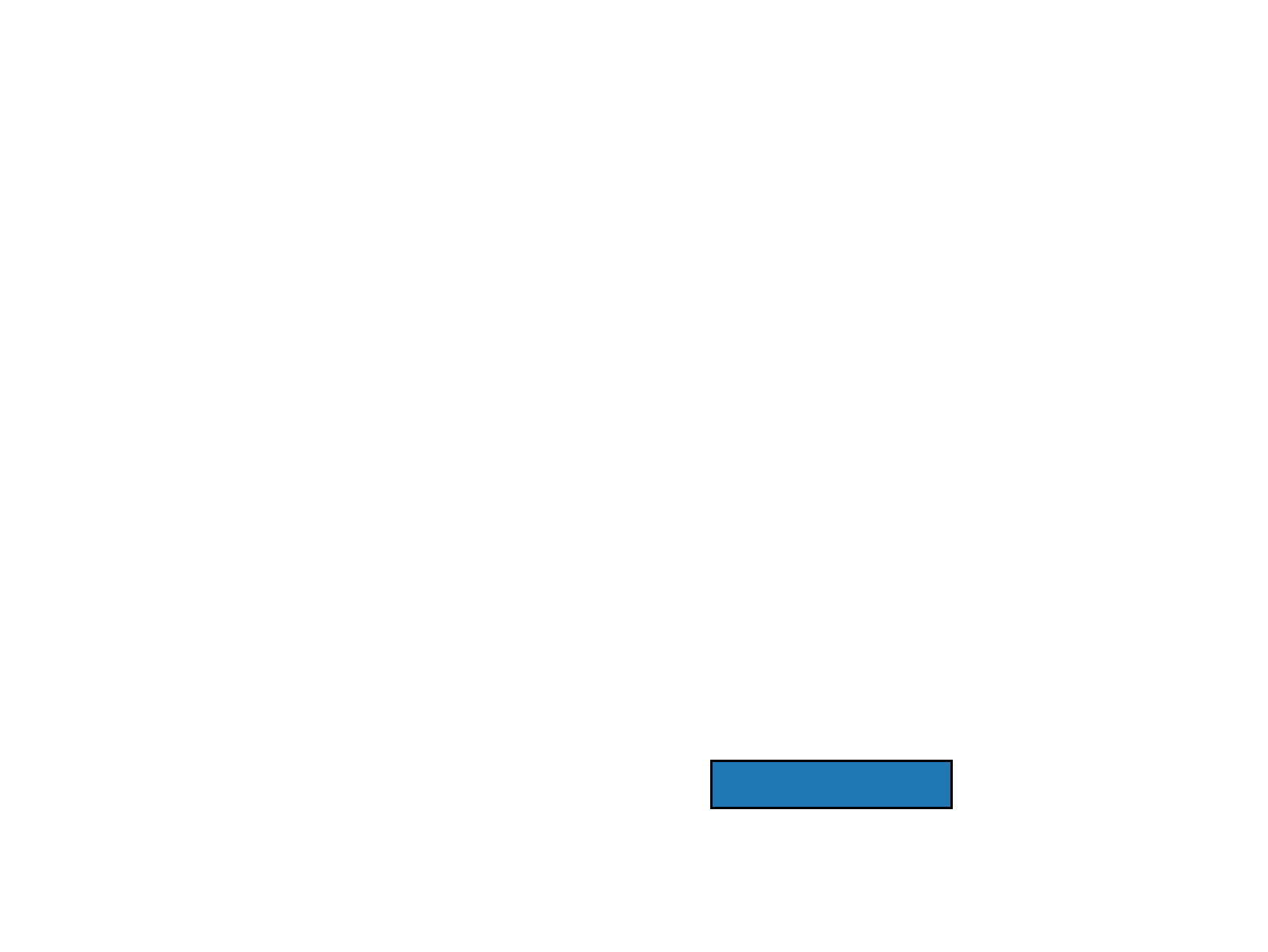}}
      \caption{Partial model with uncertain elements}\label{fig:partial-model-modes3}
    \end{minipage}\hfill
    \begin{minipage}{.48\linewidth}
      \setlength{\abovedisplayskip}{0pt}%
      \setlength{\belowdisplayskip}{0pt}%
      {\footnotesize
        \begin{align*}
          \interpretationVariable{1}{P}{Z}&\coloneq1
          &\interpretationVariable{\symbolClass{C}{i}(v)}{P}{Z}&\coloneq\Int{P}(\symbolClass{C}{i})(Z(v))\\
          \interpretationVariable{0}{P}{Z}&\coloneq0
          &\interpretationVariable{\symbolClass{R}{j}(v_1, v_2)}{P}{Z}&\coloneq\Int{P}(\symbolRef{R}{j})(Z(v_1), Z(v_2))\\
          \interpretationVariable{\neg \varphi}{P}{Z}&\coloneq\mathrlap{\true - \interpretationVariable{\varphi}{P}{Z}}\quad\qquad
          &\interpretationVariable{v_1 = v_2}{P}{Z}&\coloneq\Int{P}(\symbolEquals)(Z(v_1), Z(v_2))
        \end{align*}
        \begin{align*}
          \interpretationVariable{\varphi_1 \vee \varphi_2}{P}{Z}&\coloneq\max\bigl\{\interpretationVariable{\varphi_1}{P}{Z}, \interpretationVariable{\varphi_2}{P}{Z}\bigr\}\\
          \interpretationVariable{\varphi_1 \wedge \varphi_2}{P}{Z}&\coloneq\min\bigl\{\interpretationVariable{\varphi_1}{P}{Z}, \interpretationVariable{\varphi_2}{P}{Z}\bigr\}\\
          \interpretationVariable{\exists v: \varphi}{P}{Z}&\coloneq\max_{o \in \Obj{P}}\bigl\{\min\bigl\{\Int{P}(\symbolExists)(o), \interpretationVariable{\varphi}{P}{Z, v\mapsto o}\bigr\}\bigr\}\\
          \interpretationVariable{\forall v: \varphi}{P}{Z}&\coloneq\min_{o \in \Obj{P}}\bigl\{\max\bigl\{1 - \Int{P}(\symbolExists)(o), \interpretationVariable{\varphi}{P}{Z, v\mapsto o}\bigr\}\bigr\}
        \end{align*}}
      \caption{3-valued semantics of logic predicates}\label{fig:fol-semantics}
    \end{minipage}
\end{figure}

\begin{example}
    For the metamodel of \autoref{fig:metamodel}, \(\{ \mathsf{Train}, \mathsf{Segment}, \mathsf{Turnout}\} \subsetneq \Sigma \) are unary class predicates, and \( \{ \mathsf{location}, \mathsf{occupiedBy}, \mathsf{connectedTo}, \mathsf{straight}, \mathsf{divergent}\} \subsetneq \Sigma\) are binary %
	predicates. 
	
    \autoref{fig:partial-model-modes3} shows a partial model \(P = \langle \Obj{P}, \Int{P}, \Scope{P} \rangle\) conforming to the \modes{} metamodel. Objects are drawn as boxes with the values of the interpretations \(\Int{P}(\symbolClass{C}{i})\) of the class symbols written inside, while edges are drawn as arrows labelled with the relation symbols \(\symbolClass{R}{j}\) and the equlity symbol \(\symbolEquals\). Solid edges correspond to \(\true\) logic values, dashed edges are \(\unk\) logic values, and \(\false\) logic values are omitted. Uncertain existence \(\symbolExists\) is shown with a dashed outline.

    The switching direction of the turnouts \(\mathsf{tu_0}\) and \(\mathsf{tu_1}\) is unknown. Additionally, there are no concrete trains on the track, but a \emph{multi-object} \(\mathsf{tr_{new}}\) represents all trains and their potential locations. Formally, \(\Int{P}(\symbolExists)(\mathrm{s_0}) = \Int{P}(\symbolEquals)(\mathrm{s_0}, \mathrm{s_0}) = \Int{P}(\dslStyle{Segment})(\mathrm{s_0}) = \Int{P}(\dslStyle{connectedTo})(\mathrm{s_0}, \mathrm{tu_0}) = \true\), but \(\Int{P}(\dslStyle{connectedTo})(\mathrm{tu_0}, \mathrm{s_1}) = \unk\). Because \(\Int{P}(\symbolExists)(\mathrm{tr_{new}}) = \Int{P}(\symbolExists)(\mathrm{tr_{new}}, \mathrm{tr_{new}}) = \unk\), \(\mathrm{tr_{new}}\) is a \emph{multi-object} that may stand for any number of \dslStyle{Train} instances (even \(0\)). The \dslStyle{location} of \(\mathrm{tr_{new}}\) is also uncertain, new trains may be located on any \dslStyle{Segment} or \dslStyle{Turnout}.
    
    The scope \(\Scope{P} = \{\lVar{x}_1 \le 3, \lVar{x}_2 = 0\}\) is shown in the lower right corner of the figure. In \autoref{ex:fol-predicates}, this scope will restrict the number of \dslStyle{Train} objects in the model to ensure its well-formedness.
\end{example}

Runtime snapshots of a system are captured by \emph{concrete (instance) models}, which contain no uncertainty or multi-objects and truth values are restricted to \(\true\) and \(\false\). %

\begin{definition}
  A partial model \(M = \langle\Obj{M}, \Int{M}, \Scope{M}\rangle\) is \emph{concrete} if
  \begin{itemize}
	\item \(\Int{M}\) contains only \(\true\) and \(\false\) values, i.e., \(\Int{M}(\sigma)(\mathbf{o}) \in \{\true, \false\}\) for all \(\sigma \in \Sigma\) and \(\mathbf{o} \in \Obj{M}^{\alpha(\sigma)}\);
	\item all objects surely exist, i.e., \(\Int{M}(\symbolExists)(o) = \true\) for all \(o \in \Obj{M}\);
	\item the interpretation of the equality symbol \(\symbolEquals\) matches the usual equality of objects, i.e., for all \(o_{1}, o_{2} \in \Obj{M}\) \(\Int{M}(\symbolEquals)(o_{1}, o_{2}) = \true\) if \(o_{1} = o_{2}\), \(\false\) otherwise; and
    \item \(\Scope{M}\) is satisfiable, i.e., there is some valuation \(k\colon \LVars \to \mathbb{Z}\) such that \(k \vDash \Scope{M}\).
  \end{itemize}
\end{definition}

\begin{example}
    \autoref{fig:instance-model-modes3} shows a concrete model \(M\) conforming to the \modes{} metamodel. Only \(\true\) and \(\false\) logic values appear in the interpretation. The associated scope \(\Scope{M}\) has a single solution \(\{\lVar{x}_1 \mapsto 2, \lVar{x}_2 \mapsto 0\} \vDash \Scope{M}\).
\end{example}

Concrete models are obtained from partial models by a series of \emph{refinements}~\cite{Salay2012}, which add further information by setting unknown logic values \(\unk\) to \(\true\) or \(\false\), while known \(\true\) and \(\false\) values remain unchanged. This is captured by the refinement relation \(X \refined Y \coloneqq (X = \unk) \vee (X = Y)\). During model generation, refinements are carried out until a concrete model is reached.

\begin{definition}\label{dfn:refinement}
  The function \(\abs\colon \Obj{Q} \to \Obj{P}\) is an \emph{abstraction function} from the partial model \(Q = \langle \Obj{Q}, \Int{Q}, \Scope{Q} \rangle\) to the partial model \(P = \langle \Obj{P}, \Int{P}, \Scope{P} \rangle\) (written as \(P \refined_{\abs} Q\)) if
  \begin{itemize}
	\item for all \(\sigma \in \Sigma\), \(q_{1}, \ldots, q_{\alpha(\sigma)} \in \Obj{Q}\), logic values in \(\Int{Q}\) are the refinements of the corresponding values in \(\Int{P}\), i.e., \(\Int{P}(\sigma)(\abs(q_{1}), \ldots, \abs(q_{\alpha(\sigma)})) \refined \Int{Q}(\sigma)(q_{1}, \ldots, q_{\alpha(\sigma)}) \);
	\item surely existing object do not disappear, i.e., for all \(p \in \Obj{P}\), \(\Int{P}(\symbolExists)(p) = \true\) implies that there is some \(q \in \Obj{Q}\) with \(\abs(q) = p\); and
	\item solutions of \(\Scope{Q}\) are also solutions of \(\Scope{P}\), i.e., \(\Scope{Q} \vDash \Scope{P}\).
  \end{itemize}
  \(Q\) is a \emph{refinement} or \(P\) (written as \(P \refined Q\)) if \(P \refined_{\abs} Q\) for some \(\abs\colon \Obj{Q} \to \Obj{P}\).
\end{definition}

\begin{example}
    The concrete model \(M\) in \autoref{fig:instance-model-modes3} is a refinement of the partial model \(P\) in \autoref{fig:partial-model-modes3}: \(P \refined_{\abs} M\). The abstraction function maps \(\abs(\mathrm{tr_0}) = \mathrm{tr_{new}}\) and \(\abs(\mathrm{tr_1}) = \mathrm{tr_{new}}\), i.e., newly added trains are refinements of the train multi-object. Any other object is mapped by \(\abs\) to itself, i.e., the identities of the rest of the objects remained unchained. We may also see that the directions the turnouts were set, e.g., \(\Int{P}(\dslStyle{connectedTo})(\mathrm{tu_0}, \mathrm{s_1}) = \unk \refined \Int{M}(\dslStyle{connectedTo})(\mathrm{tu_0}, \mathrm{s_1}) = \false\) and \(\Int{P}(\dslStyle{connectedTo})(\mathrm{tu_0}, \mathrm{s_2}) = \unk \refined \Int{M}(\dslStyle{connectedTo})(\mathrm{tu_0}, \mathrm{s_2}) = \true\). Moreover, \(\Scope{M} \vDash \Scope{P}\).
\end{example}

Note that refinement is associative: if \(P_1 \refined_{\abs_1} P_2\) and \(P_2 \refined_{\abs_2} P_3\), then \(P_1 \refined_{\abs_1 \circ \abs_2} P_3\). Thus, it is possible to gradually add information during model generation with several refinement steps \(P_0 \refined P_1 \refined P_2 \refined \cdots \refined M\) to arrive at a concrete model \(M\).

\subsection{First-order Logic Predicates for Queries Over Graph Models} 

The formal definitions of metamodel and instance model enable the formulation of first-order logic (FOL) predicates, which can be evaluated as \emph{graph queries} over the logic structure of an instance model.
Informally, base predicates check either for equality or for the existence of certain objects and references of a respective type (predicate) in the underlying runtime model. Then complex predicates are derived by traditional FOL connectives (e.g., not, exists, forall, and, or).

\begin{definition}
	A \emph{first-order logic predicate} (or \emph{query}) \(\varphi\), where \(v_{1}, \ldots, v_{n}\) denote free variables (not appearing in any quantifiers) of \(\varphi\) can be evaluated over a partial model \(P\) along a \emph{variable binding} \(Z\colon \{v_1,\ldots,v_n\} \rightarrow \mathcal{O}_P \) (denoted as \(\interpretationVariable{\varphi}{P}{Z}\)) to return \(\true\), \(\false\) or \(\unk\) as shown in \autoref{fig:fol-semantics}.
\end{definition}

Because concrete models contain only \(\true\) and \(\false\) logic values, any predicate \(\varphi\) evaluates to either \(\true\) or \(\false\) in a concrete model. Hence, on concrete models, we can run queries and obtain their match sets without uncertainties.

\begin{definition}
  In a concrete model, \emph{predicate / query evaluation} aims to find a variable binding \(Z\colon \{v_1,\ldots,v_n\} \rightarrow \mathcal{O}_M\) for a predicate $\varphi$ that maps all free variables of the predicate to objects of \(M\) such that the predicate evaluates to \emph{true}, i.e., \(\interpretationVariable{\varphi}{M}{Z} = \true\). 
\end{definition}

\begin{definition}
  The \emph{match set} of a query predicate \(\varphi\) with free variables \(v_{1}, \ldots, v_{n}\) is the set
  \(\mathit{Matches}(M, \varphi) = \bigl\{Z\colon \{v_{1}, \ldots, v_{n}\} \to \dslObjectSetI{M} \mid \interpretationVariable{\varphi}{M}{Z} = 1\bigr\}\).
  One element in this set is called a \emph{match}, while \(M\#\varphi = |\mathit{Matches}(M, \varphi)|\) denotes the size of the match set.
\end{definition}

Note that in our context, a match of a query will typically represent a violation of a well-formedness constraint of the domain or a hazardous situation with respect to a safety property.

\begin{example}
    Consider the graph query \(\varphi_{\mathrm{CT}}\) for the ``close trains'' hazard formalized as a FOL expression \(\varphi_{\mathrm{CT}}\) in \autoref{fig:query-predicate}. In the concrete model \(M\) in \autoref{fig:instance-model-modes3}, \(\interpretationVariable{\varphi_{\mathrm{CT}}}{M}{s \mapsto \mathrm{s_2}, e \mapsto \mathrm{s_3}} = \true\). In the partial model \(P\) in \autoref{fig:partial-model-modes3}, \(\interpretationVariable{\varphi_{\mathrm{CT}}}{P}{s \mapsto \mathrm{s_2}, e \mapsto \mathrm{s_3}} = \false\) due to the uncertain existence of the \(\mathrm{tr_{new}}\) multi-object. In \(M\), \(\varphi_{\mathrm{CT}}\) has two matches \(\mathit{Matches}(M, \varphi_{\mathrm{CT}}) = \{\{s \mapsto \mathrm{s_2}, e \mapsto \mathrm{s_3}\}, \{s \mapsto \mathrm{s_2}, e \mapsto \mathrm{s_3}\}\}\). Therefore, \(M\#\varphi_{\mathrm{CT}} = 2\).
\end{example}

\subsection{Well-formedness and Scope Constraints} 

In order to make static analysis of data-driven monitors more precise, we can add additional domain-specific information into models as constraints to exclude impossible or irrelevant runtime snapshots from consideration.

A domain metamodel is frequently complemented in practice with \emph{well-formedness constraints} to restrict the possible relationships between domain concepts. The constraints, such as \textbf{type hierarchy}, \textbf{type compliance}, \textbf{multiplicity}, \textbf{inverse relation} and \text{containment hierarchy} constraints, can be captured by FOL predicates~\cite{Semerath2018}.

Additionally, numerical \emph{scope constraints} restrict the sizes of models to conform with allocation requirement in monitor programs and guide the analysis toward models that are relevant in practice (e.g., the size of the model and the ratios between the number of objects of given types match realistic scenarios).

In this work, we are interested in the timing analysis of monitors that take well-formed models conforming to numerical constraints as their input. Therefore, we will require model to conform to \emph{theories} formed by graph predicates.

\begin{definition}
    A \emph{theory} over a signature \(\langle \Sigma, \alpha \rangle\) is a pair \(\mathcal{T} = \langle \Phi, \lVar{r} \rangle\), where
    \begin{itemize}
        \item \(\Phi = \{ \varphi_1, \ldots, \varphi_{\lvert \Phi \rvert} \}\) is a finite set of graph predicates over \(\langle \Sigma, \alpha \rangle\); and
        \item \(\lVar{r}\colon \Phi \to \LVars\) maps graph predicates to linear equation variables.
    \end{itemize}
\end{definition}

\begin{definition}\label{dfn:compatible}
    A concrete model \(M = \langle \Obj{M}, \Int{M}, \Scope{M} \rangle\) is \emph{compatible} with the theory \(\mathcal{T} = \langle \Phi, \lVar{r} \rangle\) (written as \(M \vDash \mathcal{T}\)) if \(\Scope{M} \vDash \lVar{r}(\varphi_i) = M\#\varphi_i\) for all \(\varphi_i \in \Phi\).
\end{definition}

Theories, along with refinement, enable to constrain the number of graph predicate matches via partial models. In particular, if \(\varphi_i\) is an \emph{error predicate} that should never match well-formed models, the presence of exactly \(0\) matches should be enforced by the model scope \(\Scope{P}\).

\begin{proposition}[\cite{Marussy2020}]\label{thm:background:forward-refinement}
    Let \(P = \langle \Obj{P}, \Int{P}, \Scope{P} \rangle\) be a partial model, \(\mathcal{T} = \langle \Phi, \lVar{r} \rangle\) be a theory and \(\varphi_i \in \Phi\) be a graph predicate. If \(\Scope{P} \vDash \lVar{r}(\varphi_i) \le U\) (resp.~\(\Scope{P} \vDash \lVar{r}(\varphi_i) \ge L\)), then \(M\#\varphi_i \le U\) (resp.~\(M\#\varphi_i \ge L\)) holds for all concrete models \(P \refined M\) compatible with \(\mathcal{T}\), i.e., \(M\) satisfies the upper (resp.~lower) bound imposed on the number of \(\varphi_i\) matches.
\end{proposition}

\begin{example}\label{ex:fol-predicates}
    Consider the FOL predicates
    \begin{align*}
        \varphi_{\mathrm{Train}}(v_1) &= \dslStyle{Train}(v_1) \text,
        & \varphi_{\mathrm{connectedTo}}(v_{1}, v_{2}) &= \dslStyle{connectedTo}(v_{1}, v_{2}) \wedge \neg \dslStyle{connectedTo}(v_{2}, v_{1}) \text,
    \end{align*}
    and the theory \(\mathcal{T} = \langle \{ \varphi_{\mathrm{Train}}, \varphi_{\mathrm{connectedTo}} \}, \lVar{r} \rangle\), where \(\lVar{r}(\varphi_{\mathrm{Train}}) = \lVar{x}_1\) and \(\lVar{r}(\varphi_{\mathrm{connectedTo}}) = \lVar{x}_2\).
    
   The predicate \(\varphi_{\mathrm{Train}}\) selects all \dslStyle{Train} instances. Thus, the linear inequality \((\lVar{x}_1 \le 3) \in \Scope{P}\) in the partial model \(P\) in \autoref{fig:partial-model-modes3} corresponds to the scope constraint that there should be no more than 3 \dslStyle{Train} instances in the model.
   
   The predicate \(\varphi_{\mathrm{connectedTo}}\) selects \dslStyle{connectedTo} links that do not have a corresponding link in the reverse direction. Since railway tracks can be traversed in both directions, we enforce a symmetric \dslStyle{connectedTo} relation by a well-formedness constraint encoded as \((\lVar{x}_2 = 0) \in \Scope{P}\) in \(P\).
   
   The concrete model \(M\) in \autoref{fig:instance-model-modes3} conforms to the theory \(\mathcal{T}\). As shown in \autoref{thm:background:forward-refinement}, \(M\) obeys the scope and well-formedness constraints prescribed in \(\Scope{P}\), since \(P \refined M\).
\end{example}

\subsection{Model Generation}\label{sec:model-generation}

Automated synthesis of domain-specific graph models has been actively researched in the field of model-based software engineering~\cite{Semerath2018, Brottier2006, Fleurey2004}. Hereby, we revisit some  core concepts.

\begin{definition}
    A \emph{model generation task} is of the form \(\max \sum_{\lVar{x}_j \in \LVars} c_j \cdot \lVar{x}_j \textit{ subject to } \langle P_{\textrm{init}}, \mathcal{T} \rangle\), where
    \begin{itemize}
    	\item \( \Sigma = \{ \symbolClass{C}{1}, \ldots, \symbolClass{C}{m_{\dslStyle{C}}}, \symbolRef{R}{1}, \ldots, \symbolRef{R}{m_{\dslStyle{R}}}, \symbolExists, \symbolEquals \} \) is a metamodel;
    	\item the partial model \(P_{\textrm{init}} = \langle \Obj{P_{\textrm{init}}} \Int{P_{\textrm{init}}} \Scope{P_{\textrm{init}}} \rangle\) over \(\Sigma\) is the \emph{initial partial model};
    	\item \(\mathcal{T} = \langle \Phi, \lVar{r} \rangle\) is a theory of well-formedness and scope constraints over \(\Sigma\); and
    	\item \(\sum_{\lVar{x}_j \in \LVars} c_j \cdot \lVar{x}_j\) is the \emph{objective}.
    \end{itemize}
\end{definition}

\begin{definition}\label{dfn:solutions}
    The solutions of the model generation task are concrete models that are refinements of the initial partial model and are compatible with the theory: 
    \begin{align*} 
    	\mathit{solutions}(\Sigma, P_{\textrm{init}}, \mathcal{T}) = \{ M \mid \textnormal{\(M\) is a concrete instance of the metamodel \(\Sigma\), \(P_{\textrm{init}} \refined M\), \(M \vDash \mathcal{T}\)} \}.
    \end{align*}
\end{definition}

According to \autoref{thm:background:forward-refinement}, such concrete models \(M\) satisfy the numerical constraints given in scope \(\Scope{P_{\textrm{init}}}\) that restrict the number of matches \(M\#\varphi_i\) of graph predicates \(\varphi_i \in \Phi\). If no structural information is available about the sought models, we may set \(\Obj{P_{\textrm{init}}} = \{\mathit{new}\}\) and \(\Int{P_{\textrm{init}}}(\symbolClass{C}{i})(\mathit{new}) = \Int{P_{\textrm{init}}}(\symbolClass{R}{j})(\mathit{new}, \mathit{new}) = \Int{P_{\textrm{init}}}(\symbolExists)(\mathit{new}) = \Int{P_{\textrm{init}}}(\symbolEquals)(\mathit{new}, \mathit{new}) = \unk\) for all \(\symbolClass{C}{i}, \symbolRef{R}{j} \in \Sigma\) to obtain the maximally uncertain initial partial model with a single multi-object. Otherwise, \(P_{\textrm{init}}\) contains the known parts of the model and multi-objects may serve as placeholders for objects to be added.

\begin{definition}
    In the model generation task \(\max \sum_{\lVar{x}_j \in \LVars} c_j \cdot \lVar{x}_j \textit{ subject to } \langle P_{\textrm{init}}, \mathcal{T} \rangle\), the \emph{cost function} \(g(M) = \max \sum_{\lVar{x}_j \in \LVars} c_j \cdot \lVar{x}_j \textit{ subject to } \Scope{M}\) of the model generation task can be computed by solving an ILP problem.
    A concrete model \(M \in \mathit{solutions}(\Sigma, P_{\textrm{init}}, \mathcal{T})\) is an \emph{optimal solution} of the task if \(g(M') \le g(M)\) for all \(M' \in \mathit{solutions}(\Sigma, P_{\textrm{init}}, \mathcal{T})\), i.e., its cost is maximal.
\end{definition}

Our work relies on the model generator presented in~\cite{Semerath2018,Marussy2020} which was proved to be complete and sound in~\cite{Varro2018}. Informally, it is able to derive all concrete (instance) models in a domain (up to a designated size defined by the scopes) which satisfy the constraints by exploring a \emph{state space} of possible partial models along refinements.

\section{Timing Analysis of Query-Based Monitors}
\label{sec:realtime}

Estimating the WCET of query-based monitors is a highly complex task which involves multiple classic challenges of timing analysis.
The runtime model of the system is a continuously changing data structure that captures an up to date snapshot of the underlying running system. Hence, it is not sufficient to analyze execution time on a single input model, but all models possible at runtime must be considered.

However, the space of possible models is enormous. For example, in a metamodel with \(3\) reference types, there may be up to \(2^{3\cdot25\cdot25} = 2^{1875}\) models with \(25\) objects. Thus, explicit enumeration of graph models is intractable, which necessitates the use of abstractions.

Another major challenge is that \emph{query execution time is heavily data-dependent}, i.e., the same control flow of a query program may have substantially different run times based upon the structural characteristics of the underlying graph model.
Assuming some constraints on model size (e.g., capped by available memory) and some general restrictions on model scope (e.g., there are more segments than trains in any real model), a key open challenge is \emph{how to provide a model where the execution time of a particular query program is maximal}. In this work, we provide \emph{witness models} that maximize an estimate of the execution time, which aids in WCET analysis and in identifying bottlenecks in query execution.

Moreover, a single model may be represented in memory in several isomorphic ways (\autoref{sec:graph-model-assumptions}). During the runtime evolution of the graph, a particular snapshot might be reached in any of its possible in-memory representations. Thus, WCET estimation even for a single concrete input model must tackle the dependency of execution paths on the data representations. As a single model of \(n\) objects has \(n!\) possible in-memory representations even if we only consider inserting the objects into a single continuous linear array of \(n\) elements, explicit enumeration is again intractable.

\subsection{Comparison of Timing Analysis Approaches}

\begin{table}
    \caption{WCET analysis approaches for data-driven runtime monitor programs}
    \centering
    {\footnotesize\begin{tabular}{@{}lll@{}}
        \toprule
        & Inputs & Outputs \\
        \midrule
        CL & \(\overbrace{\text{HW description} + \text{Query program}}^{\text{Low-level analysis inputs}}\) & WCET estimate \\
        \color{gray}VAL & \color{gray}\(\text{HW description} + \overbrace{\text{Query program} + \text{Concrete model} + \text{Memory image}}^{\text{Value analysis inputs}}\) & \color{gray}\(\begin{array}{@{}l@{}}\text{WCET estimate} \\ \text{for single memory image}\end{array}\) \\
        \midrule
        \(\text{DS}_{M}\) & \(\text{HW description} + \overbrace{\text{Query program} + \mathrlap{\text{Concrete model}}\hphantom{\textbf{Partial model} + \textbf{Constraints}}}^{\textbf{Domain-specific high-level analyis inputs}}\) & WCET est.~for single model \\
        \(\text{DS}_{\Sigma}\) & \(\text{HW description} + \text{Query program} + \textbf{Constraints}\) & \(\begin{array}{@{}l@{}}\text{WCET est.~for all valid models}\\ +\;\textbf{Witness model}\end{array}\) \\
        \(\text{DS}_{P}\) & \(\text{HW description} + \text{Query program} + \textbf{Partial model} + \textbf{Constraints}\) & \(\begin{array}{@{}l@{}}\text{WCET est.~for all refinements}\\ +\;\textbf{Witness model}\end{array}\) \\
        \bottomrule
    \end{tabular}}\label{tab:scenarios}
\end{table}

\autoref{tab:scenarios} illustrates the existing and proposed approaches of WCET analysis for query programs. \emph{Classical} (CL) analysis is based on binary code of query program and the characteristics of the hardware platform, but does not consider structure and the well-formedness of the runtime models.

\emph{Value analysis} (VAL) can derive more precise WCET estimates for executing a query on a single memory image (comprised of a single concrete model). However, it is unable to consider 
equivalent in-memory representations of the same concrete model (i.e., different parts of the model allocated to different spatial locations), or to cover all possible consistent concrete models, thus it is unsuitable for the analysis of data-driven monitors.%

To alleviate this issue, we propose three \emph{domain-specific} (DS) WCET analysis methods for data-driven monitors. We introduce the concept of \emph{witness models}, which are consistent models that are \emph{feasible inputs of the graph query program} and \emph{maximize the WCET estimate for all models within the given scope}. They serve as representative data to calculate WCET for \emph{any} model within the scope.

\begin{itemize}
    \item First, we estimate WCET for a \emph{single concrete model} (\(\text{DS}_{M}\)). The estimate is valid for all in-memory representations of a given concrete model \(M\).

    \item In the next case, the set of possible runtime snapshots is specified with \emph{metamodel} \(\langle \Sigma, \alpha \rangle\) along with well-formedness and scope constraints (\(\text{DS}_{\Sigma}\)). This WCET estimate is valid for all possible runtime snapshots within the memory limits of the system, i.e., for all consistent instances of the metamodel up to the size specified by the scope constraints.

    \item Thirdly, an \emph{initial partial model} \(P\) may specify the set of possible runtime snapshots (\(\text{DS}_{P}\)) including static (known and concrete) and dynamic (uncertain at design time) parts of the runtime model. The WCET estimate is valid for all possible refinements \(P \refined M\) of \(P\).
\end{itemize}
 
\begin{figure}[tb]
	\centering
		\resizebox{0.75\linewidth}{!}{
		\upshape
		\sffamily
		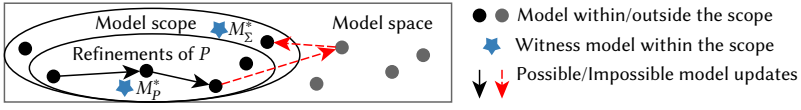}
	\caption{Classification of query input models and model updates from the perspective of WCET analysis}
	\label{fig:model-space} 
\end{figure}

\autoref{fig:model-space} sketches the \emph{model space} of runtime graph models (represented with dots), i.e., the set of all input models. Possible changes made to a model at runtime (depicted as arrows) result in a new model. To obtain a safe and tight WCET estimate for query programs, we make some assumptions about realistic (and consistent) models captured in the form of a \emph{model scope}. If an initial partial model \(P\) is provided, the analysis is further restricted to its (valid) refinements, thus inconsistent models are considered to be unrealistic. The witness model \(M^{*}_{\Sigma}\) for the consistent instances of the metamodel and \(M^{*}_{P}\) for the refinements of \(P\) are depicted as blue stars in \autoref{fig:model-space}.

The witness models \(M^{*}_P\) may aid in iteratively refining the model scope.  If the partial model corresponds to a situation that is impossible at runtime, it indicates that the model scope was specified in a too general way. We may exclude such situations by refining the partial model \(P\)~\cite{Varro2018}. However, care must be taken to avoid excluding feasible inputs and overfitting the WCET esimate. If the witness model is a feasible input, it may be inspected to study the characteristics and bottlenecks of the graph query program.

\begin{figure}[t]
	\begin{minipage}[c]{.47\textwidth}
		\subfigure[Witness model \(M^{*}\) yields execution estimate of 1309 for query \infigure{closeTrains}\label{fig:instancemodel-generated}]{
		\includegraphics[width=.96\linewidth]{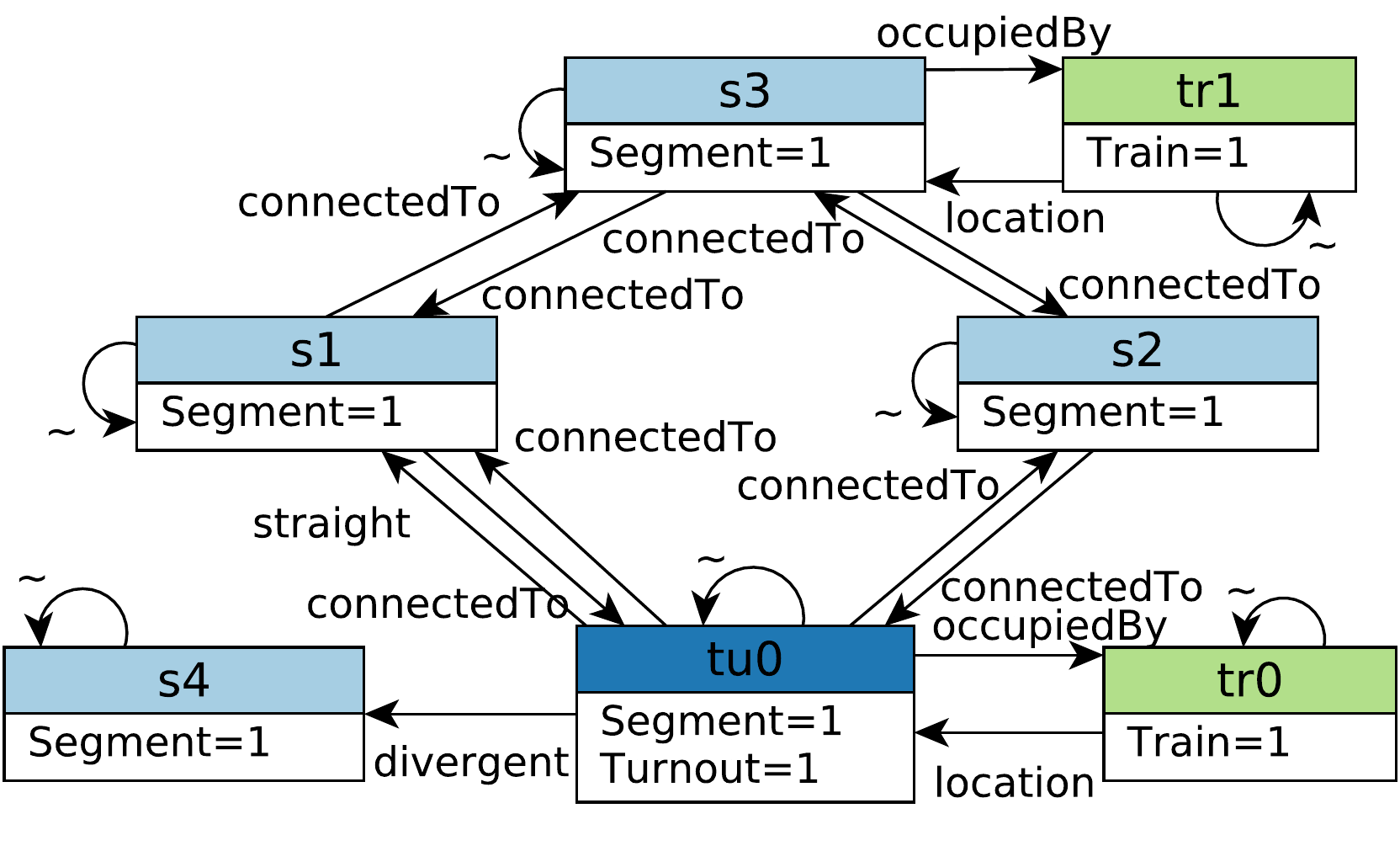}}
	\end{minipage}
	\hfill
	\begin{minipage}[c]{.47\textwidth}
		\subfigure[Malformed model \(M^\prime\) yields execution estimate of 1325 for query \infigure{closeTrains} \label{fig:instancemodel-synthetic}
		]{\includegraphics[width=\linewidth]{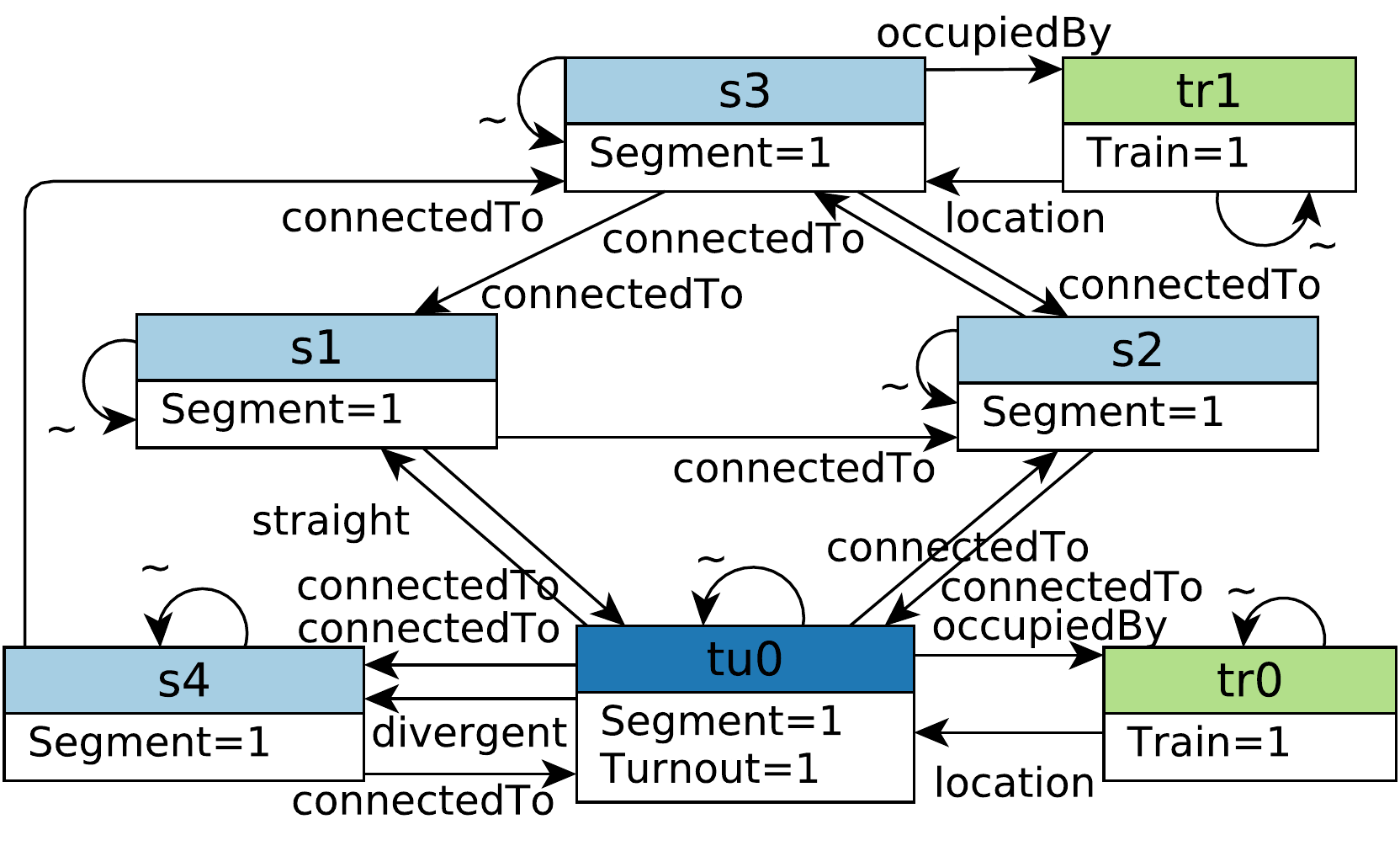}}
	\end{minipage}
	\caption{Illustrating model generation problems for witness models} %
	\label{fig:modes-wcet-instance-models}
\end{figure}

\begin{example}
  \autoref{fig:instancemodel-generated} shows the witness model \(M^{*}\) for the WCET of the \dslStyle{closeTrains} query for  well-formed models with up to 7 objects in total (as model scope), out of which up to 2 are \dslStyle{Train} instances. The corresponding WCET estimate is 1309 systicks. The model \(M'\) in \autoref{fig:instancemodel-synthetic} has the same number of elements, but with a higher execution time estimate of 1325 systicks. However, \(M'\) lies outside the model scope, because it is malformed due to non-symmetric \dslStyle{connectedTo} references (e.g., \(\mathrm{s1}\) is \dslStyle{connectedTo} \(\mathrm{s2}\) but not vice versa).
  
  Classical~(CL) WCET estimation techniques cannot exclude \(M'\) from the analysis and they would return a higher WCET estimate, while our novel \(\text{DS}_{\Sigma}\) technique can restrict the analysis to the model scope to return the correct estimate of 1309 systicks along with the witness \(M^{*}\).
\end{example}

\subsection{Architectural Overview}

\begin{figure}[tb]
	\centering
	\resizebox{.8\linewidth}{!}{
		\def\svgwidth{\linewidth}{
		\upshape
		\footnotesize
		\sffamily
		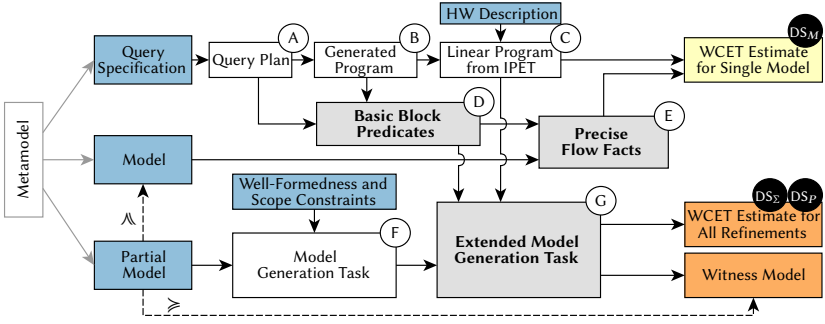}}
	\caption{Workflow of WCET estimation for query-based monitors}
	\label{fig:approach-workflow-modelgen}
\end{figure}

\autoref{fig:approach-workflow-modelgen} presents the high-level description of our design time tasks to obtain a WCET estimate in the \(\text{DS}_{M}\), \(\text{DS}_{\Sigma}\) and \(\text{DS}_{P}\) scenarios. The high-level inputs of the process include the \emph{query specification}, the target \emph{hardware description}, and the \emph{well-formedness and scope constraints} of the domain.

First, a \emph{query plan}~(A) is constructed from the query specification, based on which the \emph{query program}~(B) is generated according to \autoref{sec:query-program-structure}. Our approach is complementary to IPET-based WCET estimators and leverages the results of high- and low-level analysis in form of the CFG (possibly after some loop unrolling) and the corresponding linear program~(C).

In case of WCET estimation for a concrete model \(M\) (\(\text{DS}_{M}\)), \(M\) is also provided as an input. Based on the query plan and the generated montitor code, \emph{basic block predicates}~(D) are derived, whose matches in the concrete model \(M\) correspond to executions of basic blocks in the monitor program. We leverage these matches to construct \emph{precise flow facts}~(E) for IPET analysis in \autoref{sec:bb-predicates}. The resulting flow facts and WCET estimate consider all possible in-memory representations of \(M\).

For WCET estimation for any valid instance of a metamodel \(\langle \Sigma, \alpha \rangle\) (\(\text{DS}_{\Sigma}\)), the initial partial model \(P_{\text{init}}\) is constructed according to \autoref{sec:model-generation}. When a partial model \(P\) is already provided as input (\(\text{DS}_{P}\)), it replaces \(P_{\text{init}}\) as the initial partial model. Hence, along with the \(\mathcal{T}\) of well-formedness and scope constraints, we obtain a \emph{model generation task}~(F), whose solutions are partial models within the analyzed model scope. We incorporate the basic block predicates~(D) into an extended theory \(\mathcal{T}'\) in \autoref{sec:witness-generation}, which forms an \emph{extended model generation task}~(G) for witness model generation along with the linear program~(C). \emph{Witness models} are systematically generated using a graph solver~\cite{Semerath2018,Marussy2020} as solutions of such tasks along refinements \(P_{\text{init}} \refined M\) of the initial partial model. The cost associated with the witness model \(M^{*}\), which is a solution of the IPET linear program~(C) extended with domain-specific flow facts~(E), is a safe and tight WCET estimate.

\subsection{Approximating Execution Time with Graph Predicates}\label{sec:bb-predicates}

To derive precise flow facts for WCET analysis of a graph query program with concrete input model \(M\) and characterize its data-dependent execution time, we construct a \emph{basic block predicate} \(\psi_{\mathit{bb}}\) for each basic block \(\mathit{bb} \in \mathit{BB}\) of the query program. Free variables of \(\psi_{\mathit{bb}}\) correspond to program variables (bound by \texttt{for} loops). Due to the structure of the code generated from the query plans (\autoref{sec:query-based-monitoring:generated-code}), each execution of \(\mathit{bb}\) corresponds to a match \(Z \in \mathit{Matches}(M, \psi_{{\mathit{bb}}})\) of \(\psi_{\mathit{bb}}\) in \(M\).

For loop headers, we construct an additional \(\psi'_{\mathit{bb}_{\ell,h}}\) where the matches of \(\psi'_{\mathit{bb}_{\ell,h}}\) represent executions of the loop \(\ell\) where the loop condition holds, while the matches of \(\psi_{\mathit{bb}_{\ell,h}}\) correspond to the executions where the loop exits.

\begin{figure}
	\begin{minipage}[t][][b]{.52\textwidth}
		\LinesNumbered
		\SetAlFnt{\scriptsize}
		\SetAlCapFnt{\small \normalfont \sffamily}
		\SetAlCapNameFnt{\small \normalfont \sffamily}
		\RestyleAlgo{ruled}
		\begin{algorithm}[H]
			\caption{Basic block predicate construction}
			\label{alg:derive-predicates}
			\SetArgSty{upshape}
			\SetKwProg{Fn}{Function}{ is}{end}
			\Fn{\upshape DerivePredicates(\(\mathit{BB}, \mathit{lineTrace}, \mathit{planTrace}\)) }{
			    \(\Psi = \emptyset\)\\
			    \For{\(\mathit{bb} \in \mathit{BB}\)}{
    			    \(\psi_{\mathit{bb}} = \true\)\label{ln:derive-predicates:init}\\
    			    Let \(\mathit{ln}_b\)--\(\mathit{ln}_e\) be the source lines associated with \(\mathit{bb}\) in \(\mathit{lineTrace}\)\label{ln:derive-predicates:line-trace}\\
			        \For{\texttt{if} and \texttt{for} statements \(\mathit{st}\) containing \(\mathit{ln}_b\)--\(\mathit{ln}_e\)\label{ln:derive-predicates:loop-start}}{
			            \(\psi_{\mathit{bb}} = \psi_{\mathit{bb}} \wedge \textup{StatementToLogic}(\mathit{st}, \mathit{planTrace})\)\label{ln:derive-predicates:loop-end}\\
			        }
			        Let  \(v_1, \ldots, v_m\) be the free variables of \(\psi_{\mathit{bb}}\) \\
			        \(\Psi = \Psi \cup \{ \psi_{\mathit{bb}} \}\)\\
			        \If{\(\mathit{bb}\) is the header of the loop \(\ell\)\label{ln:derive-predicates:header-start}}{
			            \(\psi'_{\mathit{bb}} = \psi_{\mathit{bb}} \wedge \textup{StatementToLogic}(\ell, \mathit{planTrace})\)\\
			            Let  \(v_1, \ldots, v_{m + 1}\) be the free variables of \(\psi'_{\mathit{bb}}\)  \\
			            \(\Psi = \Psi \cup \{ \psi'_{\mathit{bb}} \}\)\label{ln:derive-predicates:header-end}
		            }
			    }
			    \Return \(\Psi\)
			}
		\end{algorithm}

	\end{minipage}\hfill
	\begin{minipage}[t][][b]{.47\textwidth}
		\LinesNumbered
		\SetAlFnt{\scriptsize}
		\SetAlCapFnt{\small \normalfont \sffamily}
		\SetAlCapNameFnt{\small \normalfont \sffamily}
		\RestyleAlgo{ruled}
		\begin{algorithm}[H]
			\caption{Translate search plan to logic}
			\label{alg:statement-to-logic}
			\SetArgSty{upshape}
			\SetKwProg{Fn}{Function}{ is}{end}
			\Fn{\upshape StatementToLogic(\(\mathit{st}, \mathit{planTrace}\)) }{
			    Determine the constraint implemented by \(\mathit{st}\) from \(\mathit{planTrace}\)\\
	            \uIf{\(\mathit{st}\) implements \textbf{extend} \(\exists v_i : \dslStyle{C}(\underline{v_i})\)} {
		            \Return \(\dslStyle{C}(v_i)\)
	            }
	            \uElseIf{\(\mathit{st}\) implements \textbf{extend} \(\exists v_j : \dslStyle{R}(v_i, \underline{v_j})\)} {
		            \Return \(\dslStyle{R}(v_i, v_j)\)
	            }
	            \uElseIf{\(\mathit{st}\) implements \textbf{check} \(\dslStyle{C}(v_i)\)} {
		            \Return \(\dslStyle{C}(v_i)\)
	            }
	            \uElseIf{\(\mathit{st}\) implements \textbf{check} \(\dslStyle{R}(v_i, v_j)\)} {
		            \Return \(\dslStyle{R}(v_i, v_j)\)
	            }
	            \uElseIf{\(\mathit{st}\) implements \textbf{check} \(v_i = v_j\)} {
		            \Return \(v_i = v_j\)
	            }
	            \uElseIf{\(\mathit{st}\) implements \textbf{check} \(\neg \dslStyle{C}(v_i)\)} {
		            \Return \(\neg \dslStyle{C}(v_i)\)
	            }
	            \uElseIf{\(\mathit{st}\) implements \textbf{check} \(\neg \dslStyle{R}(v_i, v_j)\)} {
		            \Return \(\neg \dslStyle{R}(v_i, v_j)\)
	            }
	            \ElseIf{\(\mathit{st}\) implements \textbf{check} \(\neg (v_i = v_j)\)} {
		            \Return \(\neg (v_i = v_j)\)
	            }
			}
		\end{algorithm}
	\end{minipage}\par
	\LinesNumbered
	\SetAlFnt{\scriptsize}
	\SetAlCapFnt{\small \normalfont \sffamily}
	\SetAlCapNameFnt{\small \normalfont \sffamily}
	\RestyleAlgo{ruled}
	\begin{algorithm}[H]
		\caption{Precise flow fact construction for a single concrete model \(M\)}
		\label{alg:precise-flow-facts}
		\SetArgSty{upshape}
		\SetKwProg{Fn}{Function}{ is}{end}
		\Fn{\upshape PreciseFlowFacts(\(\mathit{BB}, \mathit{lineTrace}, \mathit{planTrace}, \mathit{CFG} = \langle V, E, s, t, w, \mathit{tr} \rangle, \lVar{f}, M\)) }{
		    \(\Scope{\text{flow}} = \emptyset, \Psi = \textup{DerivePredicates}(\mathit{BB}, \mathit{lineTrace}, \mathit{planTrace})\)\label{ln:precise-flow-facts:init}\\
		    \For{\(\mathit{bb} \in \mathit{BB}\)}{
		        \lIf{\(\mathit{bb}\) is a loop header}{
		            \(\Scope{\text{flow}} = \Scope{\text{flow}} \cup \bigl\{\sum_{e = \langle n_1, n_2 \rangle \in E, \mathit{tr}(n_1) = \mathit{bb}} \lVar{f}(e) = M\#\psi_{\mathit{bb}} + M\#\psi'_{\mathit{bb}}\bigr\}\)\label{ln:precise-flow-facts:header}
	            }
	            \lElse{
		            \(\Scope{\text{flow}} = \Scope{\text{flow}} \cup \bigl\{\sum_{e = \langle n_1, n_2 \rangle \in E, \mathit{tr}(n_1) = \mathit{bb}} \lVar{f}(e) = M\#\psi_{\mathit{bb}}\bigr\}\)\label{ln:precise-flow-facts:other}
	            }
		    }
		    \Return \(\Scope{\text{flow}}\)
		}
	\end{algorithm}
\end{figure}

\autoref{alg:derive-predicates} takes a data-driven monitor program generated from a graph query and constructs the basic block predicates. In addition to the set of basic blocks \(\mathit{BB}\), the algorithm requires traceability information \(\mathit{lineTrace}\) (that connects basic blocks to source code lines) and \(\mathit{planTrace}\) (that connects source code lines to \textbf{extend} and \textbf{check} constraints in the query plan). 
The \(\mathit{lineTrace}\) is extracted from the IPET analysis tool (based on debug information in the compiled executable), while \(\mathit{planTrace}\) is the output of the query code generator.

As state-of-the-art WCET analysis tools~\cite{Ballabriga2006} do not recommend analyzing programs compiled with advanced optimizations, we did not assess programs that use optimization. Therefore, source line, as well as \textbf{extend} and \textbf{check} constraint information in \(\mathit{lineTrace}\) remains valid after compilation. However, the following algorithms can be extended to support compiler optimizations, as long as a compiled basic block still corresponds to a single constraint and the control flow remains structured (comprised on loops and conditionals).

In line~\ref{ln:derive-predicates:init}, \(\psi_{\mathit{bb}}\) is initialized to true, which has a single (trivial) match in any model to reflect that blocks not implementing any query plan constraints will be executed exactly once. Then, in line~\ref{ln:derive-predicates:line-trace}, we traverse \(\mathit{lineTrace}\) to extract the source lines corresponding to \(\mathit{bb}\). The loop in lines~\ref{ln:derive-predicates:loop-start}--\ref{ln:derive-predicates:loop-end} processes all \texttt{if} and \texttt{for} statements enclosing the source lines for \(\mathit{bb}\). As a result, \(\psi_{\mathit{bb}}\) becomes the conjunction of atomic predicates, which correspond to the query plan constraints implemented by the processed statements. Lastly, in lines~\ref{ln:derive-predicates:header-start}--\ref{ln:derive-predicates:header-end}, if \(\mathit{bb}\) is a loop header, we also add the atomic predicate corresponding to the loop itself to obtain \(\psi'_{\mathit{bb}}\), which characterizes executions when the loop condition holds.

\autoref{alg:statement-to-logic} implements translation of \texttt{for} and \texttt{if} statements to atomic logical predicates. The algorithm traverses \(\mathit{planTrace}\) to process the corresponding query plan constraint. For \textbf{extend} constraints (usually associated with \texttt{for} loops), the existential quantifier \(\exists\) is removed so that the constraint introduces a new free variable to \(\psi_{\mathit{bb}}\). \textbf{Check} constraints (associated with \texttt{if} statements) are returned as-is, because all their variables are already introduced by some enclosing \textbf{extend} operation. Thus, for a basic block \(\mathit{bb}\) enclosed by \(m\) statements with \textbf{extend} constraints will have \(\psi_{\mathit{bb}}\) with free variables \(v_{1}, \ldots, v_{m}\). If \(\mathit{bb}\) is a loop header, \(\psi'_{\mathit{bb}}\) has an additional \(v_{m+1}\) free variable.

\begin{example}
  \autoref{lst:ct-code} shows the generated source code of the \texttt{closeTrains} graph query, while \autoref{fig:ct-cfg} show the corresponding CFG (without any loop unrolling). The \(\mathrm{lineTrace}\) information is depicted as line numbers next to the CFG nodes, and comments above the control structures contain \(\mathrm{planTrace}\). The basic block \(\bb{9}\) corresponds to the loop header in line 17. Collecting the query plan constraints from the control structures enclosing line 17 with \autoref{alg:derive-predicates}, we find that
  {\footnotesize \begin{align*}
	\psi_{\bb{9}} &= \dslStyle{Train}(t) \wedge \dslStyle{location}(t, s) \wedge \dslStyle{connectedTo}(s, m) \\
	\psi'_{\bb{9}} &= \dslStyle{Train}(t) \wedge \dslStyle{location}(t, s) \wedge \dslStyle{connectedTo}(s, m) \wedge \dslStyle{connectedTo}(m, e).
  \end{align*}}
  In the concrete model \(M^{*}\) in \autoref{fig:instancemodel-generated}, executions of \(\bb{9}\) are represented by the 4 matches \(\mathit{Matches}(M, \psi_{\bb{9}}) = \{\{t \mapsto \mathrm{tr0}, s \mapsto \mathrm{tu0}, m \mapsto \mathrm{s1}\}, \{t \mapsto \mathrm{tr0}, s \mapsto \mathrm{tu0}, m \mapsto \mathrm{s2}\}, \{t \mapsto \mathrm{tr1}, s \mapsto \mathrm{s3}, m \mapsto \mathrm{s1}\}, \{t \mapsto \mathrm{tr0}, s \mapsto \mathrm{s3}, m \mapsto \mathrm{s2}\}\}\) of \(\psi_{\bb{9}}\), as well as the 8 matches of \(\psi'_{\bb{9}}\) obtained by extending each \(\psi_{\bb{9}}\) match by the two possible segments \dslStyle{connectedTo} the value of \(m\) as the value of the variable \(e\). Each match describes the values of the program variables when entering \(\bb{9}\).
\end{example}

\autoref{alg:precise-flow-facts} constructs precise domain-specific flow facts for a concrete model \(M\). In addition to the basic blocks \(\mathit{BB}\) and the traceability information, the algorithm reads the control flow graph \(\mathit{CFG} = \langle V, E, s, t, w, \mathit{tr} \rangle\) and the function \(\lVar{f}\colon E \to \LVars\) associating CFG edges with linear equation variables. Line \ref{ln:precise-flow-facts:init} initializes the empty system of linear equations \(\Scope{\text{flow}}\) and constructs the basic block predicates \(\Psi\). Leveraging the CFG traceability function \(\mathit{tr}\colon E \to \mathit{BB}\), expressions \(\sum_{e = \langle n_1, n_2 \rangle \in E, \mathit{tr}(n_1) = \mathit{bb}} \lVar{f}(e)\) are built, which represent the number of times a basic block \(\mathit{bb}\) is executed (\autoref{sec:background:ipet}). For a loop header, this number is equal to the number of \(\psi_{\mathit{bb}}\) and \(\psi'_{\mathit{bb}}\) matches in \(M\) (line~\ref{ln:precise-flow-facts:header}), while for other blocks, only matches of \(\psi_{\mathit{bb}}\) are counted (line~\ref{ln:precise-flow-facts:other}).

The resulting set of linear equations \(\Scope{\text{flow}}\) serve as flow facts in IPET analysis. More precisely, by incorporating \(\Scope{\text{flow}}\) into the analysis, we may obtain a safe and tight estimate for the execution time of a graph query program on the concrete model \(M\).

\begin{proposition}\label{prop:concrete-model-estimate}
  Let \(\tau\) be the execution time of the query program \(\texttt{q}\) on the concrete model \(M\),
  \begin{align*}
	\mathit{CL} &= \max \sum_{\lVar{x}_{i} \in \LVars} c_{i} \cdot \lVar{x}_{i} \textit{ subject to } \Scope{\text{IPET}} \text,
	& \mathit{DS}_{M} &= \max \sum_{\lVar{x}_{i} \in \LVars} c_{i} \cdot \lVar{x}_{i} \textit{ subject to } \Scope{\text{IPET}} \cup \Scope{\text{flow}} \text,
  \end{align*}
  where \(\mathit{CL}\) is the classical IPET estimated obtained from \(\texttt{q}\), and \(\mathit{DS}_{M}\) is the domain-specific estimate with flow facts derived from \(M\) using \autoref{alg:precise-flow-facts}. Then \(\tau \le \mathit{DS}_{M} \le \mathit{CL}\). (See proof in \autoref{sec:appendix}.)
\end{proposition}

\begin{figure}
	\LinesNumbered
	\SetAlFnt{\scriptsize}
	\SetAlCapFnt{\small \normalfont \sffamily}
	\SetAlCapNameFnt{\small \normalfont \sffamily}
	\RestyleAlgo{ruled}
	\begin{algorithm}[H]
		\caption{Witness generation task construction for a partial model \(P\)}
		\label{alg:witness-generation}
		\SetArgSty{upshape}
		\SetKwProg{Fn}{Function}{ is}{end}
		\Fn{\upshape WintessGenerationProblem(\(\mathit{BB}, \mathit{lineTrace}, \mathit{planTrace}, \mathit{CFG} = \langle V, E, s, t, w, \mathit{tr} \rangle, P = \langle \Obj{P}, \Int{P}, \Scope{P} \rangle, \mathcal{T} = \langle \Phi, \lVar{r} \rangle\)) }{
		    Construct the IPET analysis \(\max \sum_{\lVar{x}_i \in \LVars} c_i \cdot \lVar{x}_i \textit{ subject to } \Scope{\text{IPET}}\) for \(\mathit{CFG}\) with \(\lVar{f}\colon E \to \LVars\) (w.l.o.g.~\(\Ran \lVar{f} \cap \Ran \lVar{r} = \emptyset\))\label{ln:witness-generation:ipet}\\
		    \(\lVar{r}' = \lVar{r}\), \(\Scope{\text{merge}} = \emptyset\), \(\Psi = \textup{DerivePredicates}(\mathit{BB}, \mathit{lineTrace}, \mathit{planTrace})\)\label{ln:witness-generation:init}\\
		    \For{\(\mathit{bb} \in \mathit{BB}\)}{
		        \uIf{\(\mathit{bb}\) is a loop header}{
		            Let \(\lVar{x}, \lVar{x}' \in \LVars\) be two fresh variables not yet appearing in \(\Ran \lVar{f} \cup \Ran \lVar{r}'\)\\
		            \(\lVar{r}' = \lVar{r}' \cup \{ \psi_{\mathit{bb}} \mapsto \lVar{x}, \psi_{\mathit{bb}}' \mapsto \lVar{x}'\}\), \(\Scope{\text{merge}} = \Scope{\text{merge}} \cup \bigl\{\lVar{x} + \lVar{x}' -\sum_{e = \langle n_1, n_2 \rangle \in E, \mathit{tr}(n_1) = \mathit{bb}} \lVar{f}(e) = 0\}\)\label{ln:witness-generation:header}
	            }
	            \Else{
		            Let \(\lVar{x} \in \LVars\) be a fresh variable not yet appearing in \(\Ran \lVar{f} \cup \Ran \lVar{r}'\)\\
    	            \(\lVar{r}' = \lVar{r}' \cup \{ \psi_{\mathit{bb}} \mapsto \lVar{x} \}\),
    	            \(\Scope{\text{merge}} = \Scope{\text{merge}} \cup \bigl\{\lVar{x} - \sum_{e = \langle n_1, n_2 \rangle \in E, \mathit{tr}(n_1) = \mathit{bb}} \lVar{f}(e) = 0\bigr\}\)\label{ln:witness-generation:other}
	            }
		    }
		    \(\Scope{P'} = \Scope{P} \cup \Scope{\text{IPET}} \cup \Scope{\text{merge}}\), \(P' = \langle \Obj{P}, \Int{P}, \Scope{P'} \rangle\), \(\Phi' = \Phi \cup \Psi\), \(\mathcal{T}' = \langle \Phi', \lVar{r}' \rangle\)\label{ln:witness-generation:extend}\\
		    \Return \(\max \sum_{\lVar{x}_i \in \LVars} c_i \cdot \lVar{x}_i \textit{ subject to } \langle P', \mathcal{T}' \rangle\)
		}
	\end{algorithm}
\end{figure}

\subsection{Witness Generation of Worst-Case Execution Time}\label{sec:witness-generation}

To estimate the WCET of some query program \texttt{q} over a set of models, we specify the \emph{model scope} of interest as the \(\mathit{solutions}(P, \mathcal{T})\) of a model generation task. We construct an extended model generation problem \(\max \sum_{\lVar{x}_{i} \in \LVars} c_{i} \cdot \lVar{x}_{i} \textit{ subject to } \langle P', \mathcal{T}' \rangle\) in \autoref{alg:witness-generation}, where \(P \refined P'\) extends \(P\) with the results of IPET analysis and \(\mathcal{T}'\) incorporates the basic block predicates obtained from \texttt{q}. We use the notation \(\Ran \lVar{f}\) to denote the \emph{range} (possible values) of the function \(\lVar{f}\).

In line~\ref{ln:witness-generation:ipet}, the algorithm builds an IPET integer program based on the provided \(\mathit{CFG}\). Then, in line~\ref{ln:witness-generation:init}, we invoke \autoref{alg:derive-predicates} to obtain the basic block predicates \(\Psi\). The extended theory \(\mathcal{T} = \langle \Phi', \lVar{r}' \rangle\) is comprised of the predicates \(\Phi\) from the original theory \(\mathcal{T} = \langle \Phi, \lVar{r} \rangle\) and the basic block predicates \(\Psi\). Function \(\lVar{r}'\) extends \(\lVar{r}\) by assigning a variable \(\lVar{r}'(\psi_{\mathit{bb}})\) for each basic block predicate \(\psi_{\mathit{bb}}\) and a variable \(\lVar{r}'(\psi'_{\mathit{bb}})\) for each loop header predicate \(\psi'_{\mathit{bb}}\). By Definition~\ref{dfn:compatible}, in any concrete model \(M\) compatible with \(\mathcal{T}'\), we have \(\Scope{M} \vDash \lVar{r}'(\psi_{\mathit{bb}}) = M\#\psi_{\mathit{bb}}\) for each basic block \(\mathit{bb} \in \mathit{BB}\) and \(\Scope{M} \vDash \lVar{r}'(\psi'_{\mathit{bb}}) = M\#\psi'_{\mathit{bb}}\) for each loop header in addition to any constraints prescribed by the original theory \(\mathcal{T}\). Without loss of generality, we assume that  newly assigned variables are fresh (i.e., do not appear in either the original partial model scope \(\Scope{P}\) or in the IPET analysis \(\Scope{\text{IPET}}\)) and the variables of \(\Scope{\text{IPET}}\) are distinct from \(\Scope{P}\).

Lines~\ref{ln:witness-generation:header} and~\ref{ln:witness-generation:other} of \autoref{alg:witness-generation} correspond to lines~\ref{ln:precise-flow-facts:header} and~\ref{ln:precise-flow-facts:other} of \autoref{alg:precise-flow-facts}. However, we use the associated variables \(\lVar{r}'(\psi_{\mathit{bb}})\) and \(\lVar{r}'(\psi'_{\mathit{bb}})\) instead of the raw match counts \(M\#\psi_{\mathit{bb}}\) and \(M\#\psi'_{\mathit{bb}}\), so that the linear equations \(\Scope{\text{merge}}\) hold for any concrete model \(M \vDash \mathcal{T}\). Lastly, in line~\ref{ln:witness-generation:extend}, we assemble the extended model generation task and the partial model \(P' = \langle \Obj{P}, \Int{P}, \Scope{P'} \rangle\), where \(\Scope{P'}\) contains the original scope \(\Scope{P}\), the IPET linear equations \(\Scope{\text{IPET}}\), and the \(\Scope{\text{merge}}\) equations that merge the original model generation task and the IPET analysis together. The objective on the extended model generation task coincides with that of the IPET linear program.

The resulting extended model generation task provides a safe and tight estimate for the query program WCET
with theory \(\mathcal{T}\).

\begin{proposition}[Safety and tightness]\label{prop:partial-model-estimate}
  Let \(\tau(M)\) be the execution time of a query program \(\texttt{q}\) on a concrete model \(M\), \(P\) be partial model, \(\mathcal{T}\) be a theory, and
  \begin{align*}
	\mathit{CL} &= \max \sum_{\lVar{x}_{i} \in \LVars} c_{i} \cdot \lVar{x}_{i} \textit{ subject to } \Scope{\text{IPET}} \text,
	& \mathit{DS}_{P} &= \max \sum_{\lVar{x}_{i} \in \LVars} c_{i} \cdot \lVar{x}_{i} \textit{ subject to } \langle P', \mathcal{T}' \rangle \text,
  \end{align*}
  where \(\mathit{CL}\) is the classical IPET estimated obtained from \(\texttt{q}\), and \(\mathit{DS}_{P}\) is the domain-specific estimate based on the extended graph generation problem form \autoref{alg:witness-generation}. Then \(\tau(M) \le \mathit{DS}_{P} \le \mathit{CL}\) for all \(M \in \mathit{solutions}(P, \mathcal{T})\). (See proof of propositions in \autoref{sec:appendix}.)
\end{proposition}

Optimal solutions are \emph{witness models}, which maximize the domain-specific of WCET for concrete refinements of the input partial model \(P\) compatible with the theory \(\mathcal{T}\). As the witness model \(M^*\) is in the model scope, it is a feasible (as opposed to spurious) input of the query program.

\begin{proposition}[Witness model]\label{prop:witness-model}
  Let \(\mathit{DS}_{M}(M)\) be the domain-specific WCET estimate of a query program \texttt{q} obtained by \autoref{alg:precise-flow-facts} for a concrete model \(M\), \(\mathit{DS}_{P}\) be the domain-specific WCET estimate of \texttt{q} for a partial model \(P\) and theory \(\mathcal{T}\) by \autoref{alg:witness-generation}, and \(M^{*}\) be the witness model for the WCET of \texttt{q}, i.e., the optimal solution of \(\mathit{DS}_{P}\). Then \(M^{*} \in \mathit{solutions}(P, \mathcal{T})\) and \(\mathit{DS}_{M}(M) \le \mathit{DS}_{M}(M^{*}) = \mathit{DS}_{P}\) for all \(M \in \mathit{solutions}(P, \mathcal{T})\).
\end{proposition}

Moreover, refinements of the partial model \(P \refined Q\) may be used to tighten the WCET estimate by reducing the model scope under discussion.

\begin{proposition}[Tightening by refinement]\label{prop:refinement-squeezing}
  Let \(\mathit{DS}_{P}(P, \mathcal{T})\) denote the domain-specific WCET estimate of a query program \texttt{q} for a partial model \(P\) and theory \(\mathcal{T}\) obtained by \autoref{alg:witness-generation} and \(P \refined Q\). Then \(\mathit{DS}_{P}(Q, \mathcal{T}) \le \mathit{DS}_{P}(P, \mathcal{T})\). In particular, if \(P = P_{\mathit{init}}\) is the initial partial model for a metamodel \(\langle \Sigma, \alpha \rangle\) from \autoref{sec:model-generation}, then we may see that the WCET estimate for any partial model conforming to the metamodel is at least as tight as the \(\mathit{DS}_{\Sigma}\) estimate for the metamodel.
\end{proposition}

\begin{figure}[tb]
	\centering
	\resizebox{\linewidth}{!}{%
	    \def\svgwidth{2\linewidth}%
		\upshape
		\sffamily
		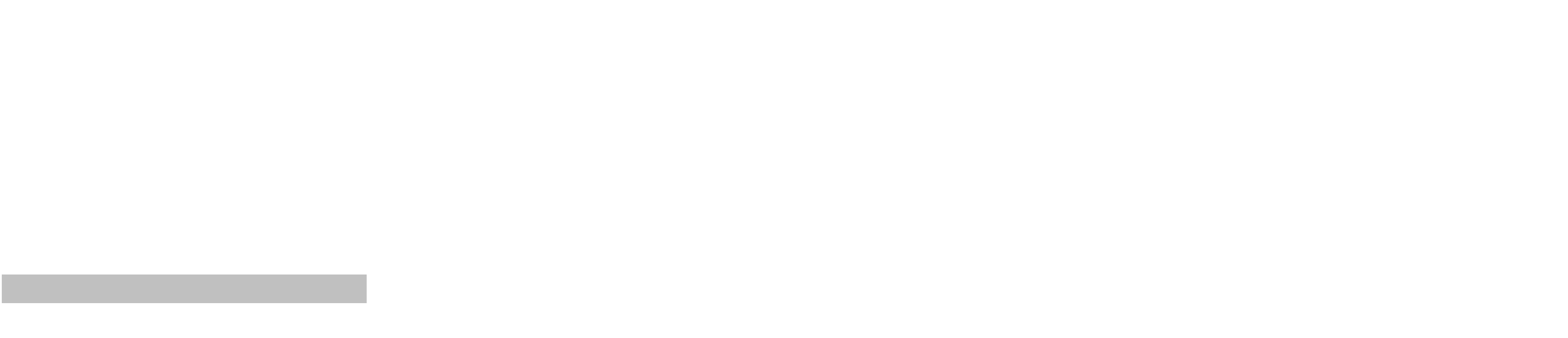}
	\caption{Execution of the graph generation task \(\max 250 \cdot \lVar{x}_{2} \textit{ subject to } \langle P', \mathcal{T}' \rangle\), where \(P' = P_{0}\), \(\mathcal{T}' = \langle \Phi', \lVar{r}' \rangle\), \(\Phi' = \{ \varphi_{\mathrm{Train}}, \varphi_{\mathrm{CT}} \}\), \(\varphi_{\mathrm{Train}} = \dslStyle{Train}(v_{1})\), \(\varphi_{\mathrm{CT}}\) is as in \autoref{fig:query-predicate}, \(\lVar{r}'(\varphi_{\mathrm{Train}}) = \lVar{x}_{1}\), and \(\lVar{r}'(\varphi_{\mathrm{CT}}) = \lVar{x}_{2}\).}\label{fig:refinement-example}
\end{figure}

\begin{example}
  \autoref{fig:refinement-example} shows a simplified execution of the graph generator. Suppose that \autoref{alg:witness-generation} has output an extended graph generation task
  \begin{equation*}
      \max 250 \cdot \lVar{x}_{2} \textit{ subject to } \langle P', \mathcal{T}' \rangle\text,
  \end{equation*}
  where \(P' = P_{0}\) as shown in \autoref{fig:refinement-example}. The extended theory is \(\mathcal{T}' = \langle \Phi', \lVar{r}' \rangle\), there \(\Phi' = \{ \varphi_{\mathrm{Train}}, \varphi_{\mathrm{CT}} \}\), \(\varphi_{\mathrm{Train}} = \dslStyle{Train}(v_{1})\), \(\varphi_{\mathrm{CT}}\) is the predicate from the \dslStyle{closeTrains} query from \autoref{fig:query-predicate}, \(\lVar{r}'(\varphi_{\mathrm{Train}}) = \lVar{x}_{1}\), and \(\lVar{r}'(\varphi_{\mathrm{CT}}) = \lVar{x}_{2}\).

  The inequalities \([0 \le \lVar{x}_{1} \le 2]\) in conjunction with the theory \(\mathcal{T}'\) prescribe a \emph{type scope} of between 0 and 2 \dslStyle{Train} instances. The objective function \(g(M) = \max_{k \vDash \Scope{M}}250 \cdot k(\lVar{x}_{2})\) corresponds to each match of \(\varphi_{\mathrm{CT}}\) (i.e., \dslStyle{closeTrains}) taking 250 clock ticks to calculate.

  Two non-isomorphic partial models \(P_{1}\), \(P_{2}\) can be obtained from \(P_{0} = P'\). In \(P_{1}\), the generator placed a train \(\mathrm{tr}_{1}\) on the middle segment \(\mathrm{s}_{2}\) of the track. Therefore, the multi-object \(\mathrm{tr}_{\mathrm{new}}\) represents at most one additional train (\(1 \le \lVar{x}_{1} \le 2\)). The \(\varphi_{\mathrm{CT}}\) predicate cannot match (\(\lVar{x}_{2} = 0\)), since it is impossible to place a new train on both \(\mathrm{s}_{1}\) and \(\mathrm{s}_{3}\). Any possible concrete refinement \(P_{1} \refined M\) of \(P_{1}\) has an objective value \(g(M) = 250 \cdot 0 = 0\).

  If we place a new train on \(\mathrm{s}_{1}\), we obtain \(P_{2}\). Placing a train on \(\mathrm{s}_{3}\) results in an isomorphic model, so it is sufficient to only consider \(P_{2}\) instead. Here, there can be \(0\) or \(1\) matches \(\varphi_{\mathrm{CT}}\) (\(0 \le \lVar{x}_{2} \le 1\)). Indeed, if we place an additional train on \(\mathrm{s}_{3}\), we obtain the concrete model \(M^{*} = P_{3}\) with a single match of \(\varphi_{\mathrm{CT}}\) (\(\lVar{x}_{2} = 1\)). The corresponding objective value is \(g(M^{*}) = 250 \cdot 1 = 250\). No larger objective value is possible by any concrete refinement of \(P_{1}\), since \(\Scope{P_{1}} \vDash \lVar{x}_{2} = 0\). Hence we may discard \(P_{1}\) along with its potential refinements, and output \(M^{*}\) as the witness model obtained as the optimal solution of model generation task.

  Assuming that the (simplified) objective function \(g\) is the domain-specific WCET estimate of the query program, \(g(M^{*}) = 250\) is our WCET estimate, which execution time bound is expected to be reached (according to the low-level IPET analysis) when executing the query program over the witness model \(M^{*}\) as input. 
  
\end{example}

\section{Evaluation}
\label{sec:evaluation}

We conducted experiments to address the following research questions related to the WCET of query programs. For each research question, we investigate the scenario (a)~\(\text{DS}_\Sigma\), where only the metamodel and the relevant well-formedness and scope constraints are known; and (b)~\(\text{DS}_P\), when an initial partial model (describing a track layout but not its runtime state) is also provided.

\bgroup\renewcommand{\labelenumi}{RQ\arabic{enumi}}
\begin{enumerate}
    
    \item How difficult is it to find witness models?
    
    \item How safe and tight are WCET estimates w.r.t.~existing approaches and real execution times? %
     
    \item How does query program complexity impact the overestimation of computed WCET bounds? 

\end{enumerate}\egroup%

RQ1 aims at determining whether our model generation based approach can find the witness model in practical time and whether it constitutes an improvement over random search. The rest of the experiments study the quality of WCET bounds, which is a key factor in the applicability of our approach. In particular, RQ2 attempts to compare our computed WCET estimates with the state of the art in challenging settings with partial runtime information, while RQ3 presents increasingly challenging query programs to our approach.

\subsection{Evaluation Overview and Setup}

\subsubsection{Queries}

To address these research questions, we use graph queries from the domain of the \modes{} CPS demonstrator~\cite{NFM2018}. This demonstrator uses high-level runtime monitoring rules captured as graph queries, and showcases synthesized monitoring programs executing these queries over the runtime graph model of the underlying running system. 
Our experiments focus only on query evaluation, and updates to the runtime model are out of scope for the current paper. Therefore, we ran the query programs on various snapshots of runtime graph models.
We evaluated the following queries adopted from~\cite{FASE2018}:
\begin{itemize}
    \item \textbf{Close trains} (\textbf{ct}): This is the query introduced in the running example of \autoref{sec:query-program-structure}. 
    \item \textbf{End of siding} (\textbf{eos}): This query finds trains that are dangerously close (one segment distance) to an end of the track.
    \item \textbf{Misaligned turnout} (\textbf{mt}): Pairs of trains and turnouts are the objectives of this query, where the train would derail if it reached the turnout because it is switched in a different direction.
    \item \textbf{Train locations} (\textbf{tl}): A simple query to find pairs of trains and segments that describe the locations of each train. 
\end{itemize}

The calculation of query search plans is out of scope of the current paper, but they were created and optimized based on the typical model statistics of runtime model snapshots in the \modes{} system.
For example, the search plan presented in \autoref{tab:close-trains-plan} is the one used by the program executing the query \infigure{Close trains}.

\subsubsection{WCET algorithms and WCET tools}

To compare the results produced by our WCET estimation approaches \(\text{DS}_\Sigma\) and \(\text{DS}_P\) with estimates produced by other tools, we used the commercial aiT~\cite{Ferdinand2004}~(version 20.10i) and the open-source OTAWA~\cite{Ballabriga2006}~(version V1.2.0) tools. For aiT, we used a \emph{high precision} configuration with pipeline-level analysis and full~(up to the determined loop bound) loop unrolling, as well as a \emph{low precision} configuration with only basic block-level analysis and no unrolling. To incorporate the results of low-level analysis into \(\text{DS}_\Sigma\) and \(\text{DS}_P\), we extracted the IPET linear equations from the low-level configuration of aiT manually, as no facility was available for automatic export or accessing the high precision system of linear equations directly. We also extracted the IPET linear equations from OTAWA, which have BB execution context information~(paths of length two).%

\subsubsection{Graph models}\label{sec:eval-graph-models}
In the following, we describe how we obtained a variety of models to assess the impact of models with different characteristics on query evaluation times.

Using the metamodel in the \modes{} case study, we generated \emph{witness models} \(M^{*}_\Sigma\) for each monitoring query and for both low-level analyses (aiT, OTAWA) such that the query is estimated to have the longest possible execution time according the low-level analysis. For all of these models, we used the same model scope inspired by the railway domain: up to 20\% of the objects can be \infigure{Trains} and up to 20\% of the objects can be \infigure{Turnouts}. The rest of the objects are \infigure{Segments}; we capped the maximum number of objects at 25.
The resulting models are syntactically valid and they can represent a realistic railway system thanks to the domain-specific well-formedness constraints.

To obtain a \emph{realistic model} \(M_\text{real}\), we manually captured a detailed runtime model snapshot of \modes{} that is similar to the one presented in \autoref{fig:instance-model-modes3} with a total of 25 objects.
Then, we removed all \infigure{Train} objects from \(M_\text{real}\) and unset all turnout directions, and used the resulting (partial) track layout \(P\) to find specific placements of trains and switching of turnouts such that the run times of the queries are maximized on the generated \(M^*_P\) witness models.

To assess the execution times of the query programs on random models, we generated models conforming to the \modes{} metamodel with up to a total of 25 objects. Due to the large space of possible graph models, representative sampling from the model space is an open question~\cite{Jackson2013,Semerath2020}. Nevertheless, we generated 250 models with the EMF random model generator\footnote{\url{https://github.com/atlanmod/mondo-atlzoo-benchmark}} (\textbf{Rand}) with up to 5 \infigure{Turnouts} and up to 5 \infigure{Trains}, but none of them represents a railway setting that can occur because they all violated well-formedness constraints due to the completely random construction.

We also generated 250 models with the VIATRA Generator (\textbf{VG}) without an optimization objective, which satisfy all well-formedness and scope constraints used for generating witness models. However, the state exploration heuristics of the generator may lead to a biased sample.

\subsubsection{Hardware setup}

We use the Infineon Relax Lite Kit-V1 Board\footnote{\url{http://www.infineon.com/xmc-dev}} to execute the query programs. This board has an XMC4500 F100-K1024 microcontroller and it is driven by a 120MHz system clock. This microcontroller is considered to be a mature industrial microcontroller and has an ARM Cortex-M4 core. For the present evaluation, the instruction cache  on the device is not used as our primary focus is on the impact of domain-specific information about high-level program flow rather than microarchitectural effects. 

The bare-metal query programs are compiled with GCC compiler for ARM version 7.2.1 with \texttt{-O0} and \texttt{-g3} flags in debug mode.
These programs run on the microcontroller while no other tasks (e.g., interrupts) are running. We rely on the cycle counter feature of the Data Watchpoint and Trace Unit in the device to extract the execution times of each query using a debugger. 
The embedded code used for the experiments as well as compiler and other  configurations are available online\footnote{%
\url{https://imbur.github.io/cps-query/}}. 

\subsection{Evaluation Results}
\label{sec:evaluation-results}
  
\begin{figure}
\begin{minipage}[c]{.44\textwidth}
    \subfigure[Measured query times over consistent models, random models, witness models, and realistic models\label{fig:boxplot-exectimes} 
    ]{ 
        \includegraphics[width=\linewidth]{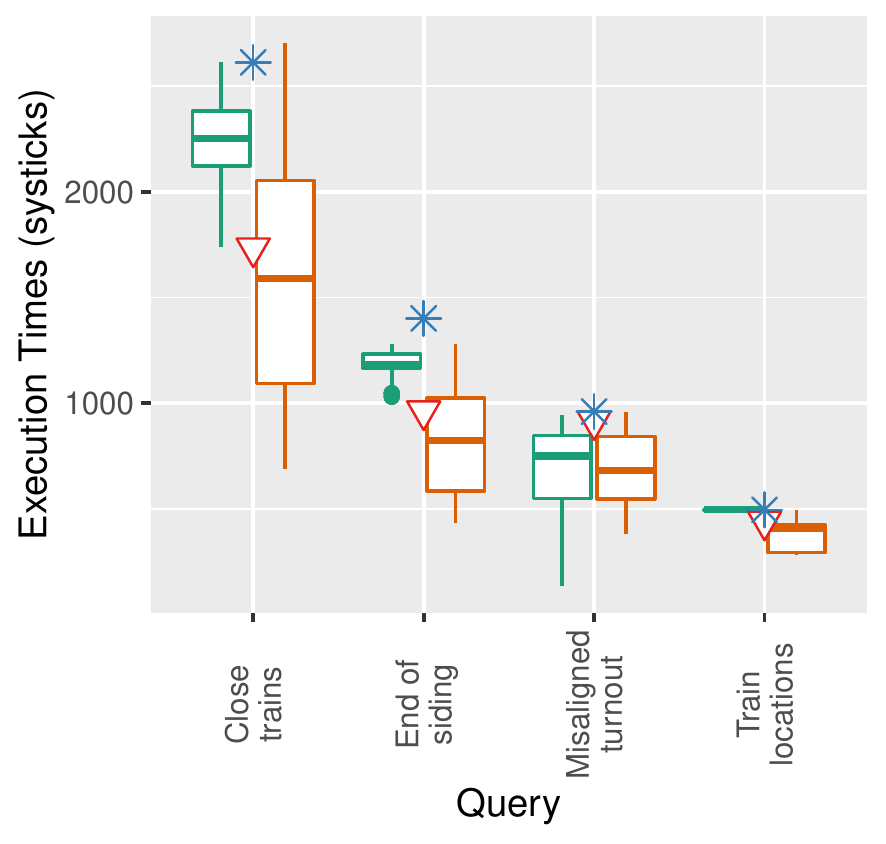}
    }
\end{minipage}
\hfill 
\begin{minipage}[c]{.43\textwidth}{
    \def\svgwidth{.55\linewidth}{
        \upshape
		\footnotesize 
		\sffamily
		\begingroup%
  \makeatletter%
  \providecommand\color[2][]{%
    \errmessage{(Inkscape) Color is used for the text in Inkscape, but the package 'color.sty' is not loaded}%
    \renewcommand\color[2][]{}%
  }%
  \providecommand\transparent[1]{%
    \errmessage{(Inkscape) Transparency is used (non-zero) for the text in Inkscape, but the package 'transparent.sty' is not loaded}%
    \renewcommand\transparent[1]{}%
  }%
  \providecommand\rotatebox[2]{#2}%
  \newcommand*\fsize{\dimexpr\f@size pt\relax}%
  \newcommand*\lineheight[1]{\fontsize{\fsize}{#1\fsize}\selectfont}%
  \ifx\svgwidth\undefined%
    \setlength{\unitlength}{129.62994003bp}%
    \ifx\svgscale\undefined%
      \relax%
    \else%
      \setlength{\unitlength}{\unitlength * \real{\svgscale}}%
    \fi%
  \else%
    \setlength{\unitlength}{\svgwidth}%
  \fi%
  \global\let\svgwidth\undefined%
  \global\let\svgscale\undefined%
  \makeatother%
  \begin{picture}(1,0.65881781)%
    \lineheight{1}%
    \setlength\tabcolsep{0pt}%
    \put(0,0){\includegraphics[width=\unitlength,page=1]{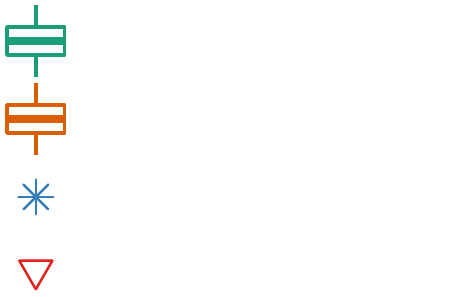}}%
    \put(0.19191335,0.54527217){\color[rgb]{0,0,0}\makebox(0,0)[lt]{\lineheight{1.25}\smash{\begin{tabular}[t]{l}Consistent models (\textbf{VG})\end{tabular}}}}%
    \put(0.18980739,0.37201015){\color[rgb]{0,0,0}\makebox(0,0)[lt]{\lineheight{1.25}\smash{\begin{tabular}[t]{l}Random models (\textbf{Rand})\end{tabular}}}}%
    \put(0.19416979,0.19874808){\color[rgb]{0,0,0}\makebox(0,0)[lt]{\lineheight{1.25}\smash{\begin{tabular}[t]{l}Witness model (\(M^*_\Sigma\))\end{tabular}}}}%
    \put(0.18980739,0.025486){\color[rgb]{0,0,0}\makebox(0,0)[lt]{\lineheight{1.25}\smash{\begin{tabular}[t]{l}Realistic model (\(M_\text{real}\))\end{tabular}}}}%
  \end{picture}%
\endgroup%
}
        \par}%
        \vspace*{25pt}
    \centering
    \subtable[Results of state space exploration during witness generation from partial model using IPET linear equations adapted from aiT low-precision\label{tab:witness-model-space} 
    ]{{\footnotesize
			\begin{tabular}{@{}lrr@{}}
				\toprule%
				Query & States visited & \% of total \\ \midrule
				Close trains & 992\,681 & 19\,\% \\
				End of siding & 878\,243 & 17\,\% \\
				Misaligned turnout & 875 & \(< 1\,\%\) \\
				Train locations & 144 & \(< 1\,\%\) \\
				\midrule
                Total & 5\,273\,100 & 100\,\% \\
				\bottomrule
			\end{tabular}\par}}
\end{minipage}
\caption{Query execution times on fully random models and realistic models}
\label{fig:measured-exectimes}
\end{figure}

\subsubsection{Research Question 1 (a)} We investigate if a witness model for a query can be obtained from simpler graph generation approaches, and we do this by measuring the execution times of queries over various models. Our results are presented in \autoref{fig:boxplot-exectimes}. The run times over models by \textbf{VG} is captured by the green boxes, while the orange ones show the run times over models by \textbf{Rand}. Each query was evaluated on the same two sets of models.
Additionally, the respective query execution time over each witness model \(M^*_\Sigma\) is added to these figures for comparison, where \(M^*_\Sigma\) is the witness model generated using the objective function built from the low-level analysis results of aiT. %
Moreover, the run time over the hand-crafted \(M_{\text{real}}\) model is also presented.

\textit{Findings.} For each consistent model considered, queries exhibited the longest observed execution times on their respective witness models. %
In fact, for two queries, \infigure{eos} and \infigure{mt}, the execution time on the witness is longer than the maximum measured execution time over any other consistent model, which highlights the importance of our witness model generation technique.

Maximum run times over models generated by \textbf{Rand} can be both higher and lower than on witness models. For example, query \infigure{ct} takes 2\% shorter over a random model than over its witness model, but the random model does not represent a realistic railway. %
On the contrary, query \infigure{eos} takes at least 8\% longer to complete on the witness model than on any model generated by \textbf{Rand}.
\textit{Therefore, computing safe and tight WCET estimates of queries which execute over well-formed models (1) is infeasible by collecting run times over random models, and (2) necessitates finding witness models by employing sophisticated model generation approaches.}

\subsubsection{Research Question 1 (b)}
All possible refinements of \(P\) constitute a potentially large model space where finding \(M^*_P\) can be challenging and requires the graph solver to apply suitable abstractions for optimization.
In our case, there are \(3^5 \cdot \sum_{i = 0}^5 \binom{20}{i} = 5\,273\,100\) models (Total row in \autoref{tab:witness-model-space}) in the space of well-formed concrete refinements of the initial partial model \(P\) containing the (selected) track layout, because each of the 5 turnouts can be in three  different states, and there can be up to 5 trains which must be located on different segments. Thus, explicit enumeration of all models is possible for such a track layout, although it is computationally expensive.

As described in \autoref{sec:eval-graph-models}, we used the graph generator to add trains to a model with an empty track layout such that the expected query run times are maximized. \autoref{tab:witness-model-space} presents the number of states explored by the graph generator compared to the space of all refinements.

\textit{Findings.} 
The number of states visited by the generator increases with the complexity of the query. For the most complex query \infigure{ct}, the generator was able to find the model with the highest estimated WCET after visiting 19\% of the model space. For the least complex \infigure{tl} query, it visited only \(144\) states, which allowed the witness generation to finish almost instantly. In conclusion, the generator explores a fraction of the state space, making it more favorable than explicit enumeration.

\subsubsection{Research Question 2} Our goal is to compare the computed WCETs obtained from different tools with our own techniques.
For this RQ, we restrict our investigation to the scenario \(\textrm{DS}_\Sigma\) where only the metamodel and the well-formedness and scope constraints are known (i.e., case \textbf{(a)}), because only our technique but not the baseline tools (aiT, OTAWA) support processing a partial model as in \(\textrm{DS}_P\) (i.e., case \textbf{(b)}). 
\autoref{tab:wcet-estimates} shows the WCET estimates for the 4 queries along with measured execution time (expressed in systicks) over the respective witness model.

\begin{table}[tb]
    \caption{Query code complexity, measured execution time, and WCET estimates in systicks}
    \label{tab:wcet-estimates}
{\footnotesize \begin{tabular}{@{}lrrrrrrr@{}}
\toprule
      & & & \multicolumn{2}{c}{\(\textbf{DS}_\Sigma\)} & \multicolumn{2}{c}{aiT} &  \\\cmidrule(lr){4-5}\cmidrule(lr){6-7}
Query & CC & \multirow{-2}{*}{\parbox{1.3cm}{\centering Exec. time\\over \(M^*_\Sigma\) \strut}}& \textbf{w/aiT} & \textbf{w/OTAWA} & low precision & high precision & OTAWA \\\midrule
Close trains   & 7   & 2652 & \textbf{\underline{3133}} & \textbf{\underline{3430}} & 3563 & 3038 & 4210  \\
End of siding  & 6   & 1395 & \textbf{1757} & \textbf{\underline{1820}} & 1757 & 1477 & 1860  \\
Misaligned t.  & 5   & 939  & \textbf{1097} & \textbf{1370} & 1097 & 987  & 1370  \\
Train locations& 3   & 489  & \textbf{592}  & \textbf{695}  & 592  & 507  & 695   \\
 \bottomrule
\end{tabular}}
\end{table}

\begin{table}[tb]
    \caption{Query code complexity, measured execution time, and WCET estimates with a partial model}
    \label{tab:wcet-p-estimates}
{\footnotesize\begin{tabular}{@{}lrrrr|rr@{}}
\toprule
      &  & & \multicolumn{2}{c|}{\(\textbf{DS}_P\)} & \multicolumn{2}{c}{\color{gray} aiT with partial memory image} \\\cmidrule(lr){4-5}\cmidrule(l){6-7}
Query & CC & \multirow{-2}{*}{\parbox{1.3cm}{\centering Exec. time\\over \(M^*_P\) \strut}}& \textbf{w/aiT} & \textbf{w/OTAWA} & {\color{gray} low precision} & {\color{gray} high precision} \\\midrule
Close trains   & 7   & 2544 & \textbf{\underline{3079}} & \textbf{\underline{3338}} &  {\color{gray} 3706} &  {\color{gray} 3091}   \\
End of siding  & 6   & 1275 & \textbf{\underline{1554}} & \textbf{\underline{1636}} &  {\color{gray} 1813} &  {\color{gray} 1523}   \\
Misaligned t.  & 5   & 969  & \textbf{1097} & \textbf{1370} &  {\color{gray} 1065} &  {\color{red!90!black} 950}   \\
Train locations& 3   & 504  & \textbf{592}  & \textbf{695}  &  {\color{gray} 586}  &  {\color{gray} 506}   \\
 \bottomrule
\end{tabular}}
\end{table} 

\textit{Findings.} 
In the case of \infigure{ct}, our WCET estimation approach produces estimates 14\% tighter than the one by aiT (low precision analysis). It is also important to point out that even without context-sensitive BB timings, our low precision approach provides only 3\% higher estimates than aiT's high precision mode, which indicates that it is able to automatically identify infeasible paths in the program based on high-level domain-specific information.
For OTAWA, improvements of the WCET estimate achieved in two cases: \infigure{ct} has a 23\%, while \infigure{eos} has a 2\% tighter estimate.
For the rest of the queries, the analysis yields the same results as aiT low precision mode or OTAWA. %

Therefore, WCET estimates by \(\text{DS}_\Sigma\) were at least as tight as those obtained by low-level IPET analysis. Thus, domain-specific analysis can improve  WCET estimates while simultaneously synthesizing witness models to study query program behavior. Conceptually, it would be possible to formulate more precise \(\text{DS}_\Sigma\) estimates by incorporating low-level analysis results from the high precision mode of aiT as shown in \autoref{sec:witness-generation}, but such equations cannot be obtained from aiT. %

\subsubsection{Research Question 3 (a)} With this RQ, we look at the impact of query complexity on the computed WCET bounds, so that we can give recommendations on where our approach offers the greatest benefits.  
The execution times of queries over \(M^*_\Sigma\) in \autoref{tab:wcet-estimates} provide a lower bound to the actual WCET (i.e., the longest possible execution time of the program over inputs which represent well-formed models in the model scope), while the CC columns shows query cyclomatic complexity.
Since the actual WCET of the program is unknown (but it must lay between the measured execution time and the WCET estimates produced by the analyses), we use the measured execution time over witness models as the baseline when discussing overestimation in WCET estimates.

\textit{Findings.}
The biggest visible advantage of \(DS_\Sigma\) is in the case of the most complex query \infigure{ct}: the overestimation is 18\% with BB timings from aiT, while the aiT low precision analysis computes a 34\% higher value. In other cases, it produces the same result as aiT, with overestimates being between 16\% (query \infigure{mt}) and 26\% (query \infigure{eos}). We come to the same conclusion using BB timings from OTAWA, although these timings are slightly more conservative. The high precision analysis available in aiT is able to leverage the microarchitectural properties and thus provide the most precise estimates with the overestimation being 14\% (observed for query \infigure{ct}). The overestimation increases with CC of the query code only in the case of high precision aiT analysis.

In general, \(\text{DS}_\Sigma\) computes a safe WCET bound and additionally provides a witness model. Moreover, it is able to discover additional infeasible paths the WCET estimate for the  most complex query, thus provide a tighter estimate.

\subsubsection{Research Question 3 (b)} The goal of this RQ is to conclude if providing an initial graph model with elements of the graph model known upfront can lower the WCET, thus provide an execution time estimate for \(M^*_P\) lower than the one computed for \(M^*_\Sigma\).
The execution times of queries over \(M^*_P\) are presented in \autoref{tab:wcet-p-estimates} similarly to \autoref{tab:wcet-estimates}. The estimates computed by \(\text{DS}_P\) are valid and safe for any possible in-memory representation of the refinements of \(P\) (concrete models containing the designated track layout).

Additionally, \autoref{tab:wcet-p-estimates} shows estimates from aiT computed by value analysis (VAL) over the initial track layout model provided to aiT as a \emph{partial memory image}, where the unknown parts of \(P\) (locations of trains and the directions of turnouts) are left uninitialized. While WCET estimates based on this input are \emph{not safe}, as they do not consider all possible in-memory representations of the refinements of \(P\), we still added these numbers to \autoref{tab:wcet-p-estimates}, because they correspond to the most likely usage of existing WCET analysis tools with partial input.

\textit{Findings.}
Comparing \(\text{DS}_\Sigma\) and \(\text{DS}_P\), we discover that providing an initial model tightens WCET estimates for the two more complex queries, but does not change the estimate for the two less complex ones. 
This is due to the nature of the initial model and query search plans. There is no arrangement of trains on the initial track layout such that the WCET estimate from \(\text{DS}_\Sigma\) for \infigure{ct} and \infigure{eos} run times is reached, whereas the evaluation of \infigure{mt} and \infigure{tl} depends less on the track layout.

\textit{Additional observations.}
For \infigure{mt} and \infigure{tl}, observed query run times on the witness models \(M^*_P\) were higher than on \(M^*_\Sigma\), which can be attributed to the placement of data in memory as mentioned in \autoref{sec:graph-model-assumptions}. Nevertheless, they were still within the WCET estimates from both \(\text{DS}_\Sigma\) and \(\text{DS}_P\). %

Unexpectedly, the partially specified data yields higher WCET estimates by both analysis modes of aiT for the two more complex queries, \infigure{ct} and \infigure{eos}, when compared with the estimates in \autoref{tab:wcet-estimates}. Thus, for these queries, it is not possible to tighten WCET estimates for partial input data even with the caveat that all objects in the partial input are statically allocated. 
We have reported our observation to the developers of aiT at AbsInt GmbH, and they have confirmed that the discrepancy in the estimates is due to the differences in the placement of data in the two binaries. %
Note that these analyses solve a less general challenge compared to \(\mathrm{DS}_P\) since  they only consider one possible physical layout of the partial data in memory (and not all possible in-memory layouts), thus they are highlighted in {\color{gray}gray} in \autoref{tab:wcet-p-estimates}.

On the other hand, aiT high-precision mode produces a \textit{lower estimate} based on the provided initial model than what is measured over \(M^*_P\) (highlighted in {\color{red}red} in \autoref{tab:wcet-p-estimates}). The underlying reason for this is that the tool relies on the exact placement of data in memory rather than the abstract graph model the data encodes. Eventually, in \(M^*_P\), the model objects were stored in the memory in different order which resulted in a higher runtime (see note about data placement in \autoref{sec:graph-model-assumptions}).

\textit{In general, providing partial input data allows for tightening WCET estimated for complex queries even in cases where value analysis with partial input data is not safe. For less complex queries, supplying a partial memory image can tighten the results of value analysis considerably, but requires committing to a specific in-memory representation of the partial data statically.}

\subsection{Threats to Validity} 

\textit{Internal validity.}
The current evaluation was performed on a device where the executing binary only included the query-based monitor, so we can assume that the measurements presented here precisely show the execution times of queries. Since the exact WCET of the program is unknown, we used the longest observed execution time to assess the overestimation of the computed WCETs. In reality, this overestimation might be lower than what is reported here, which would make our WCET estimates tighter than presented.

\noindent\textit{External validity.} We carried out the evaluation using one specific hardware and compiler, thus the presented results may not generalize to other platforms. Furthermore, the presented approach is applicable to any query-based monitor generated with \autoref{alg:code-generation}. However, evaluation of the WCET estimation techniques using additional case-studies with query-based runtime monitors from different domains could further improve the confidence in the evaluation results.

\section{Conclusions and Future Work}
\label{sec:summary}

In this paper, we presented a method to \emph{provide safe and tight WCET bounds for runtime monitoring programs derived from graph queries} to enable their use in real-time systems. We provided a static WCET estimate by incorporating low-level analysis results from traditional IPET-based tools and high-level domain-specific constraints into the objective function of an advanced graph solver. In addition to a tight WCET estimate, the result also entails a witness graph model where the query-based monitoring program execution time is expected to be the longest. %

We carried out extensive evaluation of our approach on an industry-grade hardware platform using a variety of graph models as inputs for query programs, and assessed the tightness of computed WCET by comparing it to the results produced by two different tools. We constructed witness models for highest estimated execution times of queries as well as random graph models as inputs for graph query programs as an attempt to showcase high execution times. While we have no formal guarantee that worst-case timing behavior is exhibited on witness models as inputs, in all our experiments, the longest execution times were always measured on such witness models.

In the short run, the proposed approach can be improved by passing the results of high-precision IPET analysis (including CFG unrolling and pipeline analysis) to the graph solver, while the evaluation of the approach should be done on a different hardware platforms as well. As a part of a long-term future research agenda, our approach could be extended to provide witness models with specific data placement in memory where the execution time equals to the WCET of the program.

\begin{acks}
This work has been partially supported by NSERC RGPIN-04573-16 project and the MEDA scholarship program. The research reported in this paper and carried out at the BME has been supported by the NRDI Fund based on the charter of bolster issued by the NRDI Office under the auspices of the Ministry for Innovation and Technology. The authors would like to thank AbsInt GmbH for providing us a license for the aiT timing analyzer, and also would like to thank Martin Sicks for the extensive introduction to the tool. Similarly, we are thankful to Julien Forget and Cl{\'e}ment Ballabriga for helping with using OTAWA.
\end{acks}

\bibliographystyle{ACM-Reference-Format}
\bibliography{rt-queries-modelgen}


\begin{thebibliography}{67}


\ifx \showCODEN    \undefined \def \showCODEN     #1{\unskip}     \fi
\ifx \showDOI      \undefined \def \showDOI       #1{#1}\fi
\ifx \showISBNx    \undefined \def \showISBNx     #1{\unskip}     \fi
\ifx \showISBNxiii \undefined \def \showISBNxiii  #1{\unskip}     \fi
\ifx \showISSN     \undefined \def \showISSN      #1{\unskip}     \fi
\ifx \showLCCN     \undefined \def \showLCCN      #1{\unskip}     \fi
\ifx \shownote     \undefined \def \shownote      #1{#1}          \fi
\ifx \showarticletitle \undefined \def \showarticletitle #1{#1}   \fi
\ifx \showURL      \undefined \def \showURL       {\relax}        \fi
\providecommand\bibfield[2]{#2}
\providecommand\bibinfo[2]{#2}
\providecommand\natexlab[1]{#1}
\providecommand\showeprint[2][]{arXiv:#2}

\bibitem[\protect\citeauthoryear{Abdessalem, Panichella, Nejati, Briand, and
  Stifter}{Abdessalem et~al\mbox{.}}{2018}]%
        {BriandTest}
\bibfield{author}{\bibinfo{person}{Raja~Ben Abdessalem},
  \bibinfo{person}{Annibale Panichella}, \bibinfo{person}{Shiva Nejati},
  \bibinfo{person}{Lionel~C. Briand}, {and} \bibinfo{person}{Thomas Stifter}.}
  \bibinfo{year}{2018}\natexlab{}.
\newblock \showarticletitle{Testing Autonomous Cars for Feature Interaction
  Failures Using Many-Objective Search}. In \bibinfo{booktitle}{\emph{33rd
  ACM/IEEE International Conference on Automated Software Engineering}}.
  \bibinfo{pages}{143--154}.
\newblock
\showISBNx{9781450359375}


\bibitem[\protect\citeauthoryear{Abella et~al\mbox{.}}{Abella
  et~al\mbox{.}}{2015}]%
        {Abella2015}
\bibfield{author}{\bibinfo{person}{Jaume Abella} {et~al\mbox{.}}}
  \bibinfo{year}{2015}\natexlab{}.
\newblock \showarticletitle{{WCET analysis methods: Pitfalls and challenges on
  their trustworthiness}}.
\newblock \bibinfo{journal}{\emph{10th IEEE International Symposium on
  Industrial Embedded Systems - Proceedings}} (\bibinfo{year}{2015}),
  \bibinfo{pages}{39--48}.
\newblock
\showISBNx{9781467377119}


\bibitem[\protect\citeauthoryear{Ballabriga, Cass{\'{e}}, Rochange, and
  Sainrat}{Ballabriga et~al\mbox{.}}{2010}]%
        {Ballabriga2006}
\bibfield{author}{\bibinfo{person}{Cl{\'{e}}ment Ballabriga},
  \bibinfo{person}{Hugues Cass{\'{e}}}, \bibinfo{person}{Christine Rochange},
  {and} \bibinfo{person}{Pascal Sainrat}.} \bibinfo{year}{2010}\natexlab{}.
\newblock \showarticletitle{{OTAWA: An Open Toolbox for Adaptive WCET
  Analysis}}.
\newblock In \bibinfo{booktitle}{\emph{LNCS}}. Vol.~\bibinfo{volume}{6399}.
\newblock
\showISBNx{364216255X}
\showISSN{03029743}


\bibitem[\protect\citeauthoryear{Ballabriga, Forget, and Lipari}{Ballabriga
  et~al\mbox{.}}{2017}]%
        {Ballabriga2017}
\bibfield{author}{\bibinfo{person}{Cl{\'{e}}ment Ballabriga},
  \bibinfo{person}{Julien Forget}, {and} \bibinfo{person}{Giuseppe Lipari}.}
  \bibinfo{year}{2017}\natexlab{}.
\newblock \showarticletitle{Symbolic {WCET} computation}.
\newblock \bibinfo{journal}{\emph{ACM Trans. Embedded Comput. Syst.}}
  \bibinfo{volume}{17}, \bibinfo{number}{2} (\bibinfo{year}{2017}).
\newblock
\showISSN{15583465}


\bibitem[\protect\citeauthoryear{Bartocci et~al\mbox{.}}{Bartocci
  et~al\mbox{.}}{2018}]%
        {Bartocci2018}
\bibfield{author}{\bibinfo{person}{Ezio Bartocci} {et~al\mbox{.}}}
  \bibinfo{year}{2018}\natexlab{}.
\newblock \showarticletitle{{Specification-Based Monitoring of Cyber-Physical
  Systems: A Survey on Theory, Tools and Applications}}.
\newblock In \bibinfo{booktitle}{\emph{Lectures on Runtime Verification}}.
  \bibinfo{pages}{135--175}.
\newblock


\bibitem[\protect\citeauthoryear{Blair, Bencomo, and France}{Blair
  et~al\mbox{.}}{2009}]%
        {BlairBF09}
\bibfield{author}{\bibinfo{person}{Gordon~S. Blair}, \bibinfo{person}{Nelly
  Bencomo}, {and} \bibinfo{person}{Robert~B. France}.}
  \bibinfo{year}{2009}\natexlab{}.
\newblock \showarticletitle{Models@run.time}.
\newblock \bibinfo{journal}{\emph{{IEEE} Computer}} \bibinfo{volume}{42},
  \bibinfo{number}{10} (\bibinfo{year}{2009}), \bibinfo{pages}{22--27}.
\newblock


\bibitem[\protect\citeauthoryear{Brottier, Fleurey, Steel, Baudry, and
  Le~Traon}{Brottier et~al\mbox{.}}{2006}]%
        {Brottier2006}
\bibfield{author}{\bibinfo{person}{Erwan Brottier}, \bibinfo{person}{Franck
  Fleurey}, \bibinfo{person}{Jim Steel}, \bibinfo{person}{Benoit Baudry}, {and}
  \bibinfo{person}{Yves Le~Traon}.} \bibinfo{year}{2006}\natexlab{}.
\newblock \showarticletitle{Metamodel-based test generation for model
  transformations: an algorithm and a tool}. In \bibinfo{booktitle}{\emph{2006
  17th International Symposium on Software Reliability Engineering}}.
  \bibinfo{pages}{85--94}.
\newblock


\bibitem[\protect\citeauthoryear{B{\'{u}}r, Szil{\'{a}}gyi, V{\"{o}}r{\"{o}}s,
  and Varr{\'{o}}}{B{\'{u}}r et~al\mbox{.}}{2018}]%
        {FASE2018}
\bibfield{author}{\bibinfo{person}{M{\'{a}}rton B{\'{u}}r},
  \bibinfo{person}{G{\'{a}}bor Szil{\'{a}}gyi}, \bibinfo{person}{Andr{\'{a}}s
  V{\"{o}}r{\"{o}}s}, {and} \bibinfo{person}{D{\'{a}}niel Varr{\'{o}}}.}
  \bibinfo{year}{2018}\natexlab{}.
\newblock \showarticletitle{Distributed graph queries for runtime monitoring of
  cyber-physical systems}.
\newblock In \bibinfo{booktitle}{\emph{LNCS}}. Vol.~\bibinfo{volume}{10802}.
  \bibinfo{pages}{111--128}.
\newblock
\showISBNx{9783319893624}
\showISSN{16113349}


\bibitem[\protect\citeauthoryear{Burmester, Giese, Hirsch, and
  Schilling}{Burmester et~al\mbox{.}}{2004}]%
        {Burmester2004}
\bibfield{author}{\bibinfo{person}{Sven Burmester}, \bibinfo{person}{Holger
  Giese}, \bibinfo{person}{Martin Hirsch}, {and} \bibinfo{person}{Daniela
  Schilling}.} \bibinfo{year}{2004}\natexlab{}.
\newblock \showarticletitle{Incremental design and formal verification with
  UML/RT in the FUJABA real-time tool suite}. In
  \bibinfo{booktitle}{\emph{Proceedings of the International Workshop on
  Specification and Validation of UML Models for Real Time and Embedded
  Systems, SVERTS2004}}. Citeseer.
\newblock


\bibitem[\protect\citeauthoryear{Burmester, Giese, Seibel, and Tichy}{Burmester
  et~al\mbox{.}}{2005}]%
        {Burmester2005worst}
\bibfield{author}{\bibinfo{person}{Sven Burmester}, \bibinfo{person}{Holger
  Giese}, \bibinfo{person}{Andreas Seibel}, {and} \bibinfo{person}{Matthias
  Tichy}.} \bibinfo{year}{2005}\natexlab{}.
\newblock \showarticletitle{Worst-case execution time optimization of story
  patterns for hard real-time systems}. In \bibinfo{booktitle}{\emph{3rd
  International Fujaba Days}}. \bibinfo{pages}{71--78}.
\newblock


\bibitem[\protect\citeauthoryear{Cass{\'e} and Sainrat}{Cass{\'e} and
  Sainrat}{2006}]%
        {Hugues2006}
\bibfield{author}{\bibinfo{person}{Hugues Cass{\'e}} {and}
  \bibinfo{person}{Pascal Sainrat}.} \bibinfo{year}{2006}\natexlab{}.
\newblock \showarticletitle{{OTAWA}, a framework for experimenting {WCET}
  computations}.
\newblock \bibinfo{journal}{\emph{3rd European Congress on Embedded Real-Time}}
  \bibinfo{number}{January} (\bibinfo{year}{2006}), \bibinfo{pages}{1--8}.
\newblock


\bibitem[\protect\citeauthoryear{Choi and Kim}{Choi and Kim}{1996}]%
        {Kong-RimChoi1996}
\bibfield{author}{\bibinfo{person}{Kong-Rim Choi} {and}
  \bibinfo{person}{Kyung-Chang Kim}.} \bibinfo{year}{1996}\natexlab{}.
\newblock \showarticletitle{T*-tree: a main memory database index structure for
  real time applications}. In \bibinfo{booktitle}{\emph{3rd International
  Workshop on Real-Time Computing Systems and Applications}}.
  \bibinfo{pages}{81--88}.
\newblock
\showISBNx{0-8186-7626-4}


\bibitem[\protect\citeauthoryear{Chu and Jaffar}{Chu and Jaffar}{2011}]%
        {Chu2011}
\bibfield{author}{\bibinfo{person}{Duc~Hiep Chu} {and} \bibinfo{person}{Joxan
  Jaffar}.} \bibinfo{year}{2011}\natexlab{}.
\newblock \showarticletitle{{Symbolic simulation on complicated loops for WCET
  path analysis}}. In \bibinfo{booktitle}{\emph{Proc. of the 9th ACM
  International Conference on Embedded Software, EMSOFT'11}}.
  \bibinfo{publisher}{IEEE}, \bibinfo{pages}{319--328}.
\newblock
\showISBNx{9781450307147}


\bibitem[\protect\citeauthoryear{Claire et~al\mbox{.}}{Claire
  et~al\mbox{.}}{2017}]%
        {Maiza2017}
\bibfield{author}{\bibinfo{person}{Maiza Claire} {et~al\mbox{.}}}
  \bibinfo{year}{2017}\natexlab{}.
\newblock \showarticletitle{{The W-SEPT Project: Towards Semantic-Aware WCET
  Estimation}}. In \bibinfo{booktitle}{\emph{17th International Workshop on
  Worst-Case Execution Time Analysis (WCET 2017)}}, Vol.~\bibinfo{volume}{57}.
  \bibinfo{publisher}{Schloss Dagstuhl--Leibniz-Zentrum fuer Informatik},
  \bibinfo{address}{Dagstuhl, Germany}, \bibinfo{pages}{9:1--9:13}.
\newblock
\showISBNx{978-3-95977-057-6}
\showISSN{2190-6807}


\bibitem[\protect\citeauthoryear{Colin and Bernat}{Colin and Bernat}{2002}]%
        {Colin2002}
\bibfield{author}{\bibinfo{person}{Antoine Colin} {and}
  \bibinfo{person}{Guillem Bernat}.} \bibinfo{year}{2002}\natexlab{}.
\newblock \showarticletitle{Scope-tree: A program representation for symbolic
  worst-case execution time analysis}. In \bibinfo{booktitle}{\emph{Proceedings
  14th Euromicro Conference on Real-Time Systems. Euromicro RTS 2002}}. IEEE,
  \bibinfo{pages}{50--59}.
\newblock


\bibitem[\protect\citeauthoryear{{Cucu-Grosjean}, {Santinelli}, {Houston},
  {Lo}, {Vardanega}, {Kosmidis}, {Abella}, {Mezzetti}, {Quiñones}, and
  {Cazorla}}{{Cucu-Grosjean} et~al\mbox{.}}{2012}]%
        {Grosjean2012}
\bibfield{author}{\bibinfo{person}{L. {Cucu-Grosjean}}, \bibinfo{person}{L.
  {Santinelli}}, \bibinfo{person}{M. {Houston}}, \bibinfo{person}{C. {Lo}},
  \bibinfo{person}{T. {Vardanega}}, \bibinfo{person}{L. {Kosmidis}},
  \bibinfo{person}{J. {Abella}}, \bibinfo{person}{E. {Mezzetti}},
  \bibinfo{person}{E. {Quiñones}}, {and} \bibinfo{person}{F.~J. {Cazorla}}.}
  \bibinfo{year}{2012}\natexlab{}.
\newblock \showarticletitle{Measurement-Based Probabilistic Timing Analysis for
  Multi-path Programs}. In \bibinfo{booktitle}{\emph{2012 24th Euromicro
  Conference on Real-Time Systems}}. \bibinfo{pages}{91--101}.
\newblock


\bibitem[\protect\citeauthoryear{Dou, Bianculli, and Briand}{Dou
  et~al\mbox{.}}{2018}]%
        {BriandMonitors}
\bibfield{author}{\bibinfo{person}{Wei Dou}, \bibinfo{person}{Domenico
  Bianculli}, {and} \bibinfo{person}{Lionel Briand}.}
  \bibinfo{year}{2018}\natexlab{}.
\newblock \showarticletitle{Model-Driven Trace Diagnostics for Pattern-Based
  Temporal Specifications}. In \bibinfo{booktitle}{\emph{21th ACM/IEEE
  International Conference on Model Driven Engineering Languages and Systems}}.
  \bibinfo{pages}{278--288}.
\newblock
\showISBNx{9781450349499}


\bibitem[\protect\citeauthoryear{Drusinsky}{Drusinsky}{2000}]%
        {Drusinsky2000}
\bibfield{author}{\bibinfo{person}{Doron Drusinsky}.}
  \bibinfo{year}{2000}\natexlab{}.
\newblock \showarticletitle{{The temporal rover and the ATG rover}}.
\newblock \bibinfo{journal}{\emph{Lecture Notes in Computer Science (including
  subseries Lecture Notes in Artificial Intelligence and Lecture Notes in
  Bioinformatics)}}  \bibinfo{volume}{1885} (\bibinfo{year}{2000}),
  \bibinfo{pages}{323--330}.
\newblock
\showISBNx{3540410309}
\showISSN{16113349}


\bibitem[\protect\citeauthoryear{Emery}{Emery}{2011}]%
        {Emery2011}
\bibfield{author}{\bibinfo{person}{Daniel Emery}.}
  \bibinfo{year}{2011}\natexlab{}.
\newblock \showarticletitle{Headways on high speed lines}. In
  \bibinfo{booktitle}{\emph{9th World Congress on Railway Research}}.
  \bibinfo{pages}{22--26}.
\newblock


\bibitem[\protect\citeauthoryear{Ermedahl, Sandberg, Gustafsson, Bygde, and
  Lisper}{Ermedahl et~al\mbox{.}}{2007}]%
        {Ermedahl2007loop}
\bibfield{author}{\bibinfo{person}{Andreas Ermedahl}, \bibinfo{person}{Christer
  Sandberg}, \bibinfo{person}{Jan Gustafsson}, \bibinfo{person}{Stefan Bygde},
  {and} \bibinfo{person}{Bj{\"o}rn Lisper}.} \bibinfo{year}{2007}\natexlab{}.
\newblock \showarticletitle{Loop bound analysis based on a combination of
  program slicing, abstract interpretation, and invariant analysis}. In
  \bibinfo{booktitle}{\emph{7th International Workshop on Worst-Case Execution
  Time Analysis (WCET'07)}}. Schloss Dagstuhl-Leibniz-Zentrum f{\"u}r
  Informatik.
\newblock


\bibitem[\protect\citeauthoryear{Ferdinand and Heckmann}{Ferdinand and
  Heckmann}{2004}]%
        {Ferdinand2004}
\bibfield{author}{\bibinfo{person}{Christian Ferdinand} {and}
  \bibinfo{person}{Reinhold Heckmann}.} \bibinfo{year}{2004}\natexlab{}.
\newblock \showarticletitle{aiT: Worst-Case Execution Time Prediction by Static
  Program Analysis}. In \bibinfo{booktitle}{\emph{Building the Information
  Society}}, \bibfield{editor}{\bibinfo{person}{Ren{\`e} Jacquart}} (Ed.).
  \bibinfo{publisher}{Springer US}, \bibinfo{address}{Boston, MA},
  \bibinfo{pages}{377--383}.
\newblock
\showISBNx{978-1-4020-8157-6}


\bibitem[\protect\citeauthoryear{Fischer, Niere, Torunski, and
  Z{\"u}ndorf}{Fischer et~al\mbox{.}}{1998}]%
        {Fischer1998}
\bibfield{author}{\bibinfo{person}{Thorsten Fischer}, \bibinfo{person}{J{\"o}rg
  Niere}, \bibinfo{person}{Lars Torunski}, {and} \bibinfo{person}{Albert
  Z{\"u}ndorf}.} \bibinfo{year}{1998}\natexlab{}.
\newblock \showarticletitle{Story diagrams: A new graph rewrite language based
  on the unified modeling language and java}. In
  \bibinfo{booktitle}{\emph{International Workshop on Theory and Application of
  Graph Transformations}}. Springer, \bibinfo{pages}{296--309}.
\newblock


\bibitem[\protect\citeauthoryear{Fleurey, Steel, and Baudry}{Fleurey
  et~al\mbox{.}}{2004}]%
        {Fleurey2004}
\bibfield{author}{\bibinfo{person}{Franck Fleurey}, \bibinfo{person}{Jim
  Steel}, {and} \bibinfo{person}{Benoit Baudry}.}
  \bibinfo{year}{2004}\natexlab{}.
\newblock \showarticletitle{Validation in model-driven engineering: testing
  model transformations}. In \bibinfo{booktitle}{\emph{Proceedings. 2004 First
  International Workshop on Model, Design and Validation, 2004.}}
  \bibinfo{pages}{29--40}.
\newblock


\bibitem[\protect\citeauthoryear{Gallagher}{Gallagher}{2006}]%
        {Gallagher2006}
\bibfield{author}{\bibinfo{person}{Brian Gallagher}.}
  \bibinfo{year}{2006}\natexlab{}.
\newblock \showarticletitle{Matching structure and semantics: A survey on
  graph-based pattern matching}.
\newblock \bibinfo{journal}{\emph{AAAI FS}}  \bibinfo{volume}{6}
  (\bibinfo{year}{2006}), \bibinfo{pages}{45--53}.
\newblock


\bibitem[\protect\citeauthoryear{Giese, Tichy, Burmester, Sch{\"{a}}fer, and
  Flake}{Giese et~al\mbox{.}}{2003}]%
        {Giese2003}
\bibfield{author}{\bibinfo{person}{Holger Giese}, \bibinfo{person}{Matthias
  Tichy}, \bibinfo{person}{Sven Burmester}, \bibinfo{person}{Wilhelm
  Sch{\"{a}}fer}, {and} \bibinfo{person}{Stephan Flake}.}
  \bibinfo{year}{2003}\natexlab{}.
\newblock \showarticletitle{{Towards the compositional verification of
  real-time UML designs}}.
\newblock \bibinfo{journal}{\emph{Proceedings of the ACM SIGSOFT Symposium on
  the Foundations of Software Engineering}} (\bibinfo{year}{2003}),
  \bibinfo{pages}{38--47}.
\newblock
\showISBNx{1581137435}


\bibitem[\protect\citeauthoryear{Gustafsson, Ermedahl, Sandberg, and
  Lisper}{Gustafsson et~al\mbox{.}}{2006}]%
        {Gustafsson2006}
\bibfield{author}{\bibinfo{person}{Jan Gustafsson}, \bibinfo{person}{Andreas
  Ermedahl}, \bibinfo{person}{Christer Sandberg}, {and}
  \bibinfo{person}{Bj{\"{o}}rn Lisper}.} \bibinfo{year}{2006}\natexlab{}.
\newblock \showarticletitle{{Automatic derivation of loop bounds and infeasible
  paths for WCET analysis using abstract execution}}.
\newblock \bibinfo{journal}{\emph{Proc. of Real-Time Systems Symposium}}
  (\bibinfo{year}{2006}), \bibinfo{pages}{57--66}.
\newblock
\showISBNx{0769527612}
\showISSN{10528725}


\bibitem[\protect\citeauthoryear{Hansen, Hissam, and Moreno}{Hansen
  et~al\mbox{.}}{2009}]%
        {Hansen2009}
\bibfield{author}{\bibinfo{person}{Jeffery Hansen}, \bibinfo{person}{Scott
  Hissam}, {and} \bibinfo{person}{Gabriel~A Moreno}.}
  \bibinfo{year}{2009}\natexlab{}.
\newblock \showarticletitle{Statistical-based wcet estimation and validation}.
  In \bibinfo{booktitle}{\emph{9th International Workshop on Worst-Case
  Execution Time Analysis}}. Schloss Dagstuhl-Leibniz-Zentrum f{\"u}r
  Informatik.
\newblock


\bibitem[\protect\citeauthoryear{Hartmann, Fouquet, Moawad, Rouvoy, and
  Le~Traon}{Hartmann et~al\mbox{.}}{2019}]%
        {Hartmann2019greycat}
\bibfield{author}{\bibinfo{person}{Thomas Hartmann},
  \bibinfo{person}{Fran{\c{c}}ois Fouquet}, \bibinfo{person}{Assaad Moawad},
  \bibinfo{person}{Romain Rouvoy}, {and} \bibinfo{person}{Yves Le~Traon}.}
  \bibinfo{year}{2019}\natexlab{}.
\newblock \showarticletitle{GREYCAT: Efficient what-if analytics for data in
  motion at scale}.
\newblock \bibinfo{journal}{\emph{Information Systems}}  \bibinfo{volume}{83}
  (\bibinfo{year}{2019}), \bibinfo{pages}{101--117}.
\newblock


\bibitem[\protect\citeauthoryear{Havelund}{Havelund}{2015}]%
        {Havelund2015}
\bibfield{author}{\bibinfo{person}{Klaus Havelund}.}
  \bibinfo{year}{2015}\natexlab{}.
\newblock \showarticletitle{Rule-based runtime verification revisited}.
\newblock \bibinfo{journal}{\emph{Int. J. Software Tools Technol. Trans.}}
  \bibinfo{volume}{17}, \bibinfo{number}{2} (\bibinfo{year}{2015}),
  \bibinfo{pages}{143--170}.
\newblock
\showISSN{1433-2779}


\bibitem[\protect\citeauthoryear{Havelund and Rosu}{Havelund and Rosu}{2002}]%
        {Havelund2002}
\bibfield{author}{\bibinfo{person}{Klaus Havelund} {and}
  \bibinfo{person}{Grigore Rosu}.} \bibinfo{year}{2002}\natexlab{}.
\newblock \showarticletitle{Synthesizing Monitors for Safety Properties}.
\newblock In \bibinfo{booktitle}{\emph{LNCS}}. Vol.~\bibinfo{volume}{2280}.
  \bibinfo{pages}{342--356}.
\newblock
\showISBNx{3540434194}
\showISSN{03029743}


\bibitem[\protect\citeauthoryear{Herter and Reineke}{Herter and
  Reineke}{2009}]%
        {Herter2009}
\bibfield{author}{\bibinfo{person}{J{\"o}rg Herter} {and} \bibinfo{person}{Jan
  Reineke}.} \bibinfo{year}{2009}\natexlab{}.
\newblock \showarticletitle{Making Dynamic Memory Allocation Static to Support
  WCET Analysis}. In \bibinfo{booktitle}{\emph{9th International Workshop on
  Worst-Case Execution Time Analysis (WCET'09)}}.
\newblock


\bibitem[\protect\citeauthoryear{Hou, Ozsoyoglu, and Taneja}{Hou
  et~al\mbox{.}}{1989}]%
        {Hou1989}
\bibfield{author}{\bibinfo{person}{Wen-Chi Hou}, \bibinfo{person}{Gultekin
  Ozsoyoglu}, {and} \bibinfo{person}{Baldeo~K. Taneja}.}
  \bibinfo{year}{1989}\natexlab{}.
\newblock \showarticletitle{Processing Aggregate Relational Queries with Hard
  Time Constraints}.
\newblock \bibinfo{journal}{\emph{SIGMOD Rec.}} (\bibinfo{year}{1989}), 10.
\newblock
\showISSN{0163-5808}


\bibitem[\protect\citeauthoryear{Jackson, Simko, and Sztipanovits}{Jackson
  et~al\mbox{.}}{2013}]%
        {Jackson2013}
\bibfield{author}{\bibinfo{person}{Ethan~K. Jackson}, \bibinfo{person}{Gabor
  Simko}, {and} \bibinfo{person}{Janos Sztipanovits}.}
  \bibinfo{year}{2013}\natexlab{}.
\newblock \showarticletitle{Diversely enumerating system-level architectures}.
  In \bibinfo{booktitle}{\emph{ACM International Conference on Embedded
  Software}}. \bibinfo{publisher}{IEEE}.
\newblock


\bibitem[\protect\citeauthoryear{Jantsch, Dutt, and Rahmani}{Jantsch
  et~al\mbox{.}}{2017}]%
        {Jantsch2017}
\bibfield{author}{\bibinfo{person}{Axel Jantsch}, \bibinfo{person}{Nikil Dutt},
  {and} \bibinfo{person}{Amir~M. Rahmani}.} \bibinfo{year}{2017}\natexlab{}.
\newblock \showarticletitle{Self-awareness in systems on chip—A survey}.
\newblock \bibinfo{journal}{\emph{IEEE Design \& Test}} \bibinfo{volume}{34},
  \bibinfo{number}{6} (\bibinfo{year}{2017}), \bibinfo{pages}{8--26}.
\newblock


\bibitem[\protect\citeauthoryear{J{\"{u}}rjens}{J{\"{u}}rjens}{2003}]%
        {Jurjens2003}
\bibfield{author}{\bibinfo{person}{Jan J{\"{u}}rjens}.}
  \bibinfo{year}{2003}\natexlab{}.
\newblock \showarticletitle{{Developing safety-critical systems with UML}}.
\newblock \bibinfo{journal}{\emph{Lecture Notes in Computer Science (including
  subseries Lecture Notes in Artificial Intelligence and Lecture Notes in
  Bioinformatics)}}  \bibinfo{volume}{2863} (\bibinfo{year}{2003}),
  \bibinfo{pages}{360--372}.
\newblock
\showISBNx{9783540202431}
\showISSN{16113349}


\bibitem[\protect\citeauthoryear{Knoop, Kov{\'{a}}cs, and Zwirchmayr}{Knoop
  et~al\mbox{.}}{2013}]%
        {Knoop2013}
\bibfield{author}{\bibinfo{person}{Jens Knoop}, \bibinfo{person}{Laura
  Kov{\'{a}}cs}, {and} \bibinfo{person}{Jakob Zwirchmayr}.}
  \bibinfo{year}{2013}\natexlab{}.
\newblock \showarticletitle{{WCET squeezing}}.
\newblock  (\bibinfo{year}{2013}), \bibinfo{pages}{161}.
\newblock
\showISBNx{9781450320580}


\bibitem[\protect\citeauthoryear{Kozyrev}{Kozyrev}{2016}]%
        {Kozyrev2016}
\bibfield{author}{\bibinfo{person}{V.~P. Kozyrev}.}
  \bibinfo{year}{2016}\natexlab{}.
\newblock \showarticletitle{Estimation of the execution time in real-time
  systems}.
\newblock \bibinfo{journal}{\emph{Programming and Computer Software}}
  \bibinfo{volume}{42}, \bibinfo{number}{1} (\bibinfo{year}{2016}),
  \bibinfo{pages}{41--48}.
\newblock
\showISSN{0361-7688}


\bibitem[\protect\citeauthoryear{{Law} and {Bate}}{{Law} and {Bate}}{2016}]%
        {Law2016}
\bibfield{author}{\bibinfo{person}{S. {Law}} {and} \bibinfo{person}{I.
  {Bate}}.} \bibinfo{year}{2016}\natexlab{}.
\newblock \showarticletitle{Achieving Appropriate Test Coverage for Reliable
  Measurement-Based Timing Analysis}. In \bibinfo{booktitle}{\emph{2016 28th
  Euromicro Conference on Real-Time Systems (ECRTS)}}.
  \bibinfo{pages}{189--199}.
\newblock


\bibitem[\protect\citeauthoryear{Li, Liang, Mitra, and Roychoudhury}{Li
  et~al\mbox{.}}{2007}]%
        {Li2007}
\bibfield{author}{\bibinfo{person}{Xianfeng Li}, \bibinfo{person}{Yun Liang},
  \bibinfo{person}{Tulika Mitra}, {and} \bibinfo{person}{Abhik Roychoudhury}.}
  \bibinfo{year}{2007}\natexlab{}.
\newblock \showarticletitle{Chronos: A timing analyzer for embedded software}.
\newblock \bibinfo{journal}{\emph{Science of Computer Programming}}
  \bibinfo{volume}{69}, \bibinfo{number}{1-3} (\bibinfo{year}{2007}),
  \bibinfo{pages}{56--67}.
\newblock


\bibitem[\protect\citeauthoryear{Li, Zhou, Guo, Wang, and Zhang}{Li
  et~al\mbox{.}}{2019}]%
        {Li2019}
\bibfield{author}{\bibinfo{person}{Xiaocui Li}, \bibinfo{person}{Zhangbing
  Zhou}, \bibinfo{person}{Junqi Guo}, \bibinfo{person}{Shangguang Wang}, {and}
  \bibinfo{person}{Junsheng Zhang}.} \bibinfo{year}{2019}\natexlab{}.
\newblock \showarticletitle{Aggregated multi-attribute query processing in edge
  computing for industrial IoT applications}.
\newblock \bibinfo{journal}{\emph{Computer Networks}}  \bibinfo{volume}{151}
  (\bibinfo{year}{2019}), \bibinfo{pages}{114--123}.
\newblock
\showISSN{1389-1286}


\bibitem[\protect\citeauthoryear{Li and Malik}{Li and Malik}{1997}]%
        {Li1995}
\bibfield{author}{\bibinfo{person}{Y.-T.S. Li} {and} \bibinfo{person}{Sharad
  Malik}.} \bibinfo{year}{1997}\natexlab{}.
\newblock \showarticletitle{{Performance analysis of embedded software using
  implicit path enumeration}}.
\newblock \bibinfo{journal}{\emph{IEEE T. Comput. Aid. D.}}
  \bibinfo{volume}{16}, \bibinfo{number}{12} (\bibinfo{year}{1997}),
  \bibinfo{pages}{1477--1487}.
\newblock
\showISSN{02780070}


\bibitem[\protect\citeauthoryear{Lim et~al\mbox{.}}{Lim et~al\mbox{.}}{1995}]%
        {Lim1995}
\bibfield{author}{\bibinfo{person}{Sung-Soo Lim} {et~al\mbox{.}}}
  \bibinfo{year}{1995}\natexlab{}.
\newblock \showarticletitle{An accurate worst case timing analysis for RISC
  processors}.
\newblock \bibinfo{journal}{\emph{IEEE Transactions on Software Engineering}}
  \bibinfo{volume}{21}, \bibinfo{number}{7} (\bibinfo{year}{1995}),
  \bibinfo{pages}{593--604}.
\newblock


\bibitem[\protect\citeauthoryear{Lisper}{Lisper}{2014}]%
        {Lisper2014}
\bibfield{author}{\bibinfo{person}{Bj{\"o}rn Lisper}.}
  \bibinfo{year}{2014}\natexlab{}.
\newblock \showarticletitle{SWEET--a tool for WCET flow analysis}. In
  \bibinfo{booktitle}{\emph{International Symposium On Leveraging Applications
  of Formal Methods, Verification and Validation}}. Springer,
  \bibinfo{pages}{482--485}.
\newblock


\bibitem[\protect\citeauthoryear{Martin, Alt, Wilhelm, and Ferdinand}{Martin
  et~al\mbox{.}}{1998}]%
        {Martin1998}
\bibfield{author}{\bibinfo{person}{Florian Martin}, \bibinfo{person}{Martin
  Alt}, \bibinfo{person}{Reinhard Wilhelm}, {and} \bibinfo{person}{Christian
  Ferdinand}.} \bibinfo{year}{1998}\natexlab{}.
\newblock \showarticletitle{{Analysis of loops}}.
\newblock \bibinfo{journal}{\emph{Lecture Notes in Computer Science}}
  \bibinfo{volume}{1383} (\bibinfo{year}{1998}), \bibinfo{pages}{80--94}.
\newblock
\showISBNx{3540643044}
\showISSN{16113349}


\bibitem[\protect\citeauthoryear{Marussy, Semer{\'a}th, and Varr{\'o}}{Marussy
  et~al\mbox{.}}{2020}]%
        {Marussy2020}
\bibfield{author}{\bibinfo{person}{Krist{\'o}f Marussy},
  \bibinfo{person}{Oszk{\'a}r Semer{\'a}th}, {and} \bibinfo{person}{D{\'a}niel
  Varr{\'o}}.} \bibinfo{year}{2020}\natexlab{}.
\newblock \showarticletitle{Automated Generation of Consistent Graph Models
  with Multiplicity Reasoning}.
\newblock \bibinfo{journal}{\emph{IEEE Transactions on Software Engineering}}
  (\bibinfo{year}{2020}).
\newblock
\urldef\tempurl%
\url{https://doi.org/10.1109/TSE.2020.3025732}
\showDOI{\tempurl}


\bibitem[\protect\citeauthoryear{Michais~Famelis}{Michais~Famelis}{2012}]%
        {Famelis2012}
\bibfield{author}{\bibinfo{person}{Marsha~Chechik Michais~Famelis,
  Rick~Salay}.} \bibinfo{year}{2012}\natexlab{}.
\newblock \showarticletitle{Partial models: Towards modeling and reasoning with
  uncertainty}. In \bibinfo{booktitle}{\emph{International Conference on
  Software Engineering}}. \bibinfo{publisher}{IEEE}.
\newblock


\bibitem[\protect\citeauthoryear{Ozsoyoglu and Snodgrass}{Ozsoyoglu and
  Snodgrass}{1995}]%
        {Ozsoyoglu1995}
\bibfield{author}{\bibinfo{person}{Gultekin Ozsoyoglu} {and}
  \bibinfo{person}{Richard~T. Snodgrass}.} \bibinfo{year}{1995}\natexlab{}.
\newblock \showarticletitle{Temporal and real-time databases: A survey}.
\newblock \bibinfo{journal}{\emph{IEEE Trans. Knowl. Data Eng.}}
  \bibinfo{volume}{7}, \bibinfo{number}{4} (\bibinfo{year}{1995}).
\newblock


\bibitem[\protect\citeauthoryear{Pek, Manzinger, Koschi, and Althoff}{Pek
  et~al\mbox{.}}{2020}]%
        {Pek2020}
\bibfield{author}{\bibinfo{person}{Christian Pek}, \bibinfo{person}{Stefanie
  Manzinger}, \bibinfo{person}{Markus Koschi}, {and} \bibinfo{person}{Matthias
  Althoff}.} \bibinfo{year}{2020}\natexlab{}.
\newblock \showarticletitle{Using online verification to prevent autonomous
  vehicles from causing accidents}.
\newblock \bibinfo{journal}{\emph{Nature Machine Intelligence}}
  \bibinfo{volume}{2}, \bibinfo{number}{9} (\bibinfo{year}{2020}),
  \bibinfo{pages}{518--528}.
\newblock
\showISSN{25225839}


\bibitem[\protect\citeauthoryear{Pike, Goodloe, Morisset, and Niller}{Pike
  et~al\mbox{.}}{2010}]%
        {Pike2010}
\bibfield{author}{\bibinfo{person}{Lee Pike}, \bibinfo{person}{Alwyn Goodloe},
  \bibinfo{person}{Robin Morisset}, {and} \bibinfo{person}{Sebastian Niller}.}
  \bibinfo{year}{2010}\natexlab{}.
\newblock \showarticletitle{Copilot: A Hard Real-Time Runtime Monitor}.
\newblock In \bibinfo{booktitle}{\emph{LNCS}}. Vol.~\bibinfo{volume}{6418}.
  \bibinfo{pages}{345--359}.
\newblock
\showISBNx{3642166113}
\showISSN{03029743}


\bibitem[\protect\citeauthoryear{Puschner and Schedl}{Puschner and
  Schedl}{1997}]%
        {Puschner1997}
\bibfield{author}{\bibinfo{person}{Peter~P. Puschner} {and}
  \bibinfo{person}{Anton~V. Schedl}.} \bibinfo{year}{1997}\natexlab{}.
\newblock \showarticletitle{Computing Maximum Task Execution Times - A
  Graph-Based Approach}.
\newblock \bibinfo{journal}{\emph{Real-Time Systems}} \bibinfo{volume}{13},
  \bibinfo{number}{1} (\bibinfo{year}{1997}), \bibinfo{pages}{67--91}.
\newblock
\showISSN{09226443}


\bibitem[\protect\citeauthoryear{Rierson}{Rierson}{2017}]%
        {Rierson2017}
\bibfield{author}{\bibinfo{person}{Leanna Rierson}.}
  \bibinfo{year}{2017}\natexlab{}.
\newblock \bibinfo{booktitle}{\emph{Developing Safety-Critical Software}}.
\newblock \bibinfo{publisher}{CRC Press}. 22--27 pages.
\newblock
\showISBNx{9781315218168}
\showISSN{10486690}


\bibitem[\protect\citeauthoryear{Sagiv, Reps, and Wilhelm}{Sagiv
  et~al\mbox{.}}{2002}]%
        {Sagiv2002}
\bibfield{author}{\bibinfo{person}{Mooly Sagiv}, \bibinfo{person}{Thomas Reps},
  {and} \bibinfo{person}{Reinhard Wilhelm}.} \bibinfo{year}{2002}\natexlab{}.
\newblock \showarticletitle{Parametric shape analysis via 3-valued logic}.
\newblock \bibinfo{journal}{\emph{ACM Transactions on Programming Languages and
  Systems}} \bibinfo{volume}{24}, \bibinfo{number}{3} (\bibinfo{year}{2002}),
  \bibinfo{pages}{193--298}.
\newblock


\bibitem[\protect\citeauthoryear{Salay, Famelis, and Chechik}{Salay
  et~al\mbox{.}}{2012}]%
        {Salay2012}
\bibfield{author}{\bibinfo{person}{Rick Salay}, \bibinfo{person}{Michalis
  Famelis}, {and} \bibinfo{person}{Marsha Chechik}.}
  \bibinfo{year}{2012}\natexlab{}.
\newblock \showarticletitle{Language Independent Refinement Using Partial
  Modeling}. In \bibinfo{booktitle}{\emph{FASE}}.
  \bibinfo{publisher}{Springer}.
\newblock


\bibitem[\protect\citeauthoryear{Semer{\'a}th, Farkas, Bergmann, and
  Varr{\'o}}{Semer{\'a}th et~al\mbox{.}}{2020}]%
        {Semerath2020}
\bibfield{author}{\bibinfo{person}{Oszk{\'a}r Semer{\'a}th},
  \bibinfo{person}{Rebeka Farkas}, \bibinfo{person}{G{\'a}bor Bergmann}, {and}
  \bibinfo{person}{D{\'a}niel Varr{\'o}}.} \bibinfo{year}{2020}\natexlab{}.
\newblock \showarticletitle{Diversity of graph models and graph generators in
  mutation testing}.
\newblock \bibinfo{journal}{\emph{International Journal on Software Tools for
  Technology Transfer}} \bibinfo{volume}{22}, \bibinfo{number}{1}
  (\bibinfo{year}{2020}), \bibinfo{pages}{57--78}.
\newblock


\bibitem[\protect\citeauthoryear{Semer{\'a}th, Nagy, and
  Varr{\'o}}{Semer{\'a}th et~al\mbox{.}}{2018}]%
        {Semerath2018}
\bibfield{author}{\bibinfo{person}{Oszk{\'a}r Semer{\'a}th},
  \bibinfo{person}{Andr{\'a}s~Szabolcs Nagy}, {and} \bibinfo{person}{D{\'a}niel
  Varr{\'o}}.} \bibinfo{year}{2018}\natexlab{}.
\newblock \showarticletitle{A graph solver for the automated generation of
  consistent domain-specific models}. In \bibinfo{booktitle}{\emph{40th
  International Conference on Software Engineering}}.
  \bibinfo{pages}{969--980}.
\newblock


\bibitem[\protect\citeauthoryear{Szvetits and Zdun}{Szvetits and Zdun}{2013}]%
        {Szvetits2013}
\bibfield{author}{\bibinfo{person}{Michael Szvetits} {and} \bibinfo{person}{Uwe
  Zdun}.} \bibinfo{year}{2013}\natexlab{}.
\newblock \showarticletitle{Systematic literature review of the objectives,
  techniques, kinds, and architectures of models at runtime}.
\newblock \bibinfo{journal}{\emph{Software {\&} Systems Modeling}}
  \bibinfo{volume}{15}, \bibinfo{number}{1} (\bibinfo{year}{2013}),
  \bibinfo{pages}{31--69}.
\newblock
\showISSN{1619-1366}


\bibitem[\protect\citeauthoryear{Taina and Raatikainen}{Taina and
  Raatikainen}{1996}]%
        {Taina1996}
\bibfield{author}{\bibinfo{person}{Juha Taina} {and} \bibinfo{person}{Kimmo
  Raatikainen}.} \bibinfo{year}{1996}\natexlab{}.
\newblock \showarticletitle{{Rodain: A real-time object-oriented database
  system for telecommunications}}.
\newblock \bibinfo{journal}{\emph{International Conference on Information and
  Knowledge Management, Proceedings}}  \bibinfo{volume}{Part F129290}
  (\bibinfo{year}{1996}), \bibinfo{pages}{10--14}.
\newblock
\showISBNx{0897919483}


\bibitem[\protect\citeauthoryear{Tavcar and Horvath}{Tavcar and
  Horvath}{2019}]%
        {Tavcar2019}
\bibfield{author}{\bibinfo{person}{Joze Tavcar} {and} \bibinfo{person}{Imre
  Horvath}.} \bibinfo{year}{2019}\natexlab{}.
\newblock \showarticletitle{{A review of the principles of designing smart
  cyber-physical systems for run-time adaptation: Learned lessons and open
  issues}}.
\newblock \bibinfo{journal}{\emph{IEEE Trans. Syst. Man Cybern. Syst.}}
  \bibinfo{volume}{49}, \bibinfo{number}{1} (\bibinfo{year}{2019}),
  \bibinfo{pages}{145--158}.
\newblock
\showISSN{21682232}


\bibitem[\protect\citeauthoryear{{The Eclipse Project}}{{The Eclipse
  Project}}{2021}]%
        {EMF}
{The Eclipse Project} \bibinfo{year}{2021}\natexlab{}.
\newblock \bibinfo{booktitle}{\emph{{Eclipse Modeling Framework}}}.
\newblock {The Eclipse Project}.
\newblock
\newblock
\shownote{\url{http://www.eclipse.org/emf}.}


\bibitem[\protect\citeauthoryear{Tichy, Giese, and Seibel}{Tichy
  et~al\mbox{.}}{2006}]%
        {tichy2006story}
\bibfield{author}{\bibinfo{person}{Matthias Tichy}, \bibinfo{person}{Holger
  Giese}, {and} \bibinfo{person}{Andreas Seibel}.}
  \bibinfo{year}{2006}\natexlab{}.
\newblock \showarticletitle{Story diagrams in real-time software}. In
  \bibinfo{booktitle}{\emph{Proc. of the 4th International Fujaba Days}}.
\newblock


\bibitem[\protect\citeauthoryear{Varr{\'o}, Semer{\'a}th, Sz{\'a}rnyas, and
  Horv{\'a}th}{Varr{\'o} et~al\mbox{.}}{2018}]%
        {Varro2018}
\bibfield{author}{\bibinfo{person}{D{\'a}niel Varr{\'o}},
  \bibinfo{person}{Oszk{\'a}r Semer{\'a}th}, \bibinfo{person}{G{\'a}bor
  Sz{\'a}rnyas}, {and} \bibinfo{person}{{\'A}kos Horv{\'a}th}.}
  \bibinfo{year}{2018}\natexlab{}.
\newblock \showarticletitle{Towards the Automated Generation of Consistent,
  Diverse, Scalable and Realistic Graph Models}. In
  \bibinfo{booktitle}{\emph{Graph Transformation, Specifications, and Nets (In
  Memory of Hartmut Ehrig)}}.
\newblock


\bibitem[\protect\citeauthoryear{Varr{\'{o}}, Deckwerth, Wieber, and
  Sch{\"{u}}rr}{Varr{\'{o}} et~al\mbox{.}}{2015}]%
        {Varro2015}
\bibfield{author}{\bibinfo{person}{Gergely Varr{\'{o}}},
  \bibinfo{person}{Frederik Deckwerth}, \bibinfo{person}{Martin Wieber}, {and}
  \bibinfo{person}{Andy Sch{\"{u}}rr}.} \bibinfo{year}{2015}\natexlab{}.
\newblock \showarticletitle{An algorithm for generating model-sensitive search
  plans for pattern matching on {EMF} models}.
\newblock \bibinfo{journal}{\emph{Software {\&} Systems Modeling}}
  (\bibinfo{year}{2015}), \bibinfo{pages}{597--621}.
\newblock
\showISSN{1619-1366}


\bibitem[\protect\citeauthoryear{V{\"{o}}r{\"{o}}s
  et~al\mbox{.}}{V{\"{o}}r{\"{o}}s et~al\mbox{.}}{2018}]%
        {NFM2018}
\bibfield{author}{\bibinfo{person}{Andr{\'{a}}s V{\"{o}}r{\"{o}}s}
  {et~al\mbox{.}}} \bibinfo{year}{2018}\natexlab{}.
\newblock \showarticletitle{{MoDeS3}: Model-Based Demonstrator for Smart and
  Safe Cyber-Physical Systems}.
\newblock In \bibinfo{booktitle}{\emph{NASA Formal Methods}}.
  \bibinfo{pages}{460--467}.
\newblock
\showISBNx{978-3-319-77935-5}


\bibitem[\protect\citeauthoryear{{Wenzel}, {Kirner}, {Rieder}, and
  {Puschner}}{{Wenzel} et~al\mbox{.}}{2005}]%
        {Wenzel2005}
\bibfield{author}{\bibinfo{person}{I. {Wenzel}}, \bibinfo{person}{R. {Kirner}},
  \bibinfo{person}{B. {Rieder}}, {and} \bibinfo{person}{P. {Puschner}}.}
  \bibinfo{year}{2005}\natexlab{}.
\newblock \showarticletitle{Measurement-based worst-case execution time
  analysis}. In \bibinfo{booktitle}{\emph{Third IEEE Workshop on Software
  Technologies for Future Embedded and Ubiquitous Systems (SEUS'05)}}.
  \bibinfo{pages}{7--10}.
\newblock


\bibitem[\protect\citeauthoryear{Wilhelm et~al\mbox{.}}{Wilhelm
  et~al\mbox{.}}{2008}]%
        {Wilhelm2008}
\bibfield{author}{\bibinfo{person}{Reinhard Wilhelm} {et~al\mbox{.}}}
  \bibinfo{year}{2008}\natexlab{}.
\newblock \showarticletitle{{The worst-case execution-time problem-overview of
  methods and survey of tools}}.
\newblock \bibinfo{journal}{\emph{Transactions on Embedded Computing Systems}}
  \bibinfo{volume}{7}, \bibinfo{number}{3} (\bibinfo{year}{2008}).
\newblock
\showISSN{15399087}


\bibitem[\protect\citeauthoryear{Xie, Yu, Zeng, Yang, and Liu}{Xie
  et~al\mbox{.}}{2021}]%
        {Cheng2021}
\bibfield{author}{\bibinfo{person}{Cheng Xie}, \bibinfo{person}{Beibei Yu},
  \bibinfo{person}{Zuoying Zeng}, \bibinfo{person}{Yun Yang}, {and}
  \bibinfo{person}{Qing Liu}.} \bibinfo{year}{2021}\natexlab{}.
\newblock \showarticletitle{Multilayer Internet-of-Things Middleware Based on
  Knowledge Graph}.
\newblock \bibinfo{journal}{\emph{IEEE Internet of Things Journal}}
  \bibinfo{volume}{8}, \bibinfo{number}{4} (\bibinfo{year}{2021}),
  \bibinfo{pages}{2635--2648}.
\newblock


\bibitem[\protect\citeauthoryear{Zhu, Dwyer, and Goddard}{Zhu
  et~al\mbox{.}}{2009}]%
        {Zhu2009}
\bibfield{author}{\bibinfo{person}{Haitao Zhu}, \bibinfo{person}{Matthew~B.
  Dwyer}, {and} \bibinfo{person}{Steve Goddard}.}
  \bibinfo{year}{2009}\natexlab{}.
\newblock \showarticletitle{{Predictable runtime monitoring}}.
\newblock \bibinfo{journal}{\emph{Proceedings - Euromicro Conference on
  Real-Time Systems}} \bibinfo{number}{2} (\bibinfo{year}{2009}),
  \bibinfo{pages}{173--183}.
\newblock
\showISBNx{9780769537245}
\showISSN{10683070}


\end{thebibliography}

\newpage
\appendix
\section{Proof sketches}\label{sec:appendix}
\renewcommand*{\proofname}{Proof sketch}

\bgroup\renewcommand*{\theproposition}{\ref{prop:concrete-model-estimate}}
\begin{proposition}
  Let \(\tau\) be the execution time of the query program \(\texttt{q}\) on the concrete model \(M\),
  \begin{align*}
	\mathit{CL} &= \max \sum_{\lVar{x}_{i} \in \LVars} c_{i} \cdot \lVar{x}_{i} \textit{ subject to } \Scope{\text{IPET}} \text,
	& \mathit{DS}_{M} &= \max \sum_{\lVar{x}_{i} \in \LVars} c_{i} \cdot \lVar{x}_{i} \textit{ subject to } \Scope{\text{IPET}} \cup \Scope{\text{flow}} \text,
  \end{align*}
  where \(\mathit{CL}\) is the classical IPET estimated obtained from \(\texttt{q}\), and \(\mathit{DS}_{M}\) is the domain-specific estimate with flow facts derived from \(M\) using \autoref{alg:precise-flow-facts}. Then \(\tau \le \mathit{DS}_{M} \le \mathit{CL}\).
\end{proposition}\egroup{}

\begin{proof}
  \(\tau \le \mathit{DS}_{M}\) \emph{(safety)}: Consider any execution path \(\pi\) of \texttt{q}. Because \(\mathit{CL}\) is safe, there is a solution \(k_{\pi}\colon \LVars \to \mathbb{Z}\) such that \(k_{\pi}(\lVar{f}(e)) = \pi\#e\) for all edges \(e \in E\) of the CFG of \(\texttt{q}\) and \(k_{\pi} \vDash \Scope{\text{IPET}}\).

  Consider the linear equations in \(\Scope{\text{flow}}\). We have a single linear equation for each basic block \(\mathit{bb} \in \mathit{BB}\). If \(\mathit{bb}\) is a loop header, then every execution of \(\mathit{bb}\) is represented by either a match of \(\psi_{\mathit{bb}}\) of \(\psi'_{\mathit{bb}}\). Thus, \(\sum_{e = \langle n_1, n_2 \rangle \in E, \mathit{tr}(n_1) = \mathit{bb}} \pi\#e = \sum_{e = \langle n_1, n_2 \rangle \in E, \mathit{tr}(n_1) = \mathit{bb}} k_{\pi}(\lVar{f}(e)) = M\#\psi_{\mathit{bb}} + M\#\psi'_{\mathit{bb}}\), which means the corresponding linear equation in \(\Scope{\text{flow}}\) holds.
  Otherwise, every execution of \(\mathit{bb}\) is represented by a match of \(\psi_{\mathit{bb}}\). Thus, \(\sum_{e = \langle n_1, n_2 \rangle \in E, \mathit{tr}(n_1) = \mathit{bb}} \pi\#e = \sum_{e = \langle n_1, n_2 \rangle \in E, \mathit{tr}(n_1) = \mathit{bb}} k_{\pi}(\lVar{f}(e)) = M\#\psi_{\mathit{bb}}\), which means the corresponding linear equation in \(\Scope{\text{flow}}\) holds.

  Therefore, we have \(k_{\pi} \vDash \Scope{\text{IPET}} \cup \Scope{\text{flow}}\). We also have  \(\tau \le \mathit{CL} \le \mathit{DS}_{M}\) using the safety of \(\mathit{CL}\).

  \(\mathit{DS}_{M} \le \mathit{CL}\) \emph{(tightness)}: Assume that \(\mathit{DS}_{M} > \mathit{CL}\). Then there is some \(k \vDash \Scope{\text{IPET}} \cup \Scope{\text{flow}}\) such that \(g(k) = \sum_{\lVar{x}_{i}} k(\lVar{x}_{i}) > \mathit{CL} = g(k^{*})\), where \(k^{*} \vDash \Scope{\text{IPET}}\) is the optimal solution of the classical IPET integer program with \(g(k^{*}) \ge g(k')\) for all \(k' \vDash \Scope{\text{IPET}}\). However, \(k \vDash \Scope{\text{IPET}}\) (because \(\Scope{\text{IPET}} \cup \Scope{\text{flow}} \vDash \Scope{\text{IPET}}\)), which means \(k^{*}\) cannot be optimal. Thus the assumption cannot hold.
\end{proof}

Before we prove \autoref{prop:partial-model-estimate}, we turn our attention to \autoref{prop:witness-model}. Let \(P = \langle \Obj{P}, \Int{P}, \Scope{P} \rangle\) be a partial model, \(\mathcal{T} = \langle \Phi, \lVar{r} \rangle\) be a theory. Moreover, let \(\Scope{\text{IPET}}\), \(\Scope{\text{merge}}\), \(\Scope{P'} = \Scope{P} \cup \Scope{\text{IPET}} \cup \Scope{\text{merge}}\), \(P' = \langle \Obj{P}, \Int{P}, \Scope{P'} \rangle\), \(\Phi' = \Phi \cup \Psi\), \(\mathcal{T}' = \langle \Phi', \lVar{r}' \rangle \) be the IPET linear equations, the merging linear equations, the scope, the partial model, the predicates, and the theory output by \autoref{alg:witness-generation} for the WCET estimation of a graph query program \texttt{q}, respectively. Also let \(g_{\text{IPET}}(k) = \sum_{e \in E} w(e) \cdot \lVar{f}(e) \) denote the objective function of the IPET analysis (for the \(\mathit{CFG} = \langle V, E, s, t, w, \mathit{tr} \rangle\) of \texttt{q} and the function \(\lVar{f}\colon E \to \LVars\)), \(g_{\text{witness}}(M) = \max_{k \vDash \Scope{M}} g_{\text{IPET}}(k)\) denote the objective function of the witness generation task, and let \(\Scope{\text{flow}}(M)\) and \(\mathit{DS}_{M}(M)\) denote the flow facts and the WCET estimate obtained for \texttt{q} by \autoref{alg:precise-flow-facts} for the concrete model \(M\). We will use the following lemmas:

\begin{lemma}\label{lem:solution-original}
  If \(M = \langle \Obj{M}, \Int{M}, \Scope{M} \rangle \in \mathit{solutions}(P', \mathcal{T}')\), then \(M \in \mathit{solutions}(P, \mathcal{T})\).
\end{lemma}

\begin{proof}
  \(M \in \mathit{solutions}(P', \mathcal{T}')\) means that \(P' \refined M\) and \(M \vDash \mathcal{T}'\). We will show that we also have \(P \refined M\) and \(M \vDash \mathcal{T}\).

  \(P \refined M\): Consider the abstraction function \(\abs\colon \Obj{M} \to \Obj{P}\) from \(P' \refined_{\abs} M\). Since \(P\) has the same objects \(\Obj{P}\) and interpretation \(\Int{P}\) as \(P'\), \(\abs\) will serve as the abstraction function for \(P \refined_{\abs} M\), too. Because \(\Scope{P'} = \Scope{P} \cup \Scope{\text{IPET}} \cup \Scope{\text{merge}} \supseteq \Scope{P}\) and \(\Scope{M} \vDash \Scope{P'}\) due to \(P' \refined M\), we also have \(\Scope{M} \vDash \Scope{P}\) as required by Definition~\ref{dfn:refinement}.

  \(M \vDash \mathcal{T}\): Notice that \(\Phi' = \Phi \cup \Psi \supseteq \Phi\). \(M \vDash \mathcal{T}'\) means that \(\Scope{M} \vDash \lVar{r} = M\#\varphi\) for all \(\varphi \in \Phi'\). Thus \(\Scope{M} \vDash \lVar{r} = M\#\varphi\) holds for all \(\varphi \in \Phi \subseteq \Phi'\), which means \(M \vDash \mathcal{T}\).
\end{proof}

\begin{lemma}\label{lem:solution-value}
  If \(M \in \mathit{solutions}(P', \mathcal{T}')\), then \(\mathit{DS}_{M}(M) \ge g_{\text{witness}}(M)\).
\end{lemma}

\begin{proof}
  We need to show that
  \begin{equation*}
    \textstyle\mathit{DS}_{M}(M) = \max_{k \vDash \Scope{\text{IPET}} \cup \Scope{\text{flow}}(M)} g_{\text{IPET}}(k) \ge g_{\text{witness}}(M) = \max_{k \vDash \Scope{M}} g_{\text{IPET}}(k) \text,
  \end{equation*}
  for which it suffices to demonstrate that every solution \(k \vDash \Scope{M}\) of the scope associated with \(M\) is also a solution \(k \vDash \Scope{\text{IPET}} \cup \Scope{\text{flow}}(M)\) of the IPET integer program, i.e., \(\Scope{M} \vDash \Scope{\text{IPET}} \cup \Scope{\text{flow}}(M)\). We proceed by case analysis on the linear equations \(\mathit{eq} \in \Scope{\text{IPET}} \cup \Scope{\text{flow}}(M)\):
  \begin{itemize}
    \item If \(\mathit{eq} \in \Scope{\text{IPET}}\), then \(\Scope{M} \vDash \Scope{P'} = \Scope{P} \cup \Scope{\text{IPET}} \cup \Scope{\text{merge}} \supseteq \Scope{\text{IPET}} \ni \mathit{eq}\) implies \(\Scope{M} \vDash \mathit{eq}\).
    \item For each \(\mathit{eq} = [ \sum_{e = \langle n_1, n_2 \rangle \in E, \mathit{tr}(n_1) = \mathit{bb}} \lVar{f}(e) = M\#\psi_{\mathit{bb}} + M\#\psi'_{\mathit{bb}} ] \in \Scope{\text{flow}}(M)\), where \(\mathit{bb}\) is a loop header, we have \(\Scope{M} \vDash \lVar{r}'(\psi_{\mathit{bb}}) = M\#\psi_{\mathit{bb}}\) and \(\Scope{M} \vDash \lVar{r}'(\psi'_{\mathit{bb}}) = M\#\psi'_{\mathit{bb}}\), because \(M \vDash \mathcal{T}'\). We also have \(\Scope{M} \vDash \Scope{P'} \supseteq \Scope{\text{merge}} \ni [\lVar{r}'(\psi_{\mathit{bb}}) + \lVar{r}'(\psi'_{\mathit{bb}}) - \sum_{e = \langle n_1, n_2 \rangle \in E, \mathit{tr}(n_1) = \mathit{bb}} \lVar{f}(e) = 0]\). Therefore \(\Scope{M} \vDash \mathit{eq}\) follows by substitution and rearranging.
    \item For each \(\mathit{eq} = [ \sum_{e = \langle n_1, n_2 \rangle \in E, \mathit{tr}(n_1) = \mathit{bb}} \lVar{f}(e) = M\#\psi_{\mathit{bb}} ] \in \Scope{\text{flow}}(M)\), where \(\mathit{bb}\) is not a loop header, we have \(\Scope{M} \vDash \lVar{r}'(\psi_{\mathit{bb}}) = M\#\psi_{\mathit{bb}}\), because \(M \vDash \mathcal{T}'\). We also have \(\Scope{M} \vDash \Scope{P'} \supseteq \Scope{\text{merge}} \ni [\lVar{r}'(\psi_{\mathit{bb}}) - \sum_{e = \langle n_1, n_2 \rangle \in E, \mathit{tr}(n_1) = \mathit{bb}} \lVar{f}(e) = 0]\). Therefore \(\Scope{M} \vDash \mathit{eq}\) follows by substitution and rearranging.\qedhere
  \end{itemize}
\end{proof}

\begin{definition}
  The function \(\mathit{rename}\colon \LVars \to \LVars\) is a \emph{renaming of variables} if it is bijective (i.e., there is some \(\mathit{rename}^{-1}\colon \LVars \to \LVars\) such that \(\mathit{rename}^{-1} \circ \mathit{rename}\) is the identity function on \(\LVars\)). A renaming of variables is \emph{stationary w.r.t.} the theory \(\mathcal{T} = \langle \Phi, \lVar{r} \rangle\) if \(\mathit{rename}(\lVar{r}(\varphi)) = \lVar{r}(\varphi)\) for all \(\varphi \in \Phi\). The partial model \(P^{\mathit{rename}} = \langle \Obj{P}, \Int{P}, \Scope{P}^{\mathit{rename}} \rangle\) is a \emph{renaming} of \(P = \langle \Obj{P}, \Int{P}, \Scope{P} \rangle\), where \(\Scope{P}^{\mathit{rename}}\) is obtained from \(\Scope{P}\) by replacing each variable \(\lVar{x} \in \LVars\) with \(\mathit{rename}(\lVar{x})\).
\end{definition}

\begin{lemma}\label{lem:extend-solution}
  If \(M = \langle \Obj{M}, \Int{M}, \Scope{M} \rangle \in \mathit{solutions}(P, \mathcal{T})\), then there is some \(M' = \langle \Obj{M'}, \Int{M'}, \Scope{M'} \rangle \in \mathit{solutions}(P', \mathcal{T}')\) such that \(M^{\mathit{rename}} \refined M'\) and \(\mathit{DS}_{M}(M) = \mathit{DS}_{M}(M') = g_{\text{witness}}(M')\) for some renaming of variables \(\mathit{rename}\colon \LVars \to \LVars\) stationary w.r.t.~\(\mathcal{T}\).
\end{lemma}

\begin{proof}
  W.l.o.g.\ we may assume that the variable of \(\Scope{M}\) as disjoint from those of \(\Scope{\text{IPET}}(M)\) and \(\Scope{\Psi}(M)\) and \(\mathit{rename}\) is the identity function (\(M^{\mathit{rename}} = M\)). Otherwise, we can pick \(\mathit{rename}\) to make the variables of \(\Scope{M}^{\mathit{rename}}\) disjoint from those of \(\Scope{\text{IPET}}\), \(\Scope{\text{flow}}(M)\) and \(\Scope{\Psi}(M)\).

  Let \(\Obj{M'} = \Obj{M}\), \(\Int{M'} = \Int{M}\), and \(\Scope{M'} = \Scope{M} \cup \Scope{\text{IPET}} \cup \Scope{\text{flow}}(M) \cup \Scope{\Psi}(M)\), where \(\Scope{\Psi}(M) = \{ \lVar{r}'(\psi) = M\#\psi \mid \psi \in \Psi\}\).
  We will show that \(M \refined_{\abs_{M}} M'\), where \(\abs_{M}\) is the identity function on \(\Obj{M}\). Since \(\Int{M'} = \Int{M}\), it suffices to show that \(\Scope{M'} \vDash \Scope{M}\), which follows from \(\Scope{M'} \supseteq \Scope{M}\).

  To show that \(M' \in \mathit{solutions}(P', \mathcal{T}')\), we must prove that (i)~\(P \refined M'\), (ii)~\(M' \vDash \mathcal{T}'\), and (iii)~\(M'\) is concrete. (i)~Consider the refinement function \(\abs_{P}\colon \Obj{M} \to \Obj{P}\) from \(P \refined_{\abs_{P}} M\). Since \(\Obj{M} = \Obj{M'}\), \(\Int{M} = \Int{M'}\), and \(\Int{P} = \Int{P'}\), to demonstrate \(P' \refined_{\abs_{P}} M\), it suffices to show that \(\Scope{M'} \vDash \Scope{P'}\). We will show for each linear (in)equality \(\mathit{eq} \in \Scope{P'} = \Scope{P} \cup \Scope{\text{IPET}} \cup \Scope{\text{merge}}\) that \(\Scope{M'} \vDash \mathit{eq}\):
  \begin{itemize}
    \item If \(\mathit{eq} \in \Scope{P}\), then \(\Scope{M'} \vDash \mathit{eq}\) because \(\Scope{M} \vDash \Scope{P} \ni \mathit{eq}\) and \(\Scope{M'} \vDash \Scope{M}\).
    \item If \(\mathit{eq} \in \Scope{\text{IPET}}\), then \(\Scope{M'} \vDash \mathit{eq}\) because \(\Scope{M'} \supseteq \Scope{\text{IPET}} \ni \mathit{eq}\).
    \item If \(\mathit{eq} = [ \lVar{r}'(\psi_{\mathit{bb}}) + \lVar{r}'(\psi'_{\mathit{bb}}) - \sum_{e = \langle n_1, n_2 \rangle \in E, \mathit{tr}(n_1) = \mathit{bb}} \lVar{f}(e) = 0 ] \in \Scope{\text{merge}}\), where \(\mathit{bb}\) is a loop header, then we have \( \{ \lVar{r}'(\psi_{\mathit{bb}}) = M\#\psi_{\mathit{bb}}, \lVar{r}'(\psi'_{\mathit{bb}}) = M\#\psi'_{\mathit{bb}} \} \subseteq \Scope{\Psi}(M)\) and we also have \([ \sum_{e = \langle n_1, n_2 \rangle \in E, \mathit{tr}(n_1) = \mathit{bb}} \lVar{f}(e) = M\#\psi_{\mathit{bb}} + M\#\psi'_{\mathit{bb}} ] \in \Scope{\text{flow}}(M)\). Then \(\Scope{M'} \supseteq \Scope{\text{flow}}(M) \cup \Scope{\Psi}(M) \vDash \mathit{eq}\) follows by substitution and rearranging.
    \item If \(\mathit{eq} = [ \lVar{r}'(\psi_{\mathit{bb}}) - \sum_{e = \langle n_1, n_2 \rangle \in E, \mathit{tr}(n_1) = \mathit{bb}} \lVar{f}(e) =  0 ] \in \Scope{\text{merge}}\), where \(\mathit{bb}\) is a not loop header, then we have \( [ \lVar{r}'(\psi_{\mathit{bb}}) = M\#\psi_{\mathit{bb}} ] \in \Scope{\Psi}(M)\) and we also have \([ \sum_{e = \langle n_1, n_2 \rangle \in E, \mathit{tr}(n_1) = \mathit{bb}} \lVar{f}(e) = M\#\psi_{\mathit{bb}} ] \in \Scope{\text{flow}}(M)\). Then \(\Scope{M'} \supseteq \Scope{\text{flow}}(M) \cup \Scope{\Psi}(M) \vDash \mathit{eq}\) follows by substitution and rearranging.
  \end{itemize}

  (ii)~To see that \(M' \vDash \mathcal{T}\), we also use case analysis for each predicate \(\varphi \in \Phi'\):
  \begin{itemize}
    \item If \(\varphi \in \Phi\), then \(\Scope{M'} \supseteq \Scope{M} \vDash \lVar{r}(\varphi) = M\#\varphi\), because \(M \vDash \mathcal{T}\). This implies \(\Scope{M'} \vDash \lVar{r}'(\varphi) = M'\#\varphi\), because \(\lVar{r}(\varphi) = \lVar{r}'(\varphi)\), and \(M\#\varphi = M'\#\varphi\) due to \(\Int{M} = \Int{M'}\).
    \item If \(\varphi \in \Psi\), then \(\Scope{M'} \supseteq \Scope{\Psi}(M) \vDash \lVar{r}'(\varphi) = M\#\varphi\), which implies \(\Scope{M'} \vDash \lVar{r}'(\varphi) = M'\#\varphi\) due to \(\Int{M} = \Int{M'}\) as above.
  \end{itemize}

  (iii)~Since \(M\) is concrete, \(\Int{M'} = \Int{M}\) contains only \(\true\) and \(\false\) logic values, and \(\Scope{M}\) has some solution~\(k\). To conclude that \(M'\) is concrete, we will construct a solution \(k'\) of \(\Scope{M'}\).

  Consider an execution path \(\pi\) of \texttt{q} on the input model \(M\) and the associated solution \(k_{\pi}\) of \(\Scope{\text{IPET}}\), where \(k_{\pi}(\lVar{f}(e)) = \pi\#e\). Recall from \autoref{prop:concrete-model-estimate} that \(k_{\pi} \vDash \Scope{\text{flow}}(M)\). Now let
  \begin{equation*}
    k'(\lVar{x}) = \begin{cases}
      k_{\pi}(\lVar{x}) & \text{if \(\lVar{x} = \lVar{f}(e)\) for some \(e \in E\),} \\
      M\#\psi & \text{if \(\lVar{x} = \lVar{r}'(\psi)\) for some \(\psi \in \Psi\),} \\
      k(\lVar{x}) & \text{otherwise.}
    \end{cases}
  \end{equation*}
  Recall that \(\{ \lVar{r}'(\psi) \mid \psi \in \Psi \}\) (i.e., the variables appearing in \(\Scope{\Psi}(M)\)) and \(\Ran \lVar{f}\) (i.e., the variables appearing in \(\Scope{\text{IPET}}\) and \(\Scope{\text{flow}}(M)\)) are disjoint from each other and from the variables appearing in \(\Scope{M}\), so this is well-defined. Now \(k' \vDash \Scope{\text{IPET}}\), \(k' \vDash \Scope{\text{flow}}(M)\), \(k' \vDash \Scope{\Psi}(M)\), and \(k' \vDash \Scope{M}\). Thus \(k' \vDash \Scope{M'} = \Scope{M} \cup \Scope{\text{IPET}} \cup \Scope{\text{flow}}(M) \cup \Scope{\Psi}(M)\).

  Notice that \(M\#\psi = M'\#\psi\) for all \(\psi \in \Psi\) implies that \(\Scope{\text{flow}}(M) = \Scope{\text{flow}}(M')\). Therefore,
  \begin{equation*}
   \textstyle \mathit{DS}_{M}(M) = \max_{k \vDash \Scope{\text{IPET}} \cup \Scope{\text{flow}}(M)} g_{\text{IPET}}(k) = \mathit{DS}_{M}(M') = \max_{k \vDash \Scope{\text{IPET}} \cup \Scope{\text{flow}}(M')} g_{\text{IPET}}(k) \text.
  \end{equation*}

  We can transform any solution \(k_{\text{IPET}} \vDash \Scope{\text{IPET}} \cup \Scope{\text{flow}}(M)\) into an equivalent solution \(k_{\text{witness}} \vDash \Scope{M'} = \Scope{M} \cup \Scope{\text{IPET}} \cup \Scope{\text{flow}}(M) \cup \Scope{\Psi}(M)\) with \(g_{\text{IPET}}(k_{\text{IPET}}) = g_{\text{IPET}}(k_{\text{witness}})\) by changing the values of some variables \(\lVar{x} \in \LVars \setminus \Ran \lVar{f}\), since the variables appearing in \(\Scope{M} \cup \Scope{\Psi}(M)\) are disjoint from the variables \(\Ran \lVar{f}\) appearing in \(\Scope{\text{IPET}} \cup \Scope{\text{flow}}(M)\) and all have a zero weight in \(g_{\text{IPET}}\). Therefore,
  \begin{equation*}
   \textstyle \mathit{DS}_{M}(M') = \max_{k \vDash \Scope{\text{IPET}} \cup \Scope{\text{flow}}(M')} g_{\text{IPET}}(k) = g_{\text{witness}}(M') = \max_{k \vDash \Scope{M'}} g_{\text{IPET}}(k) \text.\qedhere
  \end{equation*}
\end{proof}

\bgroup\renewcommand*{\theproposition}{\ref{prop:witness-model}}
\begin{proposition}[Witness model]
  Let \(\mathit{DS}_{M}(M)\) be the domain-specific WCET estimate of a query program \texttt{q} obtained by \autoref{alg:precise-flow-facts} for a concrete model \(M\), \(\mathit{DS}_{P}\) be the domain-specific WCET estimate of \texttt{q} for a partial model \(P\) and theory \(\mathcal{T}\) by \autoref{alg:witness-generation}, and \(M^{*}\) be the witness model for the WCET of \texttt{q}, i.e., the optimal solution of \(\mathit{DS}_{P}\). Then \(M^{*} \in \mathit{solutions}(P, \mathcal{T})\) and \(\mathit{DS}_{M}(M) \le \mathit{DS}_{M}(M^{*}) = \mathit{DS}_{P}\) for all \(M \in \mathit{solutions}(P, \mathcal{T})\).
\end{proposition}\egroup{}

\begin{proof}
  Consider a model \(M^{*} \in \mathit{solutions}(P', \mathcal{T}')\) that maximizes \(g_{\text{witness}}\). By Lemma~\ref{lem:solution-original}, we also have \(M^{*} \in \mathit{solutions}(P, \mathcal{T})\).

  \(\mathit{DS}_{M}(M^{*}) = \mathit{DS}_{P}\): \(\mathit{DS}_{P}\) is defined as \(\max_{M \in \mathit{solutions}(P', \mathcal{T'})} g_{\text{witness}}(M) = g_{\text{witness}}(M^{*})\). By Lemma~\ref{lem:solution-value}, \(\mathit{DS}_{M}(M^{*}) \ge g_{\text{witness}}(M^{*})\), which means \(\mathit{DS}_{M}(M^{*}) \ge g_{\text{witness}}(M^{*}) = \mathit{DS}_{P}\). Not let us apply the construction from Lemma~\ref{lem:extend-solution} to \(M^{*}\) to obtain \(M^{*\prime}\). We have \(\mathit{DS}_{M}(M^{*}) = \mathit{DS}_{M}(M^{*\prime}) = g_{\text{witness}}(M^{*\prime}) \le g_{\text{witness}}(M^{*})\), which concludes the proof.

  \(\mathit{DS}_{M}(M) \le \mathit{DS}_{M}(M^{*})\): Assume the contrary. The we have some \(M' \in \mathit{solutions}(P', \mathcal{T}')\) such that \(g_{\text{withness}}(M^{*}) = \mathit{DS}_{M}(M^{*}) < \mathit{DS}_{M}(M) = \mathit{DS}_{M}(M') = g_{\text{witness}}(M') \) by Lemma~\ref{lem:extend-solution}. This leads to a contradiction, because \(M^{*}\) is a maximizer of \(g_{\text{witness}}\).
\end{proof}

Now we can use \autoref{prop:witness-model} to prove \autoref{prop:partial-model-estimate} as follows.

\bgroup\renewcommand*{\theproposition}{\ref{prop:partial-model-estimate}}
\begin{proposition}[Safety and tightness]
  Let \(\tau(M)\) be the execution time of a query program \(\texttt{q}\) on a concrete model \(M\), \(P\) be partial model, \(\mathcal{T}\) be a theory, and
  \begin{align*}
	\mathit{CL} &= \max \sum_{\lVar{x}_{i} \in \LVars} c_{i} \cdot \lVar{x}_{i} \textit{ subject to } \Scope{\text{IPET}} \text,
	& \mathit{DS}_{P} &= \max \sum_{\lVar{x}_{i} \in \LVars} c_{i} \cdot \lVar{x}_{i} \textit{ subject to } \langle P', \mathcal{T}' \rangle \text,
  \end{align*}
  where \(\mathit{CL}\) is the classical IPET estimated obtained from \(\texttt{q}\), and \(\mathit{DS}_{P}\) is the domain-specific estimate based on the extended graph generation problem form \autoref{alg:witness-generation}. Then \(\tau(M) \le \mathit{DS}_{P} \le \mathit{CL}\) for all \(M \in \mathit{solutions}(P, \mathcal{T})\).
\end{proposition}\egroup{}

\begin{proof}
  \(\tau(M) \le \mathit{DS}_{P}\) \emph{(safety)}: By \autoref{prop:concrete-model-estimate}, \(\tau(M) \le \mathit{DS}_{M}(M)\). Also, by \autoref{prop:witness-model}, \(\mathit{DS}_{M}(M) \le \mathit{DS}_{M}(M^{*}) = \mathit{DS}_{P}\). Thus, \(\tau(M) \le \mathit{DS}_{M}(M) \le \mathit{DS}_{M}(M^{*}) = \mathit{DS}_{P}\).

  \(\mathit{DS}_{P} \le \mathit{CL}\) \emph{(tightness)}: By \autoref{prop:concrete-model-estimate}, \(\mathit{DS}_{M}(M^{*}) \le \mathit{CL}\). Also, by \autoref{prop:witness-model}, \(\mathit{DS}_{P} = \mathit{DS}_{M}(M^{*})\). Thus, \(\mathit{DS}_{P} = \mathit{DS}_{M}(M^{*}) \le \mathit{CL}\).
\end{proof}

\bgroup\renewcommand*{\theproposition}{\ref{prop:refinement-squeezing}}
\begin{proposition}[Tightening by refinement]
  Let \(\mathit{DS}_{P}(P, \mathcal{T})\) denote the domain-specific WCET estimate of a query program \texttt{q} for a partial model \(P\) and theory \(\mathcal{T}\) obtained by \autoref{alg:witness-generation} and \(P \refined Q\). Then \(\mathit{DS}_{P}(Q, \mathcal{T}) \le \mathit{DS}_{P}(P, \mathcal{T})\). In particular, if \(P = P_{\mathit{init}}\) is the initial partial model for a metamodel \(\langle \Sigma, \alpha \rangle\) from \autoref{sec:model-generation}, then we may see that the WCET estimate for any partial model conforming to the metamodel is at least as tight as the \(\mathit{DS}_{\Sigma}\) estimate for the metamodel.
\end{proposition}\egroup{}

\begin{proof}
  The model generation task objective function \(g_{\text{witness}}\) coincides for both \(\mathit{DS}_{P}(P, \mathcal{T})\) and \(\mathit{DS}_{P}(Q, \mathcal{T})\). Thus, to show that
  \begin{equation*}
    \textstyle\mathit{DS}_{P}(P, \mathcal{T}) = \max_{M' \in \mathit{solutions}(P', \mathcal{T})} g_{\text{witness}}(M') \ge \textstyle\mathit{DS}_{P}(Q, \mathcal{T}) = \max_{M' \in \mathit{solutions}(Q', \mathcal{T})} g_{\text{witness}}(M') \text,
  \end{equation*}
  we need to show that every \(M' \in \mathit{solutions}(Q', \mathcal{T})\) also satisfies \(M' \in \mathit{solutions}(P', \mathcal{T})\).

  Consider the partial models \(P' = \langle \Obj{P}, \Int{P}, \Scope{P'} \rangle\) and \(Q' = \langle \Obj{Q}, \Int{Q}, \Scope{Q'} \rangle\) obtained by \autoref{alg:witness-generation} for the respective input partial models, where \(\Scope{Q'} = \Scope{Q} \cup \Scope{\text{IPET}} \cup \Scope{\text{merge}}\) and \(\Scope{P'} = \Scope{P} \cup \Scope{\text{IPET}} \cup \Scope{\text{merge}}\). The systems on linear inequalities \(\Scope{\text{IPET}}\) and \(\Scope{\text{merge}}\) are the same for both cases, since they do not depend on the input partial model.

  We will show that \(P' \refined_{\abs_{Q}} Q'\), where \(\abs_{Q}\colon \Obj{Q} \to \Obj{P}\) is the abstraction function \(P \refined_{\abs_{Q}} Q\) from \(Q\) to \(P\). Because \(P\) and \(P'\), as well as \(Q\) and \(Q'\) have the same objects and interpretations, we only need to show \(\Scope{Q'} \vDash \Scope{P'}\). We already have \(\Scope{Q} \vDash \Scope{P}\) due to \(P \refined_{\abs_{Q}} Q\), so we conclude \(\Scope{Q} \cup \Scope{\text{IPET}} \cup \Scope{\text{merge}} \vDash \Scope{Q} \cup \Scope{\text{IPET}} \cup \Scope{\text{merge}}\).

  Now consider any \(M' \in \mathit{solutions}(Q', \mathcal{T})\). We have \(Q' \refined_{\abs_{M'}} M'\) and \(M' \vDash \mathcal{T}\) by Definition~\ref{dfn:solutions}. Then we also have \(P' \refined_{\abs{M'} \circ \abs{Q'}} M'\) by the associativity of refinement, from which we conclude \(M' \in \mathit{solutions}(P', \mathcal{T})\).
\end{proof}

\end{document}